\documentclass[aps,prd,superscriptaddress,reprint,amsmath,amssymb,nofootinbib,floatfix]{revtex4-2}
\usepackage{graphicx}
\usepackage{dcolumn}
\usepackage[utf8]{inputenc}
\usepackage{xcolor}
\usepackage[unicode=true,colorlinks,linkcolor=blue,anchorcolor=blue,urlcolor=blue,citecolor=blue,breaklinks=true]{hyperref}
\usepackage{epstopdf}
\usepackage{orcidlink}
\newcommand{\ii}{\mathrm{i}\,}
\newcommand{\ee}{\mathrm{e}}
\newcommand{\pararrow}{\mathord{\buildrel{\lower3pt\hbox{$\scriptscriptstyle\leftrightarrow$}}\over {\partial}}} 
\newcommand{\pararrowk}[1]{\mathord{\buildrel{\lower3pt\hbox{$\scriptscriptstyle\leftrightarrow$}}\over {\partial}\hspace*{-0.18em}{}^#1}\hspace*{-0.18em} \,} 
\newcommand{\mytrace}[1]{\langle #1 \rangle} 

\newcommand{\kets}[1]{| #1\rangle}

\usepackage{bm}
\usepackage{slashed}

\newcommand{\qfnu}{\affiliation{College of Physics and Engineering, Qufu Normal University, Qufu 273165, China}}

\newcommand{\hnnu}{\affiliation{Institute of Particle and Nuclear Physics, Henan Normal University, Xinxiang 453007, China}}

\begin{document}
	
	\title{Hadronic decays of possible pseudoscalar $P$-wave $D\bar{D}^\ast/\bar{D}D^\ast$ molecular state}
	
	\author{Shi-Dong Liu\,\orcidlink{0000-0001-9404-5418}}\email{liusd@qfnu.edu.cn}\qfnu
	\author{Qi Wu\,\orcidlink{0000-0002-5979-8569}}\email{wuqi@htu.edu.cn} \hnnu
	\author{Gang Li\,\orcidlink{0000-0002-5227-8296}}\email{gli@qfnu.edu.cn (Corresponding author)} \qfnu
 
	\begin{abstract}
	Recently, the new structure $G(3900)$ observed by the BESIII Collaboration in the $e^+e^-\to D\bar{D}$ was identified to be the $P$-wave $D\bar{D}^\ast/\bar{D}D^\ast$ vector molecular resonance using a unified meson exchange model. Apart from the vector $P$-wave state, a possible pseudoscalar $P$-wave molecular state of the $D\bar{D}^\ast/\bar{D}D^\ast$ [called $G_0(3900)$ for short] was also predicted, which is likely to be observed in future experiments. Within the molecular framework, we calculated the partial decay widths for a series of hadronic decays of the $G_0$, including $G_0\to \omega(\rho^0) J/\psi $, $\pi^+\pi^- \eta_c (1S)$, $\pi^+\pi^- \chi_{c1} (1P)$, and $D^0\bar{D}^0\pi^0$. Under present model parameters, the hidden-charm decay modes are dominated by the $G_0\to\omega J/\psi$ and $G_0\to\pi^+\pi^-\eta_c (1S)$, and the partial widths can reach 1 MeV and 0.1 MeV, respectively. The open-charm channel $G_0\to D^0\bar{D}^0\pi^0$ exhibits a rather small decay rate ($\sim 0.1~\mathrm{keV}$). In terms of our present predictions, we suggest BESIII and Belle II to search for the pseudoscalar $P$-wave $D\bar{D}^\ast / \bar{D}D^\ast$ molecular state with $J^{PC} = 0^{-+}$ in the hidden-charm processes $G_0 \to \omega J/\psi$ or $G_0 \to \pi^+\pi^-\eta_c(1S)$.
	\end{abstract}
	
	\date{\today}
	
	\maketitle
	\section{Introduction}\label{sec:intro}
The discovery of the $X(3872)$ by the Belle Collaboration in 2003 \cite{Belle:2003nnu} represents a significant milestone in the hadron spectroscopy, as it is the first candidate of exotic states that contain heavy quarks.
The $X(3872)$ and subsequently observed exotic states in different experiments, such as the $Z_c(3900)$ \cite{BESIII:2013ris,Belle:2013yex,D0:2018wyb}, $Z_{cs} (3985)$ \cite{BESIII:2020qkh}, $Y$ states (e.g., $Y(4230)$ \cite{BaBar:2005hhc,CLEO:2006ike,Belle:2007dxy,BESIII:2016bnd}, $Y(4500)$ \cite{BESIII:2022joj,BESIII:2022qal,BESIII:2023cmv},  and $Y(4660)$ \cite{Belle:2007umv,BESIII:2023cmv}), $Z_b(10610/10650)$ \cite{Belle:2011aa,Belle:2012glq}, $T_{cc}^+$ \cite{LHCb:2021auc,LHCb:2021vvq}, the $P_c$ family \cite{LHCb:2015yax,LHCb:2019kea,LHCb:2020jpq}, and $X(6900)$ \cite{LHCb:2020bwg,CMS:2023owd,ATLAS:2023bft} challenge our understanding of QCD, but also provide us with many special platforms to get insights into strong interactions. 
These exotic states stimulate many theoretical interpretations, including compact multiquark states, hadronic molecules, hybrids, and threshold effects (see reviews \cite{Wang:2025sic,Chen:2024eaq,Liu:2024uxn,Chen:2016qju,Chen:2022asf,Meng:2022ozq,BESIII:2020nme,Guo:2019twa,Brambilla:2019esw,Liu:2019zoy,Guo:2017jvc,Liu:2013waa,Brambilla:2010cs,Cleven:2013rkf} and references therein). 
However, no single model can fully explain all experimental observations. 
Deciphering the nature of exotic states still needs significant effort in experimental and theoretical aspects.

Among those theoretical interpretations of exotic states, the hadronic molecule model, given the analogy between the nuclei and hadronic molecules and the fact that most experimentally observed exotic states have masses near some hadron-pair threshold, is a popular and natural framework. 
All the exotic states mentioned above have corresponding molecular interpretation \cite{Wang:2025sic,Chen:2024eaq,Liu:2024uxn,Chen:2016qju,Chen:2022asf,Meng:2022ozq,BESIII:2020nme,Guo:2019twa,Brambilla:2019esw,Liu:2019zoy,Guo:2017jvc,Liu:2013waa,Brambilla:2010cs,Cleven:2013rkf,Hanhart:2025bun}. 
Generally, the $S$-wave interaction among hadrons forms a bound state more easily than other higher waves \cite{Du:2016qcr} so that the previously observed exotic resonances were usually regarded as $S$-wave molecules, e.g., the $X(3872)$ as the $D\bar{D}^*/\bar{D}D^*$ \cite{Guo:2014taa,Cleven:2013rkf,Dong:2017gaw}, $Y(4230)$ as the $D_1\bar{D}/D\bar{D}_1$ \cite{Cleven:2013mka,Guo:2013zbw,Li:2013yla,Dong:2017gaw,Wang:2013cya}, $T_{cc}^+$ as the $D^{\ast+}D^0/D^{\ast 0}D^+$ \cite{Du:2021zzh,Feijoo:2021ppq,Meng:2021jnw,Ling:2021bir}, $X(6200)$ as $J/\psi J/\psi$ \cite{Song:2024ykq} the bound state in an $S$ wave. 
However, it is also accepted that the higher-wave, especially the $P$-wave interaction, can make moderate effects on certain observables within the relevant energy region \cite{Du:2016qcr,Lin:2024qcq}.

In 2024, the BESIII Collaboration analyzed the Born cross section for the $e^+e^-\to D\bar{D}$ process with unprecedented precision and found a new structure around $3.9~\mathrm{GeV}$ \cite{BESIII:2024ths}.
This structure, called $G(3900)$, has a mass of $(3872.5\pm 14.2\pm 3.0) ~\mathrm{MeV}$ with a larger width of $(179.7\pm 14.1\pm 7.0)~\mathrm{MeV}$  \cite{BESIII:2024ths}. 
The $G(3900)$ was also observed in early experiments by the BaBar \cite{BaBar:2006qlj,BaBar:2009elc} and Belle \cite{Belle:2007qxm} Collaborations.
In Ref. \cite{Husken:2024hmi}, the $G(3900)$ structure is attributed primarily to interference effects between nearby resonances
and the opening of the $D^\ast\bar{D}$ channel, rather than a genuine resonance. 
A similar argument was also obtained in Refs. \cite{Zhang:2009gy,Zhang:2010zv,Cao:2014qna}.
The global analysis of the physical scattering amplitudes for the processes $e^+e^-\to D\bar{D}$, $D\bar{D}^\ast+c.c$, and $D^\ast \bar{D}^\ast$ by solving the Lippmann-Schwinger equation indicates the $G(3900)$ as a dynamically generated state \cite{Ye:2025ywy}.
On the contrary, in Refs. \cite{Du:2016qcr,Lin:2024qcq,Chen:2025gxe,Nakamura:2023obk,BESIII:2025wlf}, the $G(3900)$ could be identified as the $P$-wave hadronic molecule of the $D\bar{D}^\ast / \bar{DD^\ast}$, namely being a genuine resonance. 
In particular, the authors in Ref. \cite{Lin:2024qcq} established, on a unified meson-exchange model, the $P$-wave resonances by fixing the relatively mature $S$-wave interactions for the states $X(3872)$, $Z_c(3900)$, and $T_{cc}^+$.
Therefore, the existence of the $P$-wave resonances of the $D\bar{D}^\ast / \bar{DD^\ast}$ appears highly reliable. 

The novel scenario adopted in Ref. \cite{Lin:2024qcq} not only identifies the $G(3900)$ as the first $P$-wave $D\bar{D}^\ast / \bar{DD^\ast}$ state, but also predicts other possible $P$-wave hadronic molecules near the $D\bar{D}^\ast$ energy region, such as the pseudoscalar state with the quantum numbers $I^{G}(J^{PC}) = 0^+(0^{-+})$.
This possible pseudoscalar state is also predicted within the framework of the quasipotential Bethe-Salpeter equation \cite{Chen:2025gxe}.
As an analogy of the $G(3900)$, we call the pseudoscalar $D\bar{D}^\ast / \bar{D}D^\ast$ molecule $G_0(3900)$. 
In Ref. \cite{Lin:2024qcq}, the $G_0(3900)$ emerges as a resonance either below the $D\bar{D}^*$ threshold at $\Lambda=0.5$ GeV or above the $D\bar{D}^*$ threshold at $\Lambda=0.6$ GeV. However, it could be a bound state, virtual state, or resonance by varying the cutoff \cite{Chen:2025gxe}. In this work, we aim to investigate the hadronic decays of $G_0(3900)$ under the molecular state assumption.
Thanks to the quantum numbers, the decay mode of the $G_0$ into the open charmed meson pair $D\bar{D}$ is forbidden. 
Thus, in this work, we shall, using an effective Lagrangian approach, study a series of hadronic decays of the possible molecular state $G_0(3900)$, including the processes $G_0\to \omega(\rho^0) J/\psi $, $\pi^+\pi^- \eta_c (1S)$, $\pi^+\pi^- \chi_{c1} (1P)$, $D^0\bar{D}^0\pi^0$. The $G_0$ is regarded as a bound state of the $D\bar{D}^\ast / \bar{D}D^\ast$ in a $P$-wave, whose mass is specified by a binding energy $E_b$: $m_{G_0}= m_D+m_{{\bar D}^\ast} -E_{\mathrm{b}}$.

The rest of this work is organized as follows. We first give, in Sec. \ref{sec:lags}, the Lagrangians we need. Then, in Sec. \ref{sec:results} the numerical results and discussion are described in detail. Finally, a brief summary is given in Sec. \ref{sec:summary}.

\section{Theoretical Framework}\label{sec:lags}
Following the conventions in Ref. \cite{Lin:2024qcq}, the wave function of the $G_0(3900)$ (hereafter abbreviated as $G_0$) as the $D\bar{D}^\ast/\bar{D}D^\ast$ molecular state is of the form
\begin{equation}\label{eq:G0wavefunction}
	\kets{G_0} = \frac{1}{\sqrt{2}}(\kets{D\bar{D}^\ast} - \kets{\bar{D}D^\ast})\,.
\end{equation}
Here $\kets{D\bar{D}^\ast} = (\kets{D^0\bar{D}^{\ast 0}}+\kets{D^+D^{\ast -}})/\sqrt{2}$ for short. This wave function implies that the proportions of the neutral and charged components in the molecule $G_0$ are assumed to be equal. A more general form is expressed as $\kets{G_0} = (\cos\theta \kets{D^0\bar{D}^{\ast 0}} + \sin\theta \kets{D^+ D^{\ast -}} + \mathrm{c.c.})/\sqrt{2}$. The proportion value for the neutral or charged components, being equivalent to the phase angle $\theta$, can be extracted from the experimental measurements for some relevant decays of the $G_0$, for instance $G_0\to\omega(\rho^0) J/\psi$. In the case of the $X(3872)$ treated as an $S$-wave state of $D\bar{D}^\ast/\bar{D}D^\ast$, the recent LHCb measurements of $X(3872)\to\rho^0 J/\psi$ and $\omega J/\psi$ give $\theta = 28.8^\circ$ \cite{LHCb:2022jez,Zhang:2024fxy}. To date, the $G_0$ has only been predicted theoretically and lacks experimental confirmation. The predicted $G_0$ carries the quantum numbers $J^{PC}=0^{-+}$ \cite{Lin:2024qcq} so that we consider its coupling to its components in a $P$-wave and ignore the other possible higher-wave ones. The $G_0$, as the $P$-wave $D\bar{D}^\ast/\bar{D}D^\ast$ molecular state, might have different mass from that of the $X(3872)$, thereby leading to different phase angle from $28.8^\circ$. In this work, we adopt $\theta = 45^\circ$ as an illustrative example. The other phase angles are also possible. The present framework is general and applicable to any phase angle.

The effective Lagrangian could be constructed as \cite{Zhu:2021exs,Ma:2010xx}
\begin{align}\label{eq:Lag4G0}
	\mathcal{L}_{G_0} &=\frac{1}{\sqrt{2}} g_{G_0DD^\ast}G_0 (x) \int\mathrm{d}^4y\Phi(y^2)\nonumber \\
	&\times \big[ D(x+\omega_{D^\ast}y)\pararrowk{\mu}\bar{D}^\ast_\mu (x-\omega_D y) \nonumber\\
	&- \bar{D}(x+\omega_{D^\ast}y)\pararrowk{\mu}D^\ast_\mu (x-\omega_D y)\big]\,,
\end{align}
where $y$ is a relative Jacobi coordinate and $\omega_i=m_i/(m_i+m_j)$. Throughout this work, $m_i$ stands for the mass of the meson specified by the subscript unless otherwise stated.

Specially, the $\Phi(y^2)$ in Eq. \eqref{eq:Lag4G0} is a correlation function to describe the distribution of the $D$ and $\bar {D}^\ast$ in the molecular state $G_0$, and to render the Feynman diagram's ultraviolet finite; its Fourier transform reads \cite{Dong:2017gaw}
\begin{equation}
	\Phi(y^2) = \int \frac{\mathrm{d}^4p}{(2\pi)^4} \ee^{-\ii py}\tilde{\Phi} (-p^2)\,.
\end{equation}
Any form for the $\tilde{\Phi}(-p^2)$ is allowable as long as it could drop rapidly in the ultraviolet region. As widely used in considerable literature \cite{Yue:2024bvy,Zhu:2021exs,Dong:2017gaw,Xiao:2016hoa,Ma:2010xx} (and references therein), we also take the Gaussian form,
\begin{equation}\label{eq:gaussfromfactor}
	\tilde{\Phi}(p^2_E)\,\dot{=} \,\exp(-p_E^2/\Lambda^2)\,.
\end{equation} 
Here $p_E$ is the Euclidean Jacobi momentum. The cutoff $\Lambda$ is of the order of $1~\mathrm{GeV}$, whose value is process dependent.

\begin{figure}
	\centering
	\includegraphics[width=0.94\linewidth]{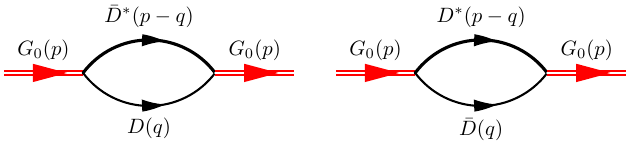}
	\caption{Mass operator of the $G_0$.}
	\label{fig:feynmandiagmassoperator}
\end{figure}

The coupling constant $g_{G_0 DD^\ast}$ can be determined by the compositeness condition $Z=0$ \cite{Weinberg:1962hj,Salam:1962ap}, where $Z$ is the wave function renormalization constant,
\begin{equation}
	Z = 1- \left. \frac{\mathrm{d} \Sigma}{\mathrm{d}p^2_{G_0}}\right|_{p_{G_0}^2=m_{G_0}^2}=0\,.
\end{equation}
For the pseudoscalar composite particle, $\Sigma$ is its mass operator, as illustrated in Fig. \ref{fig:feynmandiagmassoperator}. Using the Lagrangian in Eq. \eqref{eq:Lag4G0}, we obtain the mass operator as
\begin{align}
	\Sigma(p^2) &= g_{G_0DD^\ast}^2 \int \frac{\mathrm{d}^4q}{(2\pi)^4} \tilde{\Phi}^2[(q-\omega_D p)^2]\frac{\ii}{q^2-m_{D}^2}\nonumber\\
	& \times \frac{\ii \bar{g}_{\mu\nu}(p-q,m_{D^\ast})}{(p-q)^2-m_{D^\ast}^2}(p-2q)^\mu (p-2q)^\nu
\end{align}
with
\begin{equation}
	\bar{g}_{\mu\nu}(p,m) = -g_{\mu\nu} + \frac{p_\mu p_\nu}{m^2}\,.
\end{equation}
Since the $G_0$ has not been observed experimentally, we assume that its mass is approximately equal to those of the $X(3872)$ and the $G(3900)$ since they are all near the $D\bar{D}^*$ threshold. Thus, in the molecular picture, we could take $m_{G_0} = m_{D^0} +m_{{\bar D}^{\ast 0}} - E_\mathrm{b}$, where $E_\mathrm{b}$ is regarded as the binding energy of the $G_0$.

\begin{figure}
	\centering
	\includegraphics[width=0.94\linewidth]{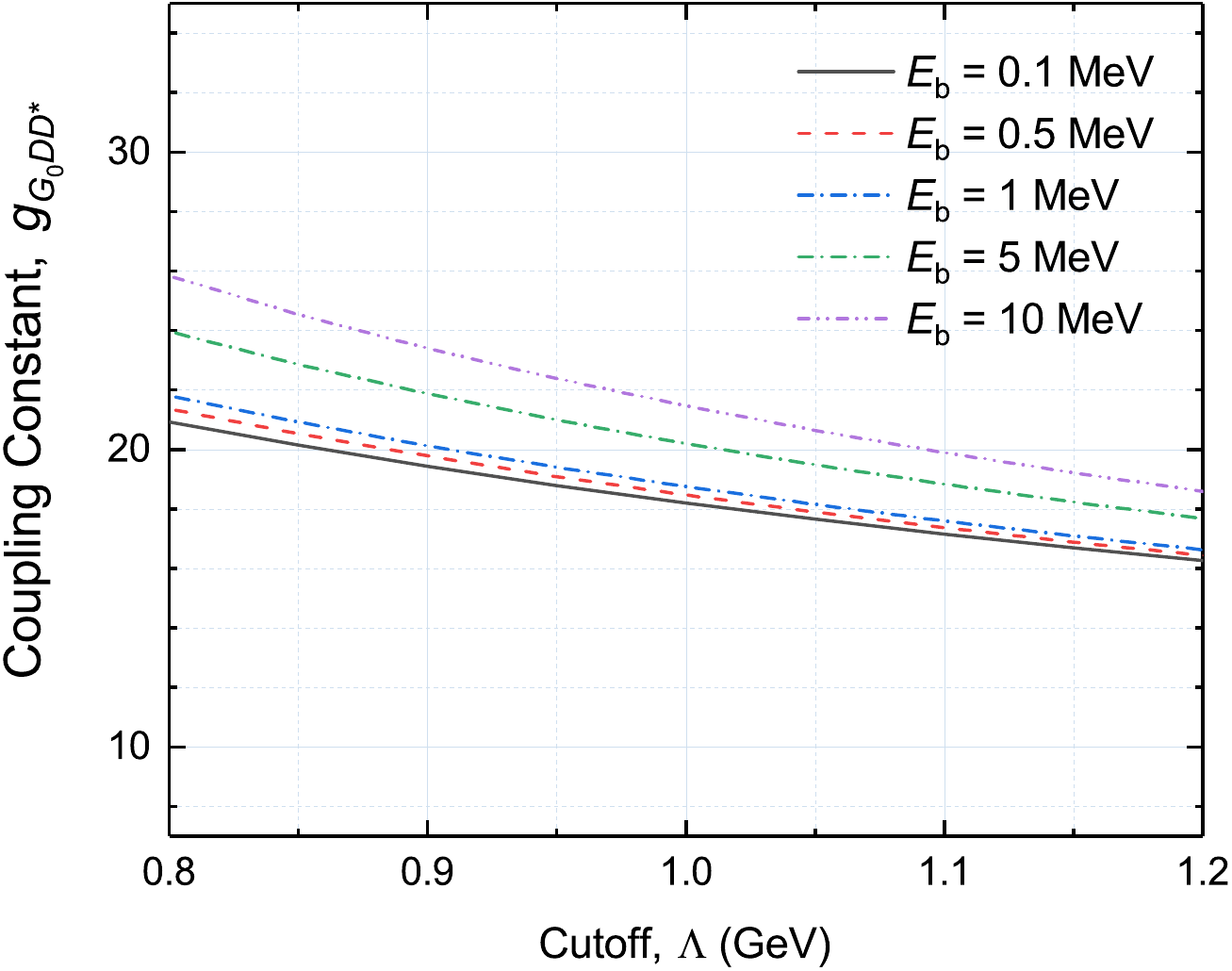}
	\caption{Cutoff($\Lambda$) dependence of the coupling constant $g_{G_0DD^\ast}$ for binding energies $E_\mathrm{b}=0.1\sim 10$ MeV.}
	\label{fig:g0couplingconstant}
\end{figure}

In Fig. \ref{fig:g0couplingconstant} the cutoff($\Lambda$) dependence of the coupling constant $g_{G_0DD^\ast}$ is shown for different binding energies ranging from 0.1 MeV to 10 MeV. It is seen that the coupling constant $g_{G_0DD^\ast}$ decreases with increasing the cutoff $\Lambda$. At a given $\Lambda$, $g_{G_0DD^\ast}$ increases as the binding energy grows. 
In terms of the wave function in Eq. \eqref{eq:G0wavefunction} we used, the effective couplings of the $G_0$ to the neutral and charged charmed meson pairs are equal accordingly. The relative size of the effective neutral and charged coupling constants needs to be determined based on future experimental results. For the $X(3872)$ treated as an $S$-wave $D\bar{D}^\ast/\bar{D}D^\ast$ state, the effective charged coupling constant is found to be about half the neutral one \cite{Zhang:2024fxy}, which is extracted from the recent LHCb experiments \cite{LHCb:2022jez}.

In this work, we shall be mainly concerned with the hidden-charm hadronic decay processes $G_0\to \omega(\rho^0) J/\psi $, $\pi^+\pi^- \eta_c (1S)$, $\pi^+\pi^- \chi_{c1} (1P)$, and the open-charm $D^0\bar{D}^0\pi^0$, based on the suggestions \cite{Lin:2024qcq}.
To evaluate the relevant Feynman diagrams, we need the effective Lagrangians of the final states with the possible charmed mesons. Under the heavy quark limit and chiral symmetry, the effective Lagrangians are constructed as \cite{Colangelo:2003sa,Casalbuoni:1996pg,Li:2021jjt}
\begin{subequations}\label{eq:LagSPPV}
	\begin{align}
		\mathcal{L}_S &= \ii g_S \mytrace{S^{(c\bar{c})}\bar{H}_a^{(\bar{c}q)}\gamma_\mu\pararrowk{\mu}\bar{H}_a^{c\bar{q}}}+\mathrm{H.c.},\\
		\mathcal{L}_P &= \ii g_P \mytrace{P^{(c\bar{c})}\bar{H}_a^{(\bar{c}q)}\gamma_\mu\bar{H}_a^{c\bar{q}}}+\mathrm{H.c.},\\
		\mathcal{L}_\mathbb{P} &= \ii g_\mathbb{P} \mytrace{H_b^{(c\bar{q})\mu}\gamma_\mu\gamma_5 \mathcal{A}_{ba}^\mu \bar{H}_a^{(c\bar{q})}},\\
		\mathcal{L}_\mathbb{V} &= \ii\beta \mytrace{H_b^{(c\bar{q})}v^\mu(-\rho_\mu)_{ba}\bar{H}_a^{(c\bar{q})}}\nonumber\\
		&+\ii\lambda\mytrace{H_b^{(c\bar{q})}\sigma^{\mu\nu}F_{\mu\nu}(\rho)_{ba}\bar{H}_a^{(c\bar{q})}}\,.
	\end{align}
\end{subequations}
Here $\mytrace{\cdots}$ means the trace over the $4\times 4$ matrices and the letters $a$ and $b$ are the light flavor indices; $S^{(c\bar{c})}$ and $P^{(c\bar{c})}$ are, respectively, the $S$- and $P$-wave charmonium multiplets,
	\begin{align}
		S^{(c\bar{c})} &= \frac{1+\slashed{v}}{2}(\psi^\mu\gamma_\mu- \eta_c\gamma_5)\frac{1-\slashed{v}}{2},\\
		P^{(c\bar{c})\mu}&=\frac{1+\slashed{v}}{2}\Big(\chi_{c2}^{\mu\alpha}\gamma_\alpha + \frac{1}{\sqrt{2}}\epsilon^{\mu\alpha\beta\sigma}v_\alpha\gamma_\beta \chi_{c1\sigma}\nonumber\\
		&+ \frac{1}{\sqrt{3}}(\gamma^\mu-v^\mu)\chi_{c0}+ h_{c}^\mu\gamma_5\Big)\frac{1-\slashed{v}}{2};
	\end{align}
$H_a^{(c\bar{q})}$ and $H_a^{(\bar{c}q)}$ denote, respectively, the ground charmed and anticharmed meson doublets with spin-parity $J^P=(0^-,\,1^-)$ \cite{Xu:2016kbn,Manohar:2000dt},
	\begin{align}
		H_a^{(c\bar{q})} = \frac{1+\slashed{v}}{2}({D}_a^{\ast\mu}\gamma_\mu +\ii {D}_a\gamma_5),\\
		H_a^{(\bar{c}q)} = (\bar{{D}}_a^{\ast\mu}\gamma_\mu +\ii \bar{{D}}_a\gamma_5)\frac{1-\slashed{v}}{2}\,,
	\end{align}
and the corresponding conjugate fields are defined as $\bar{H}_a^{(c\bar{q})} = \gamma_0H_a^{(c\bar{q})\dagger}\gamma_0$ and $\bar{H}_a^{(\bar{c}q)} = \gamma_0H_a^{(\bar{c}q)\dagger}\gamma_0$, respectively; $\mathcal{A}^\mu$ is the axial current of the light pseudoscalar fields: 
\begin{equation}
	\mathcal{A}^\mu = \frac{1}{2} (\xi^\dagger\partial^\mu \xi-\xi\partial^\mu\xi^\dagger)\approx \frac{\ii}{f_\pi} \partial^\mu \mathbb{P},
\end{equation}
where $\xi = \ee^{\ii\mathbb{P}/f_\pi}$ with the pion decay constant $f_\pi = (130.2\pm 1.2)$ MeV \cite{ParticleDataGroup:2024cfk} and $\mathbb{P}$ being a $3\times 3$ matrix of the pseudoscalar fields,
\begin{equation}
	\mathbb{P} = \begin{pmatrix}
		\frac{1}{\sqrt{2}}\pi^0+\frac{1}{\sqrt{6}} \eta & \pi^+ & K^+\\
		\pi^- & - \frac{1}{\sqrt{2}}\pi^0+\frac{1}{\sqrt{6}}\eta & K^0 \\
		K^- & \bar{K}^0 & - \frac{\sqrt{6}}{3}\eta
	\end{pmatrix};
\end{equation}
$\rho_\mu = \ii (g_V/\sqrt{2})\mathbb{V}_\mu$ and $F_{\mu\nu} = \partial_\mu\rho_\nu-\partial_\nu\rho_\mu+[\rho_\mu,\,\rho_\nu]$ with $\mathbb{V}$ being a $3\times 3$ matrix of the light vector fields,
\begin{equation}
	\mathbb{V} = \begin{pmatrix}
		\frac{1}{\sqrt{2}}(\rho^0 + \omega) & \rho^+ & K^{\ast +}\\
		\rho^- & \frac{1}{\sqrt{2}} (\omega-\rho^0)& K^{\ast 0}\\
		K^{\ast -} & \bar{K}^{\ast 0}& \phi
	\end{pmatrix}.
\end{equation}

After tracing Eq. \eqref{eq:LagSPPV}, we find
\begin{subequations}\label{eq:LagsExplicit}
	\begin{align}
		\mathcal{L}_S &= \ii g_{\psi {D}{D}}\psi_\mu\bar{{D}}^\dagger \pararrowk{\mu}{D}^\dagger\nonumber\\
		&-g_{\psi{D}{D}^\ast}\epsilon_{\mu\nu\alpha\beta}\partial^\mu\psi^\nu(\bar{{D}}^{\ast\dagger\alpha}\pararrowk{\beta}{D}^\dagger - \bar{{D}}^\dagger\pararrowk{\beta}{D}^{\ast\dagger\alpha})\nonumber\\
		&-\ii g_{\psi{D}^\ast{D}^\ast} \psi_\mu (\bar{{D}}^{\ast\dagger}_\nu\pararrowk{\mu}{D}^{\ast\dagger\nu}-\bar{{D}}^{\ast\dagger}_\nu\pararrowk{\nu}{D}^{\ast\dagger\mu}-\bar{{D}}^{\ast\dagger}_\mu\pararrowk{\nu}{D}^{\ast\dagger}_\nu)\nonumber\\
		&-\ii g_{\eta_c{D}{D}^\ast} (\bar{{D}}^{\ast\dagger}_\mu\pararrowk{\mu}{D}^\dagger +\bar{{D}}^\dagger\pararrowk{\mu}{D}^{\ast\dagger}_\mu)\nonumber\\
		& - g_{\eta_c{D}^\ast{D}^\ast} \epsilon_{\mu\nu\alpha\beta}\partial^\mu\eta_c\bar{{D}}^{\ast\dagger\alpha}\pararrowk{\nu}{D}^{\ast\dagger\beta},\\
		\mathcal{L}_P &= g_{\chi_{c0}{D}{D}}\chi_{c0}\bar{{D}}^\dagger{D}^\dagger - g_{\chi_{c0}{D}^\ast{D}^\ast}\bar{{D}}^{\ast\dagger\mu}{D}^{\ast\dagger}_\mu\nonumber\\
		&+ g_{\chi_{c1}{D}{D}^\ast} \chi_{c1}^\mu (\bar{{D}}^{\ast\dagger}_\mu{D}^\dagger - \bar{{D}}^\dagger{D}^{\ast\dagger}_\mu)\nonumber\\
		& + g_{\chi_{c2}{D}^\ast{D}^\ast}  \chi_{c2}^{\mu\nu}(\bar{{D}}^{\ast\dagger}_\mu{D}^{\ast\dagger}_\nu+\bar{{D}}^{\ast\dagger}_\nu{D}^{\ast\dagger}_\mu)\nonumber\\
		&-g_{h_c{D}{D}^\ast}h_c^\mu (\bar{{D}}^{\ast\dagger}_\mu{D}^\dagger + \bar{{D}}^\dagger{D}^{\ast\dagger}_\mu)\nonumber\\
		&+\ii g_{h_c{D}^\ast{D}^\ast}\epsilon_{\mu\nu\alpha\beta}\partial^\nu h_c^\mu{D}^{\ast\dagger}_\alpha\bar{{D}}^{\ast\dagger}_\beta,\\
		\mathcal{L}_\mathbb{P}&=\ii g_{{D}{D}^\ast\mathbb{P}}({D}_b \partial_\mu\mathbb{P}_{ba}{D}^{\ast\dagger \mu}_{a}-{D}^{\ast\mu}_b\partial_\mu\mathbb{P}_{ba}{D}^\dagger_a)\nonumber\\
		&-\frac{1}{2}g_{{D}^\ast{D}^\ast\mathbb{P}} \epsilon_{\mu\nu\alpha\beta}{D}^{\ast\mu}_b\partial_\nu\mathbb{P}_{ba}		\pararrowk{\alpha}{D}^{\ast\dagger\beta}_{a},\\
		\mathcal{L}_{\mathbb{V}}& =\ii  g_{{D}{D}\mathbb{V}}{D}_b\pararrowk{\mu}{D}^\dagger_a \mathbb{V}_{\mu ba} \nonumber\\
		&+2f_{{D}{D}^\ast\mathbb{V}}\epsilon_{\mu\nu\alpha\beta}\partial^\mu \mathbb{V}^\nu_{ba} ({D}_b\pararrowk{\alpha}{D}^{\ast\dagger\beta}_a - {D}^{\ast\beta}_b\pararrowk{\alpha}{D}^\dagger_a)\nonumber\\
		&-\ii g_{{D}^\ast{D}^\ast\mathbb{V}} \mathbb{V}_{\mu ba} {D}^{\ast\nu}_b\pararrowk{\mu}{D}^{\ast\dagger}_{\nu a}\nonumber\\
		&+ 4 \ii  f_{{D}^\ast{D}^\ast \mathbb{V}}{D}^{\ast}_{b\nu} (\partial^\mu\mathbb{V}^\nu-\partial^\nu\mathbb{V}^\mu)_{ba}{D}^{\ast\dagger\mu}_a.
	\end{align}
\end{subequations}

The coupling constants in Eq. \eqref{eq:LagsExplicit} are linked to each other by the global constants $g_{S(P,\mathbb{P})}$, $\beta$, and $\lambda$,
\begin{align}
	\frac{g_{\psi{D}{D}}}{\sqrt{m_\psi}m_{{D}}} &= \frac{g_{\psi{D}{D}^\ast}}{\sqrt{m_{D}m_{D}^\ast/m_\psi}}=\frac{g_{\psi{D}^\ast{D}^\ast}}{\sqrt{m_\psi}m_{{D}^\ast}}=2g_S,\\
	\frac{g_{\chi_{c0}{D}{D}}}{\sqrt{3m_{\chi_{c0}}}m_{{D}}} &= \frac{\sqrt{3}g_{\chi_{c0}{D}^\ast{D}^\ast}}{\sqrt{m_{\chi_{c0}}m_{{D}^\ast}}}=\frac{g_{\chi_{c1}{D}{D}^\ast}}{\sqrt{2m_{\chi_{c1}}m_{{D}}m_{{D}^\ast}}}\nonumber\\
	&=\frac{g_{\chi_{c2}{D}^\ast{D}^\ast}}{\sqrt{m_{\chi_{c2}}m_{{D}^\ast}{D}^\ast}}=	2g_P,\\
	g_{{D}^\ast{D}^\ast\mathbb{P}}	&= \frac{g_{{D}{D}^\ast\mathbb{P}}}{\sqrt{m_{D}m_{{D}^\ast}}}=\frac{2g_{\mathbb{P}}}{f_\pi}\\
	g_{{D}{D}\mathbb{V}}&=g_{{D}^\ast{D}^\ast\mathbb{V}} = \frac{\beta g_\mathbb{V}}{\sqrt{2}},\\
	f_{{D}{D}^\ast\mathbb{V}} &= \frac{f_{{D}^\ast{D}^\ast\mathbb{V}}}{m_{{D}^\ast}} = \frac{\lambda g_\mathbb{V}}{\sqrt{2}}.
\end{align}
 In terms of the vector meson dominance \cite{Colangelo:2003sa,Deandrea:2003pv}, $g_S = \sqrt{m_{\psi}}/(2m_Df_{\psi})$ and $g_P= -\sqrt{m_{\chi_{c0}}/3}/f_{\chi_{c0}}$, where $f_\psi$ and $f_{\chi_{c0}}$ are the $J/\psi$ and $\chi_{c0}$ decay constants, respectively. The $J/\psi$ decay constant $f_\psi$ can be extracted from the dielectron decay width $\Gamma_{ee}(J/\psi\to e^+e^-)$ \cite{Badalian:2009bu,Li:2012as,Liu:2023gtx}. Using the newly updated Particle Data Group (PDG) data \cite{ParticleDataGroup:2024cfk}, we obtained $f_{\psi} = (416\pm 4)~\mathrm{MeV}$, agreeing well with the value $(418\pm 9)$ MeV by the Lattice QCD \cite{Becirevic:2013bsa}, thereby $g_S=(1.13\pm0.01)~\mathrm{GeV^{-3/2}}$ \cite{Liu:2025bjm}. The $\chi_{c0}$ decay constant $f_{\chi_{c0}} = (343\pm 112)$ MeV, which was estimated in the framework of the QCD sum rules \cite{Veliev:2010gb}, and $g_P = 0.98~\mathrm{GeV^{-1/2}}$ accordingly. Using the $\Gamma[D^{\ast +}\to D^0\pi^+(D^+\pi^0)]$, we find $g_{\mathbb{P}} = 0.57\pm 0.01$  \cite{ParticleDataGroup:2024cfk}. Finally, the vector meson dominance yields $g_\mathbb{V} = m_\rho/f_\pi$ and $\beta = 0.9$ \cite{Isola:2003fh}; Comparing the form factor $B\to K^\ast$ obtained by different theoretical calculations (such as the effective chiral Lagrangian, light cone sum rules and lattice QCD) gives $\lambda = 0.56~\mathrm{GeV^{-1}}$ \cite{Isola:2003fh,Casalbuoni:1992dx} (and references therein).

\subsection{Decays of $G_0\to \omega(\rho^0) J/\psi  $}

\begin{figure}
	\centering
	\includegraphics[width=0.820\linewidth]{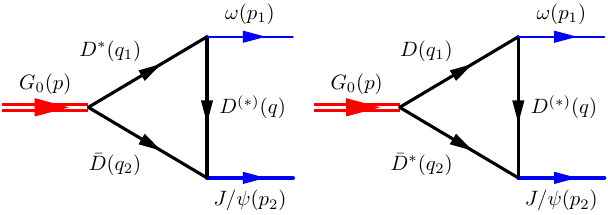}
	\caption{Triangle loops for the process $G_0\to \omega J/\psi$. The charge conjugated loops are not shown here, but included in the calculations. To get the case for $G_0\to \rho^0 J/\psi$, we only need to replace the $\omega$ with $\rho^0$.}
	\label{fig:feynmandiagjomg}
\end{figure}

With the preparations above, we could evaluate the concerned processes. The Feynman diagrams of the processes $G_0\to \omega  J/\psi$ and $G_0\to \rho^0  J/\psi $ are shown in Fig. \ref{fig:feynmandiagjomg}. Due to the different exchanged meson, each diagram corresponds to two amplitudes. The explicit expressions are as follows:
	\begin{align}\label{eq:ampJpsiOmg}
    \mathcal{M}_a^{(1)}&=\epsilon_\mu^\ast(p_2)\epsilon^{\ast}_\nu(p_1) \int\dfrac{\mathrm{d}^4q}{(2\pi)^4}\tilde{\Phi}[(q_1\omega_D - q_2\omega_{D^\ast})^2]\nonumber\\
	&\times \Big[-\frac{\ii}{\sqrt{2}}g_{G_0 D D^\ast}(q_1-q_2)^\alpha\Big][g_{\psi DD}(q-q_2)^\mu]\nonumber\\
	&\times [2f_{DD^\ast \mathbb{V}}\epsilon^{\eta\nu\phi\beta}p_{1\eta}(q+q_1)_\phi]\mathcal{S}_{\alpha\beta}(q_1,m_{D^\ast})\nonumber\\
	&\times \mathcal{S}(q_2,m_D)\mathcal{S}(q,m_D)\mathcal{F}(q,m_D)\,,\\
	\mathcal{M}_a^{(2)}&=\epsilon_\mu^\ast(p_2)\epsilon^{\ast}_\nu(p_1) \int\dfrac{\mathrm{d}^4q}{(2\pi)^4}\tilde{\Phi}[(q_1\omega_D - q_2\omega_{D^\ast})^2]\nonumber\\
	&\times \Big[-\frac{\ii}{\sqrt{2}}g_{G_0 D D^\ast}(q_1-q_2)^\alpha\Big]\nonumber\\
	&\times [g_{\psi DD^\ast}\epsilon^{\eta\mu\sigma\phi}p_{2\eta}(q-q_2)_\phi][g_{D^\ast D^\ast\mathbb{V}}g^{\beta\delta}(q+q_1)^\nu\nonumber\\
	&-4f_{D^\ast D^\ast \mathbb{V}}(g^{\beta\nu}p_1^\delta- g^{\delta\nu}p_1^\beta)]\mathcal{S}_{\alpha\beta}(q_1,m_{D^\ast})\nonumber\\
	&\times \mathcal{S}(q_2,m_D)\mathcal{S}_{\delta\sigma}(q,m_{D^\ast})\mathcal{F}(q,m_{D^\ast})\,,\\
	\mathcal{M}_b^{(1)}  &=\epsilon_\mu^\ast(p_2)\epsilon^{\ast}_\nu(p_1) \int\dfrac{\mathrm{d}^4q}{(2\pi)^4}\tilde{\Phi}[(q_1\omega_{D^\ast} - q_2\omega_{D})^2]\nonumber\\
	&\times \Big[\frac{\ii}{\sqrt{2}}g_{G_0DD^\ast}(q_2-q_1)^\xi\Big] [-g_{DD\mathbb{V}} (q+q_1)^\nu]\nonumber\\
	&\times [-g_{\psi DD^\ast}\epsilon^{\eta\mu\rho\phi} p_{2\eta}(q-q_2)_\phi] \mathcal{S}(q_1,m_D)\nonumber\\
	&\times \mathcal{S}_{\xi\rho}(q_2,m_{D^\ast})\mathcal{S}(q,m_D)\mathcal{F}(q,m_D)\,,\\
	\mathcal{M}_b^{(2)}  &=\epsilon_\mu^\ast(p_2)\epsilon^{\ast}_\nu(p_1) \int\dfrac{\mathrm{d}^4q}{(2\pi)^4}\tilde{\Phi}[(q_1\omega_{D^\ast} - q_2\omega_{D})^2]\nonumber\\
	&\times \Big[\frac{\ii}{\sqrt{2}}g_{G_0DD^\ast}(q_2-q_1)^\xi\Big][g_{\psi D^\ast D^\ast}(g^{\mu\rho}(q-q_2)^\sigma\nonumber\\
	& + g^{\mu\sigma}(q-q_2)^\rho-g^{\rho\sigma}(q-q_2)^\mu)]\nonumber\\
	&\times [-2f_{DD^\ast\mathbb{V}} \epsilon^{\eta\nu\phi\delta} p_{1\eta}(q+q_1)_\phi] \mathcal{S}(q_1,m_D)\nonumber\\
	&\times \mathcal{S}_{\xi\rho}(q_2,m_{D^\ast})\mathcal{S}_{\delta\sigma}(q,m_D)\mathcal{F}(q,m_{D^\ast})\,.
\end{align}
Here $\mathcal{S}_{\mu\nu}(p,m)$ and $\mathcal{S}(p,m)$ are the propagators of the vector and pseudoscalar mesons, respectively,
\begin{subequations}
	\begin{align}
	\mathcal{S}_{\mu\nu}(p,m)&=\frac{ \bar{g}_{\mu\nu}(p,m)}{p^2-m^2}\,,\\
	\mathcal{S}(p,m) &= \frac{1}{p^2-m^2}\,.
\end{align}
\end{subequations}
Since the exchanged mesons are not on shell, we introduce a monopole form factor to account for the off shell effect, namely,
\begin{equation}\label{eq:formfactor}
	\mathcal{F}(q,m) = \frac{m^2-\Lambda^{\prime 2}}{q^2-\Lambda^{\prime 2}}\,.
\end{equation}
Here $\Lambda'$ is parametrized as $\Lambda' = m+\alpha \Lambda_{\mathrm{QCD}}$ with $\Lambda_{\mathrm{QCD}}=0.22~\mathrm{GeV}$. The parameter $\alpha$ is usually taken to be around 1.0.

There is an important thing to notice about these two processes: the decays $G_0\to\omega J/\psi$ and $G_0\to \rho^0  J/\psi$ are forbidden due to the phase space when the $G_0$ masses $m_{G_0} \lesssim (m_{D^0}+m_{{\bar D}^{\ast 0}})=3871.69~\mathrm{MeV}$,  $m_{\omega} = 782.66~\mathrm{MeV}$, and $m_{\rho^0} = 775.26~\mathrm{MeV}$ \cite{ParticleDataGroup:2024cfk} are adopted. However, these processes can occur when the $\omega$ and $\rho^0$ mass distributions are considered and will be seen in the cascade decays $G_0\to \omega J/\psi\to \pi^+\pi^-\pi^0 J/\psi$ and $G_0\to \rho^0 J/\psi \to  \pi^+\pi^-J/\psi$, similar to the case of the $X(3872)\to\omega J/\psi$ \cite{Gamermann:2009fv,BESIII:2019qvy,ParticleDataGroup:2024cfk,LHCb:2015jfc}. Taking the $\omega$ width into account, the partial decay width for the $G_0 \to\omega J/\psi$ is expressed as \cite{Gamermann:2009fv,Wu:2021udi}
\begin{align}
	\Gamma (G_0\to\omega J/\psi) &= \frac{1}{W} \int_{(3m_{\pi})^2}^{(m_{G_0}-m_{J/\psi})^2}\mathrm{d}s f(s,m_{\omega},\Gamma_{\omega})\nonumber\\
	& \times \frac{|\mathbf{p}_\omega(s)|}{8\pi m_{G_0}^2} |\mathcal{M}_{\mathrm{tot}}(s)|^2,
\end{align}
where $W = \int_{(3m_{\pi})^2}^{(m_{G_0}-m_{J/\psi})^2}\mathrm{d}s f(s,m_{\omega},\Gamma_{\omega})$ with $f(s,m_{\omega},\Gamma_{\omega})$ being the Breit-Wigner distribution in the following form
\begin{equation}
	f(s,m_{\omega},\Gamma_{\omega}) = \frac{1}{\pi} \frac{m_\omega\Gamma_\omega}{(s-m_\omega^2)^2+m_\omega^2\Gamma_\omega^2}\,.
\end{equation}
Moreover, the momentum $\mathbf{p}_\omega$ and amplitude $\mathcal{M}_\mathrm{\mathrm{tot}}$ are obtained by replacing the $\omega$ mass with the $\sqrt{s}$. For the case of the $\rho^0$ emission, the calculations are similar.

\subsection{Decay of $G_0\to D^0 \bar{D}^0\pi^0$}

\begin{figure}
	\centering
	\includegraphics[width=0.82\linewidth]{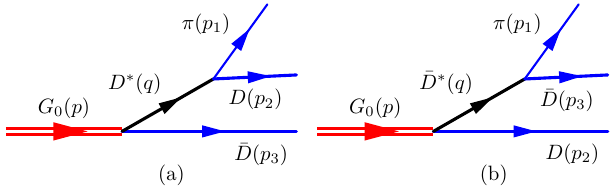}
	\caption{Tree-level diagrams for the process $G_0\to  D^0\bar{D}^0\pi$. Diagram (b) is the charge conjugated diagram of (a).}
	\label{fig:feynmandiagddpi}
\end{figure}

In Fig. \ref{fig:feynmandiagddpi} we present the tree-level Feynman diagrams of the three-body decay process $G_0\to D^0\bar{D}^0\pi^0$. One may note that in the case of $X(3872)\to D^0\bar{D}^0\pi^0$, the $D\bar{D}$ final state interaction (FSI) effect, if there is a near-threshold pole in the $D\bar{D}$ system \cite{Guo:2014hqa}, is comparable to the tree contribution. As pointed out in Ref. \cite{Guo:2014hqa}, the importance of the $D\bar{D}$ FSI depends strongly on the low-energy constant (the $C_{0A}$ in Ref. \cite{Guo:2014hqa}), which, however, is not well-known. Moreover, according to our calculated results presented below, the $G_0\to D^0\bar{D}^0 \pi^0$ is not the main decay channel. Hence, in spite of the possible importance of the $D\bar{D}$ FSI effect, we do not consider its contribution in this work\footnote{We estimated the impact of the $D\bar{D}$ FSI on the $G_0\to D^0 \bar{D}^0\pi^0$ using the value $C_{0A} \sim -2~\mathrm{fm^2}$ given in Ref. \cite{Guo:2014hqa}. At $\Lambda = 1.0~\mathrm{GeV}$, the partial decay width of the $G_0\to D^0 \bar{D}^0\pi^0$ due to the $D\bar{D}$ FSI effect was predicted to be around $0.03~\mathrm{keV}$, being the same order of magnitude as that via the tree diagrams in Fig. \ref{fig:feynmandiagddpi} (see Table \ref{tab:decaywidths})}.
Using the foregoing Lagrangians, we obtain the tree-level amplitudes in the following form:
	\begin{align}
	\mathcal{M}_a &= \Big[-\frac{\ii}{\sqrt{2}}g_{G_0DD^\ast}(q-p_3)^\mu\Big]\tilde{\Phi}[(q\omega_D-p_3\omega_{D^\ast})^2]\nonumber\\
	&\times [g_{DD^\ast\mathbb{P}} p_1^\nu] \mathcal{S}_{\mu\nu}(q,m_{D^\ast})\,,\\
	\mathcal{M}_b &=  \Big[\frac{\ii}{\sqrt{2}}g_{G_0DD^\ast}(q-p_2)^\mu\Big]\tilde{\Phi}[(q\omega_D-p_2\omega_{D^\ast})^2]\nonumber\\
	&\times [-g_{DD^\ast\mathbb{P}} p_1^\nu] \mathcal{S}_{\mu\nu}(q,m_{D^\ast})\,.
\end{align}

For three-body processes, the differential partial decay width is evaluated by the following expression
\begin{equation}\label{eq:threebodydecaywidthformula}
	\frac{\mathrm{d}\Gamma}{\mathrm{d}m_{12}\mathrm{d}m_{23}} = \frac{1}{64\pi^3}\frac{1}{m_{G_0}^3} m_{12}m_{23} |\mathcal{M}_{\mathrm{tot}}|^2 \,,
\end{equation}
where $m_{ij}$ are the invariant mass of the particles $i$ and $j$ in the final states.

\subsection{Dipionic decays of $G_0\to \pi^+\pi^-\eta_c(1S)/\chi_{c1}(1P)$}

\begin{figure}
	\centering
	\includegraphics[width=0.84\linewidth]{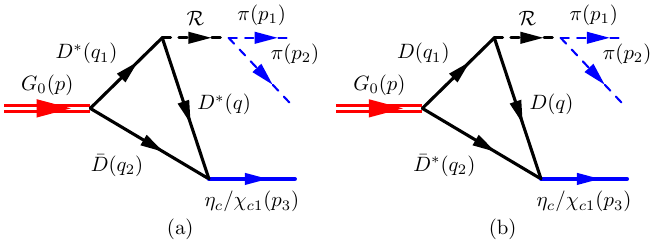}
	\caption{Triangle loops for the processes $G_0\to \pi^+\pi^-\eta_c(1S)$ and $G_0\to \pi^+\pi^-\chi_{c1}(1P)$. The $\mathcal{R}$ represents the scalar mesons $f_0(500)$ and $f_0(980)$ for the $\eta_c (1S)$, whereas for the $\chi_{c1}(1P)$ only the $f_0(500)$ is considered.}
	\label{fig:feynmandiagtwopi}
\end{figure}

Here we assume the decay process $G_0\to \pi^+\pi^-\eta_c(1S)/\chi_{c1}(1P)$ occurs via the triangle loops shown in Fig. \ref{fig:feynmandiagtwopi}, where the two pions are produced by the scalar mesons $f_0(500)$ and $f_0(980)$. In order to calculate these two decays, we also need, apart from the Lagrangians mentioned above, the interactions related to the $f_0$ and the charmed mesons \cite{Meng:2008dd,Bai:2022cfz,Chen:2011qx,Chen:2015bma}
\begin{align}
	\mathcal{L}_{f_0} = g_{f_0\mathcal{D}\mathcal{D}}f_0 DD^\dagger-g_{f_0 D^\ast D^\ast} f_0 D^\ast_\mu D^{\ast\mu\dagger} \,.
\end{align}
The coupling constants are 
\begin{subequations}
	\begin{align}
		g_{f_0(980)DD} &= \sqrt{2}g_{f_0(500)DD} =m_D g_\pi/\sqrt{3}\,,\\
		g_{f_0(980)D^\ast D^\ast} &= \sqrt{2}g_{f_0(500)D^\ast D^\ast} =m_{D^\ast} g_\pi/\sqrt{3}\,,
	\end{align}
\end{subequations}
with $g_{\pi} = 3.73$ \cite{Bai:2022cfz}. In this work, we take $m_{f_0(500)}=449~\mathrm{MeV}$, $\Gamma_{\mathrm{tot}}[f_0(500)]=550~\mathrm{MeV}$; $m_{f_0(980)}=993~\mathrm{MeV}$, $\Gamma_{\mathrm{tot}}[f_0(980)]=61.3~\mathrm{MeV}$ \cite{Bai:2022cfz}. The decay of the scalar meson $f_0$ into two pions is described by 
\begin{equation}
	\mathcal{L}_{f_0\pi\pi} = g_{f_0\pi\pi}f_0\pi\pi\,,
\end{equation}
where $g_{f_0(500)\pi\pi} = 3.25~\mathrm{GeV}$ for the $ f_0(500)$ and $g_{f_0(980)\pi\pi} = 1.13~\mathrm{GeV}$ for the $ f_0(980)$ \cite{Bai:2022cfz}.

For the decay $G_0\to\pi^+\pi^-\eta_c (1S)$, the amplitudes are
	\begin{align}
		\mathcal{M}_a &= \int \frac{\mathrm{d}^4q}{(2\pi)^4}\tilde{\Phi}[(q_1\omega_D - q_2\omega_{D^\ast})^2]\nonumber\\
		&\times \Big[-\frac{\ii}{\sqrt{2}}g_{G_0 D D^\ast}(q_1-q_2)^\alpha\Big] [-g_{\eta_c DD^\ast}(q-q_2)^\mu]\nonumber\\
		&\times [g_{f_0D^\ast D^\ast}g^{\delta\sigma}][g_{f_0\pi\pi}]\mathcal{S}_{\alpha\delta}(q_1,m_{D^\ast})\mathcal{S}(q_2,m_D) \nonumber\\
		&\times  \mathcal{S}_{\sigma \mu }(q,m_{D^\ast}) \mathcal{S}^{f_0}\mathcal{F}(q,m_{D^\ast})\,,\\
		\mathcal{M}_b &= \int \frac{\mathrm{d}^4q}{(2\pi)^4}\tilde{\Phi}[(q_1\omega_{D}^\ast - q_2\omega_{D})^2]\nonumber\\
		&\times  \Big[\frac{\ii}{\sqrt{2}}g_{G_0 D D^\ast}(q_2-q_1)^\alpha\Big][-g_{\eta_c DD^\ast}(q-q_2)^\mu]\nonumber\\
		&\times [g_{f_0DD}][g_{f_0\pi\pi}]\mathcal{S}(q_1,m_{D^\ast})\mathcal{S}_{\alpha\mu}(q_2,m_D)\nonumber\\
		&\times \mathcal{S}(q,m_D)\mathcal{S}^{f_0}\mathcal{F}(q,m_D)\,.
	\end{align}
Here the $\mathcal{S}^{f_0}$ stands for the propagators of the scalar mesons $f_0(500)$ and $f_0(980)$ in the following form
\begin{equation}
	\mathcal{S}^{f_0} = \frac{1}{m_{\pi\pi}^2-m_{f_0}^2+\ii m_{f_0}\Gamma_{f_0}},
\end{equation}
where $m_{\pi\pi}$ is the invariant mass of the final two pions and $\Gamma_{f_0}$ is the full width of the $f_0$'s we considered.

The amplitudes for the process $G_0\to\pi^+\pi^-\chi_{c1}(1P)$ can be readily obtained by replacing the $\eta_c(1S)$ vertex with the $\chi_{c1}(1P)$ vertex. The partial decay widths are determined using Eq. \eqref{eq:threebodydecaywidthformula}. For the channel $G_0\to \pi^+\pi^-\chi_{c1}(1P)$, we need summation over the $\chi_{c1} (1P)$ spin states.

\section{Results and Discussion}\label{sec:results}
In the following, we present the numerical results for the decay processes $G_0\to \omega(\rho^0) J/\psi $, $\pi^+\pi^- \eta_c (1S)$, $\pi^+\pi^- \chi_{c1} (1P)$, and $D^0\bar{D}^0\pi^0$. The computation of the loop integrals was conducted with the aid of the Schwinger parametrization.

\begin{figure}
	\centering
	\includegraphics[width=0.94\linewidth]{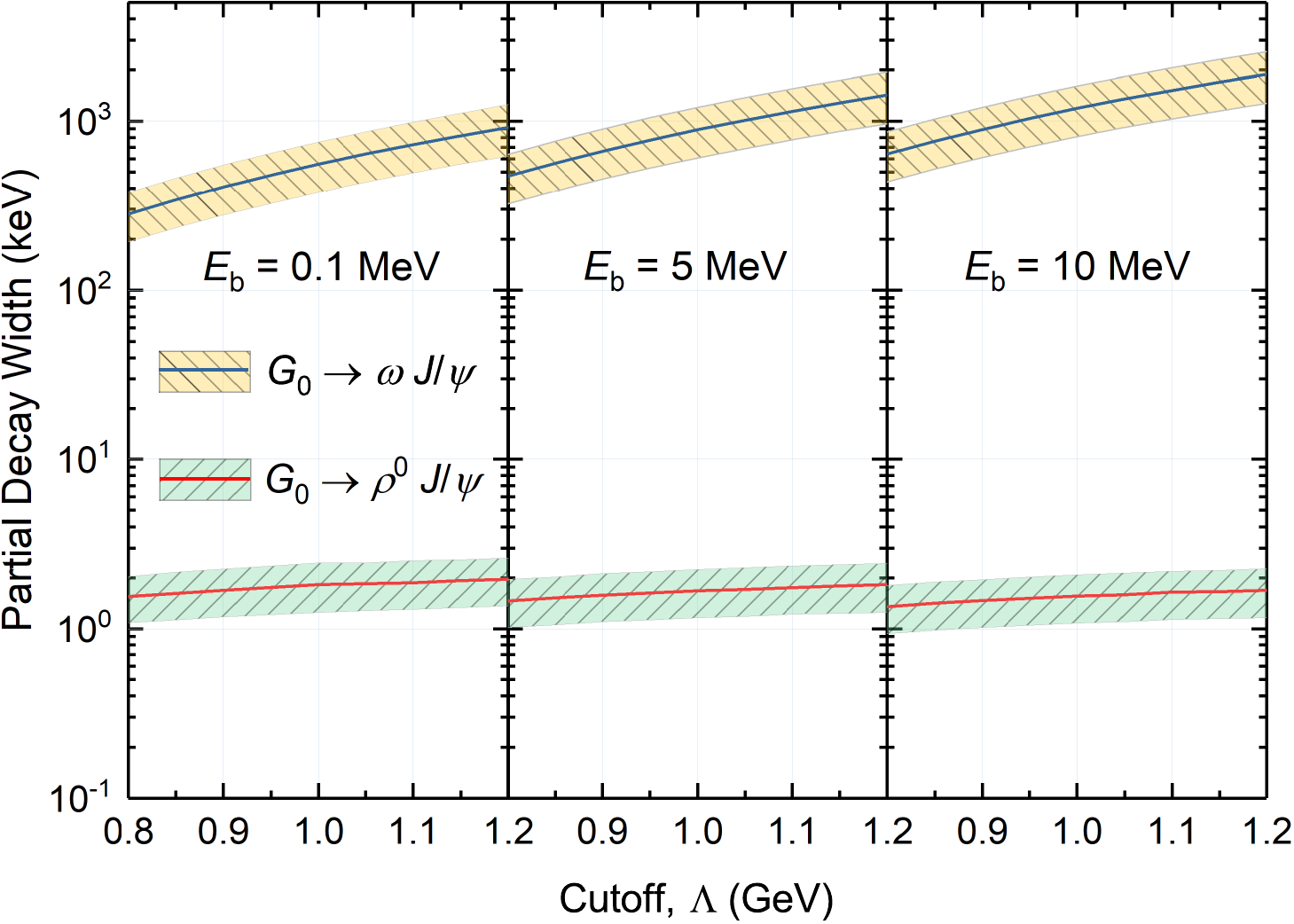}
	\caption{Partial decay widths of the processes $G_0\to\omega J/\psi$ and $G_0\to\rho^0 J/\psi$ for different binding energies $E_b = 0.1$, $5$, and $10$ MeV. The middle thick lines represent the results calculated with $\alpha=1.0$ and the shaded bands depict the variation due to the $\alpha$ values from $0.8$ (lower boundary line) to $1.2$ (upper boundary lines). }
	\label{fig:widthg0jomgrho}
\end{figure}

First, we focus our attention on the two-body hidden charm decay processes of the $G_0\to\omega J/\psi$ and $G_0\to\rho^0 J/\psi$. The transition with emission of $\omega / \rho^0$ is isospin conserved/violated so that the contributions to the $\omega/\rho^0$ transition amplitudes are given by the summation/difference between the charged and neutral charmed meson loops. In Fig. \ref{fig:widthg0jomgrho}, partial decay widths are shown for different binding energies $E_\mathrm{b}$. As seen, the partial decay width of $G_0\to\omega J/\psi$ is more sensitive to the cutoff $\Lambda$ than that of $G_0\to\rho^0 J/\psi$. With increasing the binding energy $E_\mathrm{b}$, the partial decay width of $G_0\to\omega J/\psi$ increases slightly, while the width of $G_0\to\rho^0 J/\psi$ suffers a small decline. Under our present conditions, the partial width for the isospin-conserved process is 
\begin{equation}
	\Gamma(G_0\to\omega J/\psi) = (0.2 \sim 1)~\mathrm{MeV}\,,
\end{equation}
which is about $(4\sim 20)$ times larger than $\Gamma[X(3872) \to\omega J/\psi] \approx 51~\mathrm{keV} $ \cite{ParticleDataGroup:2024cfk}. 

However, within the molecular model, the partial width for the isospin-breaking process $G_0\to\rho^0 J/\psi$ exhibits strong sensitivity to the proportion of neutral and charged components in the $G_0$ wave function. Hence, the present calculations, performed at equal proportion of the neutral and charged components, should have high uncertainties (might be not very reliable). If the isospin breaking effect for the $G_0\to\rho^0 J/\psi$ results mainly from the mass difference between the charged and neutral meson loops, our model predicts the ratio,
\begin{equation}\label{eq:rvalue}
	\mathcal{R}_{\omega/\rho} = \frac{\Gamma(G_0\to\omega J/\psi)}{\Gamma(G_0\to\rho^0 J/\psi)} = 200\sim 1000\,.
\end{equation}
This ratio should be interpreted with caution due to the substantial uncertainties in $\Gamma(G_0\to\rho^0 J/\psi)$. Nevertheless, the large ratio $\mathcal{R}_{\omega/\rho}$ implies that the contributions from the interference between the charged and neutral meson loops are rather small and the dominant source of isospin violation in $G_0$ decays should come from the different coupling strengths of neutral and charged components. It is noticed that the experimental measurements of the $X(3872)\to \omega (\rho^0) J/\psi$ indicate a large isospin violation, corresponding to a rather small ratio $\mathrm{R}_{\omega/\rho} = \Gamma(X(3872)\to\omega J/\psi) /\Gamma(X(3872)\to\rho^0 J/\psi) \approx (0.7\sim 1.7)$ \cite{Dias:2024zfh,Wu:2021udi,ParticleDataGroup:2024cfk}. Within the molecule framework, this large isospin violation for the $X(3872)$ is attributed primarily to its different effective couplings to the neutral and charged channels \cite{Zhang:2024fxy}. The large ratio predicted for the $G_0\to\omega(\rho^0) J/\psi$ in Eq. \eqref{eq:rvalue} results from that we adopted an equal couplings for the $G_0$ in the calculations. When using different neutral and charged couplings, smaller $\mathcal{R}_{\omega/\rho}$ would be predicted. The future experimental results of the ratio $R_{\omega/\rho}$ for the $G_0$  can help to constrain the effective charged and neutral couplings of the $G_0$ to the neutral and charged components.

\begin{figure}
	\centering
	\includegraphics[width=0.94\linewidth]{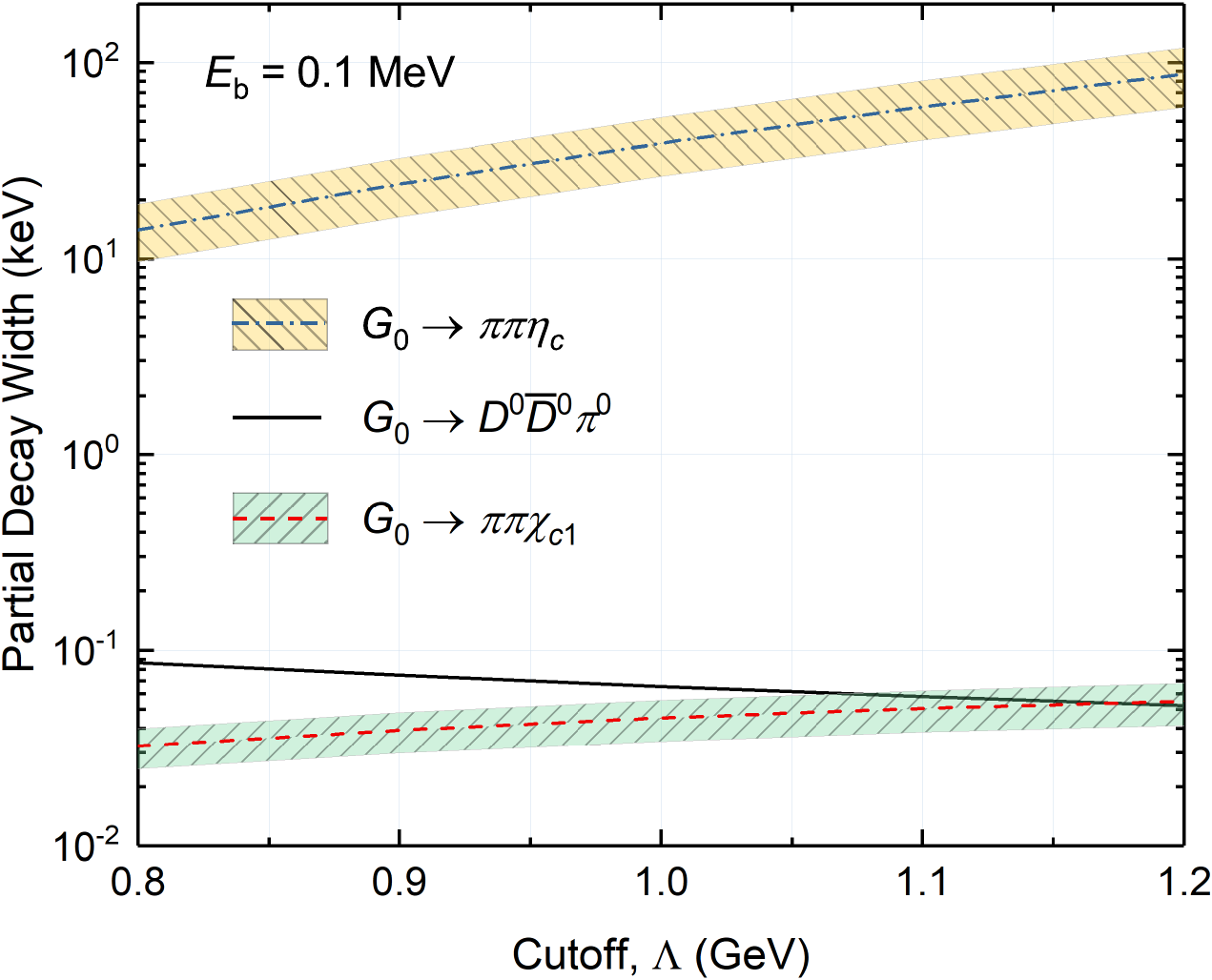}
	\caption{Partial decay widths of the processes $G_0\to\pi^+\pi^-\eta_c(1S)$, $G_0\to\pi^+\pi^-\chi_{c1}(1P)$, and $D^0\bar{D}^0\pi^0$ as a function of the cutoff $\Lambda$, obtained using the binding energy $E_\mathrm{b}=0.1~\mathrm{MeV}$. Further caption text similar to those in Fig. \ref{fig:widthg0jomgrho}.}
	\label{fig:widthg0pipietachi}
\end{figure}

Next, we consider the three-body hidden and open charm decay processes: $G_0\to \pi^+\pi^-\eta_c (1S)$, $\pi^+\pi^-\chi_{c1} (1P)$, and $D^0\bar{D}^0\pi^0$. The predicted partial widths are shown in Fig. \ref{fig:widthg0pipietachi}. Due to the limit of phase space to the open charm channel $G_0\to D^0\bar{D}^0\pi^0$, we here only show the results obtained with $E_\mathrm{b}=0.1~\mathrm{MeV}$. It is seen that the partial decay width of the $G_0\to\pi^+\pi^-\eta_c (1S)$ is
\begin{equation}
	\Gamma[G_0\to\pi^+\pi^-\eta_c (1S)] = 10\sim 120 ~\mathrm{keV}\,.
\end{equation}
In contrast, the $G_0$ decays into the $\pi^+\pi^-\chi_{c1} (1P)$ and $D^0\bar{D}^0\pi^0$ with rather small rate, less than 0.1 keV.
According to the PDG data \cite{ParticleDataGroup:2024cfk}, the partial decay widths for the $X(3872)\to\pi^+\pi^-\eta_c (1S)/\chi_{c1} (1P)$ have upper limits: $\Gamma[X(3872)\to\pi^+\pi^-\eta_c] < 166.6~\mathrm{keV}$ and $\Gamma[X(3872)\to\pi^+\pi^-\chi_{c1}] < 8.33~\mathrm{keV}$. However, the $X(3872)$ decays into the $D^0\bar{D}^0\pi^0$ by nearly $50\%$ \cite{ParticleDataGroup:2024cfk}, which is much larger than the case of $G_0$. 

\begin{figure}
	\centering
	\includegraphics[width=0.98\linewidth]{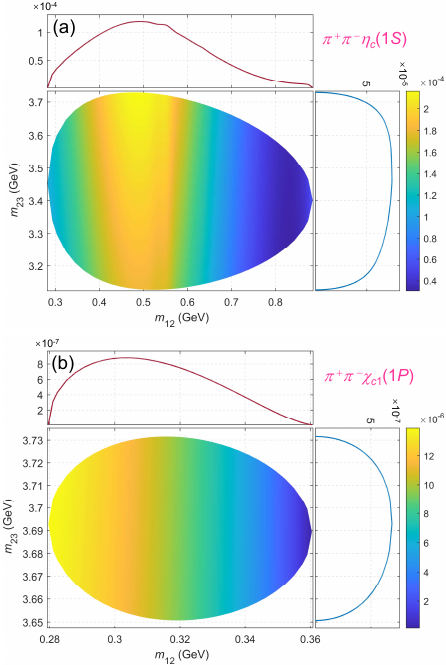}
	\caption{Distributions of the invariant mass of the final states for the processes $G_0\to\pi^+\pi^-\eta_c(1S)$ (a) and $G_0\to\pi^+\pi^-\chi_{c1}(1P)$ (b). The solid lines are the corresponding spectra projected onto the $m_{12}$ (red) and $m_{23}$ (blue) axis, namely $(\mathrm{d} \Gamma / \mathrm{d} m_{12})$ and $(\mathrm{d} \Gamma / \mathrm{d} m_{23})$, where $m_{12}$ is the $\pi\pi$ invariant mass and $m_{23}$ is the $\pi\eta_c(1S)$ ($\pi\chi_{c1}(1P)$ invariant mass. The calculations were performed using $E_\mathrm{b}=0.1~\mathrm{MeV}$, $\Lambda=1.0~\mathrm{GeV}$, and $\alpha = 1.0$.}
	\label{fig:invariantmassspectra}
\end{figure}

To facilitate comparison between the present results and the future experimental measurements of the $G_0$, the invariant mass spectra of the final particles for the processes $G_0\to\pi^+\pi^-\eta_c (1S)$ and $G_0\to\pi^+\pi^-\chi_{c1} (1P)$ are shown in Figs. \ref{fig:invariantmassspectra}(a) and (b), respectively. In view of the decay mechanism we adopt for these two three-body decays in which the two pions are produced via the intermediate mesons $f_0(500)$ and $f_0(980)$ shown in Fig. \ref{fig:feynmandiagtwopi}, the two-pion invariant mass distributions exhibit the feature structures induced by the introduced $f_0(500)$ and $f_0(980)$, as expected. The invariant mass spectra could give a direct test of the validity of the diponic decay mechanism we use here since the spectrum pattern, unlike the absolute partial width, is nearly independent of the cutoff parameters $\Lambda$ and $\alpha$.

\begin{figure}
	\centering
	\includegraphics[width=0.94\linewidth]{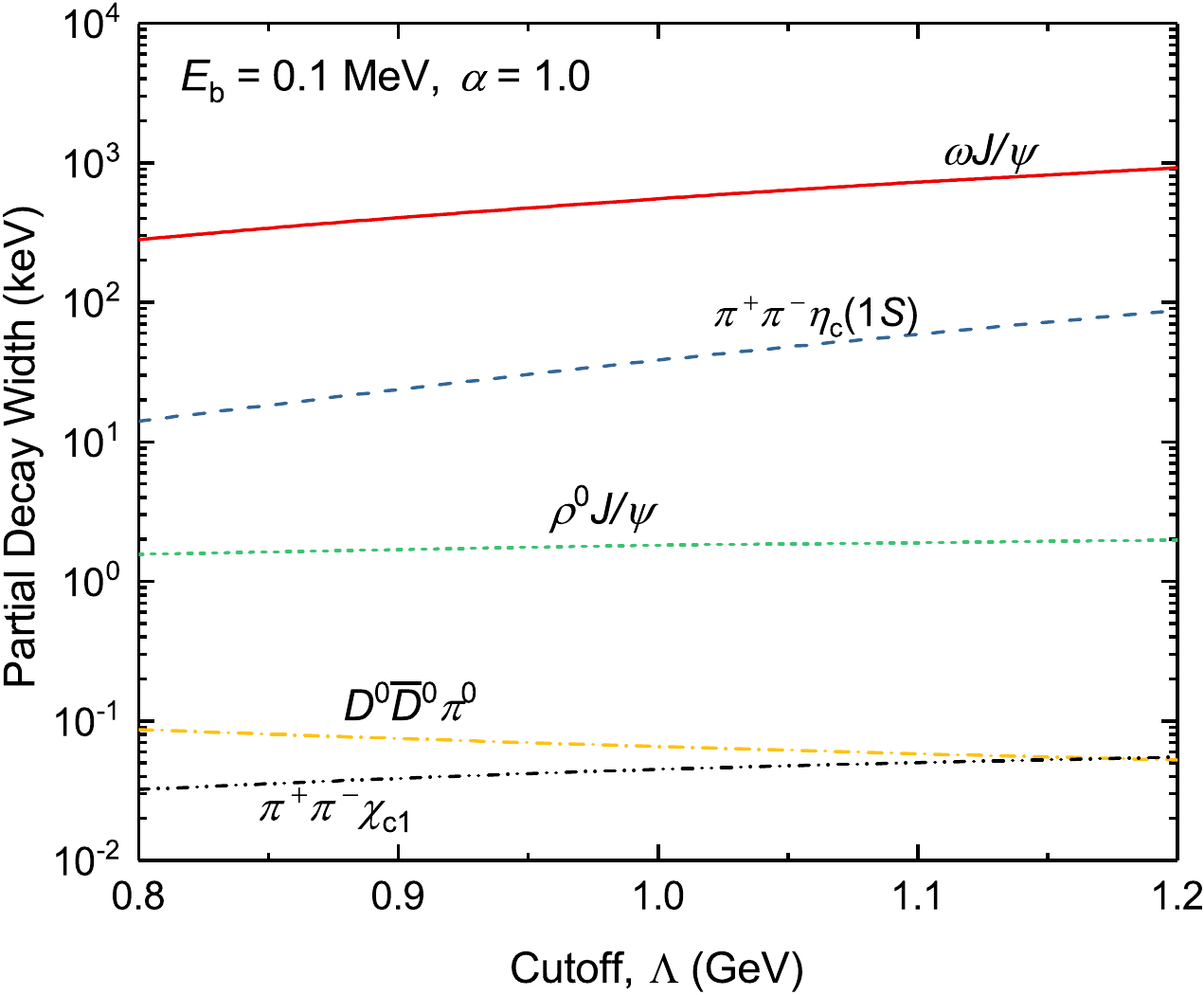}
	\caption{Partial widths of the $G_0$ decaying into different final states as indicated in the graph. The results were obtained using $E_\mathrm{b}=0.1~\mathrm{MeV}$ and $\alpha=1.0$.}
	\label{fig:widthalldecays}
\end{figure}

In Fig. \ref{fig:widthg0pipietachi}, although we do not show the results for other binding energies, for example, $E_\mathrm{b} = 5~\mathrm{MeV}$, the variation of the partial decay widths induced by the binding energy is found to be small (see Table \ref{tab:decaywidths}). For comparison, Fig. \ref{fig:widthalldecays} displays all partial decay widths for the considered hadronic processes, obtained using the cutoff parameter of $\alpha=1.0$ and the binding energy of $E_\mathrm{b}=0.1~\mathrm{MeV}$. It is seen that among the hidden-charm modes, the $G_0\to\omega J/\psi$ is the dominant channel. The open-charm channel $G_0\to D^0\bar{D}^0\pi^0$ exhibits quite small decay width. As mentioned above, the partial decay widths for the isospin-violated processes, for example, $G_0\to\rho^0 J/\psi$, are quite sensitive to the proportion of the neutral and charged components in the molecular state $G_0$. When the proportion of the neutral and charged components is unequal, the partial decay widths for the isospin-violated processes (e.g., $G_0\to\rho^0 J/\psi$ and $\pi^0\chi_{c0(c2)}$) would be enhanced greatly.

In Table \ref{tab:decaywidths}, we summarize the partial widths for the hadronic decays considered in this work. The total hidden-charm decay widths estimated in this work can reach a couple of MeV, though with significant uncertainties arising from the model cutoff parameters $\Lambda$ and $\alpha$. Future BESIII or Belle II measurements of the possible $G_0$ decay channels will provide constraints on the model parameters. In particular, precise measurement of the width ratio $\mathcal{R}_{\omega/\rho} = \Gamma(G_0\to\omega J/\psi)/ \Gamma(G_0\to\rho^0 J/\psi)$ is very helpful in determining the proportion of the neutral and charged components in the $G_0$ if it is of molecular structure.

\begin{table}
	\caption{The partial decay widths (in units of keV) of the $G_0$ into different final states we considered. The width range is due to the cutoff $\Lambda = (0.8\sim 1.2)~\mathrm{GeV}$  and $\alpha = 0.8\sim 1.2$.}
	\label{tab:decaywidths}
	\begin{ruledtabular}
		\begin{tabular}{lccc}
			$E_\mathrm{b}\,(\mathrm{MeV})$ & 0.1   & 5  & 10 \\
			\colrule
			$\omega J/\psi\,(10^{3})$  & $0.2\sim 1.3$  & $0.3\sim 1.9$  & $0.4\sim 2.6$ \\
			$\rho^0 J/\psi$ & $1.1\sim 2.6$ & $1.0\sim 2.5$  & $0.9\sim 2.3$ \\
			$\pi^+\pi^-\eta_c$& $9.6\sim 119.1$ & $12.0\sim 99.9$& $\cdots$ \\   
			$\pi^+\pi^-\chi_{c1}\,(10^{-2})$&$2.5\sim 6.8$&$2.5\sim 6.2$&$\cdots$ \\
			$D^0\bar{D}^0\pi^0$&$0.05\sim 0.09$&$\cdots$&$\cdots$\\
			\colrule
			Total (MeV) & $0.2\sim 1.4$& $0.3\sim 2.0$ &$\cdots$
		\end{tabular}
	\end{ruledtabular}
\end{table}

\section{Summary}\label{sec:summary}

Hadronic decays of the possible $P$-wave $D\bar{D}^\ast / \bar{D}D^\ast$ molecular pseudoscalar state were studied using an effective Lagrangian approach. This work was motivated by recent theoretical interpretation of the new resonance $G(3900)$ recently observed by BESIII Collaboration \cite{BESIII:2024ths} as the $P$-wave $D\bar{D}^\ast/ \bar{D}D^\ast$ vector state using a unified meson-exchange model \cite{Lin:2024qcq}. In particular, the model predicts the possible existence of other $P$-wave $D\bar{D}^\ast / \bar{D}D^\ast$ molecular states with distinct quantum numbers. In this work, we focus on the pseudoscalar state that carries the quantum numbers $J^{PC} = 0^{-+}$, already predicted by the theoretical work \cite{Lin:2024qcq,Chen:2025gxe}. Based on the suggestions in Ref. \cite{Lin:2024qcq}, the partial widths of the hidden-charm hadronic decay processes $G_0\to \omega(\rho^0) J/\psi $, $\pi^+\pi^- \eta_c (1S)$, $\pi^+\pi^- \chi_{c1} (1P)$, and the open-charm $D^0\bar{D}^0\pi^0$ are predicted.

In the current model, the hidden-charm decay modes are dominated by the $G_0\to\omega J/\psi$ and $G_0\to\pi^+\pi^-\eta_c (1S)$. For the $G_0\to\omega J/\psi$, the partial decay width can reach 1 MeV, while the partial decay width for the $G_0\to\pi^+\pi^-\eta_c (1S)$ is smaller by an order of magnitude, i.e., 0.1 MeV. Within the molecular framework, the partial decay width of the open-charm channel $G_0\to D^0\bar{D}^0\pi^0$ is predicted to be of the order of 0.1 keV, contrasting significantly with that for $X(3872)$. The isospin-violated decays, for instance, the $G_0\to\rho^0 J/\psi$ and $G_0\to\pi^0\chi_{c0(c2)}$, might also be important, if the neutral and charged components in the molecular $G_0$ are of unequal proportion, similar to the case of $X(3872)$. In terms of our present predictions, we suggest BESIII and Belle II to search for the $P$-wave $D\bar{D}^\ast / \bar{D}D^\ast$ molecular state with $J^{PC} = 0^{-+}$ in the hidden-charm processes $G_0 \to \omega J/\psi$ or $G_0 \to \pi^+\pi^-\eta_c(1S)$. Given cascade decay $G_0\to\omega J/\psi \to \pi^+\pi^-\pi^0 J/\psi$, the mass peak of $G_0(3900)$ might be found by reconstructing the final particles $\pi^+\pi^-\pi^0 J/\psi$.
	
\begin{acknowledgments}\label{sec:acknowledgements}
This work is partly supported by the National Natural Science Foundation of China under Grants No. 12475081, No. 12405093, and No. 12105153, as well as supported, in part, by National Key Research and Development Program under Grant No. 2024YFA1610504. It is also supported by Taishan Scholar Project of Shandong Province (Grant No. tsqn202103062) and the Natural Science Foundation of Shandong Province under Grants No. ZR2025MS04, No. ZR2021MA082, and No. ZR2022ZD26.
\end{acknowledgments}

	\bibliography{particlePhys.bib}

\providecommand{\noopsort}[1]{}
\begin{thebibliography}{96}%
\makeatletter
\providecommand \@ifxundefined [1]{%
 \@ifx{#1\undefined}
}%
\providecommand \@ifnum [1]{%
 \ifnum #1\expandafter \@firstoftwo
 \else \expandafter \@secondoftwo
 \fi
}%
\providecommand \@ifx [1]{%
 \ifx #1\expandafter \@firstoftwo
 \else \expandafter \@secondoftwo
 \fi
}%
\providecommand \natexlab [1]{#1}%
\providecommand \enquote  [1]{``#1''}%
\providecommand \bibnamefont  [1]{#1}%
\providecommand \bibfnamefont [1]{#1}%
\providecommand \citenamefont [1]{#1}%
\providecommand \href@noop [0]{\@secondoftwo}%
\providecommand \href [0]{\begingroup \@sanitize@url \@href}%
\providecommand \@href[1]{\@@startlink{#1}\@@href}%
\providecommand \@@href[1]{\endgroup#1\@@endlink}%
\providecommand \@sanitize@url [0]{\catcode `\\12\catcode `\$12\catcode
  `\&12\catcode `\#12\catcode `\^12\catcode `\_12\catcode `\%12\relax}%
\providecommand \@@startlink[1]{}%
\providecommand \@@endlink[0]{}%
\providecommand \url  [0]{\begingroup\@sanitize@url \@url }%
\providecommand \@url [1]{\endgroup\@href {#1}{\urlprefix }}%
\providecommand \urlprefix  [0]{URL }%
\providecommand \Eprint [0]{\href }%
\providecommand \doibase [0]{https://doi.org/}%
\providecommand \selectlanguage [0]{\@gobble}%
\providecommand \bibinfo  [0]{\@secondoftwo}%
\providecommand \bibfield  [0]{\@secondoftwo}%
\providecommand \translation [1]{[#1]}%
\providecommand \BibitemOpen [0]{}%
\providecommand \bibitemStop [0]{}%
\providecommand \bibitemNoStop [0]{.\EOS\space}%
\providecommand \EOS [0]{\spacefactor3000\relax}%
\providecommand \BibitemShut  [1]{\csname bibitem#1\endcsname}%
\let\auto@bib@innerbib\@empty
\bibitem [{\citenamefont {Choi}\ \emph {et~al.}(2003)\citenamefont {Choi},
  \citenamefont {Olsen}, \citenamefont {Abe} \emph {et~al.}}]{Belle:2003nnu}%
  \BibitemOpen
  \bibfield  {author} {\bibinfo {author} {\bibfnamefont {S.}~\bibnamefont
  {Choi}}, \bibinfo {author} {\bibfnamefont {S.}~\bibnamefont {Olsen}},
  \bibinfo {author} {\bibfnamefont {K.}~\bibnamefont {Abe}}, \emph {et~al.}
  (\bibinfo {collaboration} {Belle}),\ }\bibfield  {title} {\bibinfo {title}
  {{Observation of a Narrow Charmoniumlike State in Exclusive
  ${B}^{\ifmmode\pm\else\textpm\fi{}}\ensuremath{\rightarrow}{K}^{\ifmmode\pm\else\textpm\fi{}}{\ensuremath{\pi}}^{+}{\ensuremath{\pi}}^{\ensuremath{-}}J/\ensuremath{\psi}$
  Decays}},\ }\href {https://doi.org/10.1103/PhysRevLett.91.262001} {\bibfield
  {journal} {\bibinfo  {journal} {Phys. Rev. Lett.}\ }\textbf {\bibinfo
  {volume} {91}},\ \bibinfo {pages} {262001} (\bibinfo {year} {2003})},\
  \Eprint {https://arxiv.org/abs/hep-ex/0309032} {arXiv:hep-ex/0309032}
  \BibitemShut {NoStop}%
\bibitem [{\citenamefont {Ablikim}\ \emph {et~al.}(2013)\citenamefont
  {Ablikim}, \citenamefont {Achasov}, \citenamefont {Ai} \emph
  {et~al.}}]{BESIII:2013ris}%
  \BibitemOpen
  \bibfield  {author} {\bibinfo {author} {\bibfnamefont {M.}~\bibnamefont
  {Ablikim}}, \bibinfo {author} {\bibfnamefont {M.}~\bibnamefont {Achasov}},
  \bibinfo {author} {\bibfnamefont {X.}~\bibnamefont {Ai}}, \emph {et~al.}
  (\bibinfo {collaboration} {BESIII}),\ }\bibfield  {title} {\bibinfo {title}
  {{Observation of a Charged Charmoniumlike Structure in
  ${e}^{\mathbf{+}}{e}^{\mathbf{\ensuremath{-}}}\ensuremath{\rightarrow}{\ensuremath{\pi}}^{\mathbf{+}}{\ensuremath{\pi}}^{\mathbf{\ensuremath{-}}}J/\ensuremath{\psi}$
  at $\sqrt{s}\mathbf{=}4.26\text{ }\text{ }\mathrm{GeV}$}},\ }\href
  {https://doi.org/10.1103/PhysRevLett.110.252001} {\bibfield  {journal}
  {\bibinfo  {journal} {Phys. Rev. Lett.}\ }\textbf {\bibinfo {volume} {110}},\
  \bibinfo {pages} {252001} (\bibinfo {year} {2013})},\ \Eprint
  {https://arxiv.org/abs/1303.5949} {arXiv:1303.5949 [hep-ex]} \BibitemShut
  {NoStop}%
\bibitem [{\citenamefont {Liu}\ \emph {et~al.}(2013)\citenamefont {Liu},
  \citenamefont {Shen}, \citenamefont {Yuan} \emph {et~al.}}]{Belle:2013yex}%
  \BibitemOpen
  \bibfield  {author} {\bibinfo {author} {\bibfnamefont {Z.}~\bibnamefont
  {Liu}}, \bibinfo {author} {\bibfnamefont {C.}~\bibnamefont {Shen}}, \bibinfo
  {author} {\bibfnamefont {C.}~\bibnamefont {Yuan}}, \emph {et~al.} (\bibinfo
  {collaboration} {Belle}),\ }\bibfield  {title} {\bibinfo {title} {Study of
  ${e}^{\mathbf{+}}{e}^{\mathbf{\ensuremath{-}}}\ensuremath{\rightarrow}{\ensuremath{\pi}}^{\mathbf{+}}{\ensuremath{\pi}}^{\mathbf{\ensuremath{-}}}{J}/\ensuremath{\psi}$
  and observation of a charged charmoniumlike state at belle},\ }\href
  {https://doi.org/10.1103/PhysRevLett.110.252002} {\bibfield  {journal}
  {\bibinfo  {journal} {Phys. Rev. Lett.}\ }\textbf {\bibinfo {volume} {110}},\
  \bibinfo {pages} {252002} (\bibinfo {year} {2013})},\ \Eprint
  {https://arxiv.org/abs/1304.0121} {arXiv:1304.0121 [hep-ex]} \BibitemShut
  {NoStop}%
\bibitem [{\citenamefont {Abazov}\ \emph {et~al.}(2018)\citenamefont {Abazov},
  \citenamefont {Abbott}, \citenamefont {Acharya} \emph {et~al.}}]{D0:2018wyb}%
  \BibitemOpen
  \bibfield  {author} {\bibinfo {author} {\bibfnamefont {V.~M.}\ \bibnamefont
  {Abazov}}, \bibinfo {author} {\bibfnamefont {B.~K.}\ \bibnamefont {Abbott}},
  \bibinfo {author} {\bibfnamefont {B.~S.}\ \bibnamefont {Acharya}}, \emph
  {et~al.} (\bibinfo {collaboration} {D0}),\ }\bibfield  {title} {\bibinfo
  {title} {Evidence for ${Z}_{c}^{\ifmmode\pm\else\textpm\fi{}}(3900)$ in
  semi-inclusive decays of $b$-flavored hadrons},\ }\href
  {https://doi.org/10.1103/PhysRevD.98.052010} {\bibfield  {journal} {\bibinfo
  {journal} {Phys. Rev. D}\ }\textbf {\bibinfo {volume} {98}},\ \bibinfo
  {pages} {052010} (\bibinfo {year} {2018})},\ \Eprint
  {https://arxiv.org/abs/1807.00183} {arXiv:1807.00183 [hep-ex]} \BibitemShut
  {NoStop}%
\bibitem [{\citenamefont {Ablikim}\ \emph {et~al.}(2021)\citenamefont
  {Ablikim}, \citenamefont {Achasov}, \citenamefont {Adlarson} \emph
  {et~al.}}]{BESIII:2020qkh}%
  \BibitemOpen
  \bibfield  {author} {\bibinfo {author} {\bibfnamefont {M.}~\bibnamefont
  {Ablikim}}, \bibinfo {author} {\bibfnamefont {M.~N.}\ \bibnamefont
  {Achasov}}, \bibinfo {author} {\bibfnamefont {P.~A.}\ \bibnamefont
  {Adlarson}}, \emph {et~al.} (\bibinfo {collaboration} {BESIII}),\ }\bibfield
  {title} {\bibinfo {title} {Observation of a near-threshold structure in the
  ${K}^{+}$ recoil-mass spectra in
  ${e}^{+}{e}^{\ensuremath{-}}\ensuremath{\rightarrow}{K}^{+}({D}_{s}^{\ensuremath{-}}{D}^{*0}+{D}_{s}^{*\ensuremath{-}}{D}^{0})$},\
  }\href {https://doi.org/10.1103/PhysRevLett.126.102001} {\bibfield  {journal}
  {\bibinfo  {journal} {Phys. Rev. Lett.}\ }\textbf {\bibinfo {volume} {126}},\
  \bibinfo {pages} {102001} (\bibinfo {year} {2021})},\ \Eprint
  {https://arxiv.org/abs/2011.07855} {arXiv:2011.07855 [hep-ex]} \BibitemShut
  {NoStop}%
\bibitem [{\citenamefont {Aubert}\ \emph {et~al.}(2005)\citenamefont {Aubert}
  \emph {et~al.}}]{BaBar:2005hhc}%
  \BibitemOpen
  \bibfield  {author} {\bibinfo {author} {\bibfnamefont {B.}~\bibnamefont
  {Aubert}} \emph {et~al.} (\bibinfo {collaboration} {BaBar}),\ }\bibfield
  {title} {\bibinfo {title} {{Observation of a broad structure in the $\pi^+
  \pi^- J/\psi$ mass spectrum around 4.26-GeV/c$^2$}},\ }\href
  {https://doi.org/10.1103/PhysRevLett.95.142001} {\bibfield  {journal}
  {\bibinfo  {journal} {Phys. Rev. Lett.}\ }\textbf {\bibinfo {volume} {95}},\
  \bibinfo {pages} {142001} (\bibinfo {year} {2005})},\ \Eprint
  {https://arxiv.org/abs/hep-ex/0506081} {arXiv:hep-ex/0506081} \BibitemShut
  {NoStop}%
\bibitem [{\citenamefont {Coan}\ \emph {et~al.}(2006)\citenamefont {Coan},
  \citenamefont {Gao}, \citenamefont {Liu} \emph {et~al.}}]{CLEO:2006ike}%
  \BibitemOpen
  \bibfield  {author} {\bibinfo {author} {\bibfnamefont {T.}~\bibnamefont
  {Coan}}, \bibinfo {author} {\bibfnamefont {Y.}~\bibnamefont {Gao}}, \bibinfo
  {author} {\bibfnamefont {F.}~\bibnamefont {Liu}}, \emph {et~al.} (\bibinfo
  {collaboration} {CLEO}),\ }\bibfield  {title} {\bibinfo {title} {{Charmonium
  Decays of $Y(4260)$, $\ensuremath{\psi}(4160)$, and
  $\ensuremath{\psi}(4040)$}},\ }\href
  {https://doi.org/10.1103/PhysRevLett.96.162003} {\bibfield  {journal}
  {\bibinfo  {journal} {Phys. Rev. Lett.}\ }\textbf {\bibinfo {volume} {96}},\
  \bibinfo {pages} {162003} (\bibinfo {year} {2006})},\ \Eprint
  {https://arxiv.org/abs/hep-ex/0602034} {arXiv:hep-ex/0602034} \BibitemShut
  {NoStop}%
\bibitem [{\citenamefont {Yuan}\ \emph {et~al.}(2007)\citenamefont {Yuan} \emph
  {et~al.}}]{Belle:2007dxy}%
  \BibitemOpen
  \bibfield  {author} {\bibinfo {author} {\bibfnamefont {C.~Z.}\ \bibnamefont
  {Yuan}} \emph {et~al.} (\bibinfo {collaboration} {Belle}),\ }\bibfield
  {title} {\bibinfo {title} {{Measurement of $e^+e^-\to\pi^+\pi^- J/\psi$
  Cross-Section via Initial State Radiation at {{Belle}}}},\ }\href
  {https://doi.org/10.1103/PhysRevLett.99.182004} {\bibfield  {journal}
  {\bibinfo  {journal} {Phys. Rev. Lett.}\ }\textbf {\bibinfo {volume} {99}},\
  \bibinfo {pages} {182004} (\bibinfo {year} {2007})},\ \Eprint
  {https://arxiv.org/abs/0707.2541} {arXiv:0707.2541 [hep-ex]} \BibitemShut
  {NoStop}%
\bibitem [{\citenamefont {Ablikim}\ \emph {et~al.}(2017)\citenamefont {Ablikim}
  \emph {et~al.}}]{BESIII:2016bnd}%
  \BibitemOpen
  \bibfield  {author} {\bibinfo {author} {\bibfnamefont {M.}~\bibnamefont
  {Ablikim}} \emph {et~al.} (\bibinfo {collaboration} {BESIII}),\ }\bibfield
  {title} {\bibinfo {title} {{Precise measurement of the $e^+e^-\to
  \pi^+\pi^-J/\psi$ cross section at center-of-mass energies from 3.77 to 4.60
  GeV}},\ }\href {https://doi.org/10.1103/PhysRevLett.118.092001} {\bibfield
  {journal} {\bibinfo  {journal} {Phys. Rev. Lett.}\ }\textbf {\bibinfo
  {volume} {118}},\ \bibinfo {pages} {092001} (\bibinfo {year} {2017})},\
  \Eprint {https://arxiv.org/abs/1611.01317} {arXiv:1611.01317 [hep-ex]}
  \BibitemShut {NoStop}%
\bibitem [{\citenamefont {Ablikim}\ \emph
  {et~al.}(2022{\natexlab{a}})\citenamefont {Ablikim} \emph
  {et~al.}}]{BESIII:2022joj}%
  \BibitemOpen
  \bibfield  {author} {\bibinfo {author} {\bibfnamefont {M.}~\bibnamefont
  {Ablikim}} \emph {et~al.} (\bibinfo {collaboration} {BESIII}),\ }\bibfield
  {title} {\bibinfo {title} {{Observation of the $Y(4230)$ and a new structure
  in $e^+ e^-\to K^+K^- J/\psi$}},\ }\href
  {https://doi.org/10.1088/1674-1137/ac945c} {\bibfield  {journal} {\bibinfo
  {journal} {Chin. Phys. C}\ }\textbf {\bibinfo {volume} {46}},\ \bibinfo
  {pages} {111002} (\bibinfo {year} {2022}{\natexlab{a}})},\ \Eprint
  {https://arxiv.org/abs/2204.07800} {arXiv:2204.07800 [hep-ex]} \BibitemShut
  {NoStop}%
\bibitem [{\citenamefont {Ablikim}\ \emph
  {et~al.}(2022{\natexlab{b}})\citenamefont {Ablikim}, \citenamefont {Achasov},
  \citenamefont {Adlarson} \emph {et~al.}}]{BESIII:2022qal}%
  \BibitemOpen
  \bibfield  {author} {\bibinfo {author} {\bibfnamefont {M.}~\bibnamefont
  {Ablikim}}, \bibinfo {author} {\bibfnamefont {M.~N.}\ \bibnamefont
  {Achasov}}, \bibinfo {author} {\bibfnamefont {P.~A.}\ \bibnamefont
  {Adlarson}}, \emph {et~al.} (\bibinfo {collaboration} {BESIII}),\ }\bibfield
  {title} {\bibinfo {title} {{Study of the resonance structures in $e^{+}e^{-}
  \rightarrow \pi^{+}\pi^{-}J/\psi$ process}},\ }\href
  {https://doi.org/10.1103/PhysRevD.106.072001} {\bibfield  {journal} {\bibinfo
   {journal} {Phys. Rev. D}\ }\textbf {\bibinfo {volume} {106}},\ \bibinfo
  {pages} {072001} (\bibinfo {year} {2022}{\natexlab{b}})},\ \Eprint
  {https://arxiv.org/abs/2206.08554} {arXiv:2206.08554 [hep-ex]} \BibitemShut
  {NoStop}%
\bibitem [{\citenamefont {Ablikim}\ \emph {et~al.}(2023)\citenamefont
  {Ablikim}, \citenamefont {Achasov}, \citenamefont {Adlarson} \emph
  {et~al.}}]{BESIII:2023cmv}%
  \BibitemOpen
  \bibfield  {author} {\bibinfo {author} {\bibfnamefont {M.}~\bibnamefont
  {Ablikim}}, \bibinfo {author} {\bibfnamefont {M.}~\bibnamefont {Achasov}},
  \bibinfo {author} {\bibfnamefont {P.}~\bibnamefont {Adlarson}}, \emph
  {et~al.} (\bibinfo {collaboration} {BESIII}),\ }\bibfield  {title} {\bibinfo
  {title} {Observation of three charmoniumlike states with
  ${J}^{PC}={1}^{\ensuremath{-}\ensuremath{-}}$ in
  ${e}^{+}{e}^{\ensuremath{-}}\ensuremath{\rightarrow}{D}^{*0}{D}^{*\ensuremath{-}}{\ensuremath{\pi}}^{+}$},\
  }\href {https://doi.org/10.1103/PhysRevLett.130.121901} {\bibfield  {journal}
  {\bibinfo  {journal} {Phys. Rev. Lett.}\ }\textbf {\bibinfo {volume} {130}},\
  \bibinfo {pages} {121901} (\bibinfo {year} {2023})},\ \Eprint
  {https://arxiv.org/abs/2301.07321} {arXiv:2301.07321 [hep-ex]} \BibitemShut
  {NoStop}%
\bibitem [{\citenamefont {Wang}\ \emph {et~al.}(2007)\citenamefont {Wang} \emph
  {et~al.}}]{Belle:2007umv}%
  \BibitemOpen
  \bibfield  {author} {\bibinfo {author} {\bibfnamefont {X.~L.}\ \bibnamefont
  {Wang}} \emph {et~al.} (\bibinfo {collaboration} {Belle}),\ }\bibfield
  {title} {\bibinfo {title} {{Observation of Two Resonant Structures in
  $e^+e^-\to\pi^+\pi^-\psi(2S)$ via Initial State Radiation at Belle}},\ }\href
  {https://doi.org/10.1103/PhysRevLett.99.142002} {\bibfield  {journal}
  {\bibinfo  {journal} {Phys. Rev. Lett.}\ }\textbf {\bibinfo {volume} {99}},\
  \bibinfo {pages} {142002} (\bibinfo {year} {2007})},\ \Eprint
  {https://arxiv.org/abs/0707.3699} {arXiv:0707.3699 [hep-ex]} \BibitemShut
  {NoStop}%
\bibitem [{\citenamefont {Bondar}\ \emph {et~al.}(2012)\citenamefont {Bondar},
  \citenamefont {Garmash}, \citenamefont {Mizuk} \emph
  {et~al.}}]{Belle:2011aa}%
  \BibitemOpen
  \bibfield  {author} {\bibinfo {author} {\bibfnamefont {A.}~\bibnamefont
  {Bondar}}, \bibinfo {author} {\bibfnamefont {A.}~\bibnamefont {Garmash}},
  \bibinfo {author} {\bibfnamefont {R.}~\bibnamefont {Mizuk}}, \emph {et~al.}
  (\bibinfo {collaboration} {Belle}),\ }\bibfield  {title} {\bibinfo {title}
  {{Observation of Two Charged Bottomoniumlike Resonances in
  $\ensuremath{\Upsilon}(5S)$ Decays}},\ }\href
  {https://doi.org/10.1103/PhysRevLett.108.122001} {\bibfield  {journal}
  {\bibinfo  {journal} {Phys. Rev. Lett.}\ }\textbf {\bibinfo {volume} {108}},\
  \bibinfo {pages} {122001} (\bibinfo {year} {2012})},\ \Eprint
  {https://arxiv.org/abs/1110.2251} {arXiv:1110.2251 [hep-ex]} \BibitemShut
  {NoStop}%
\bibitem [{\citenamefont {Adachi}\ \emph {et~al.}(2012)\citenamefont {Adachi}
  \emph {et~al.}}]{Belle:2012glq}%
  \BibitemOpen
  \bibfield  {author} {\bibinfo {author} {\bibfnamefont {I.}~\bibnamefont
  {Adachi}} \emph {et~al.} (\bibinfo {collaboration} {Belle}),\ }\href@noop {}
  {\bibinfo {title} {{Evidence for a $Z_b^0(10610)$ in Dalitz analysis of
  $\Upsilon(5S)\to \Upsilon (nS)\pi^0\pi^0$}}} (\bibinfo {year} {2012}),\
  \Eprint {https://arxiv.org/abs/1207.4345} {arXiv:1207.4345 [hep-ex]}
  \BibitemShut {NoStop}%
\bibitem [{\citenamefont {Aaij}\ \emph
  {et~al.}(2022{\natexlab{a}})\citenamefont {Aaij} \emph
  {et~al.}}]{LHCb:2021auc}%
  \BibitemOpen
  \bibfield  {author} {\bibinfo {author} {\bibfnamefont {R.}~\bibnamefont
  {Aaij}} \emph {et~al.} (\bibinfo {collaboration} {LHCb}),\ }\bibfield
  {title} {\bibinfo {title} {Study of the doubly charmed tetraquark
  ${T_{cc}^+}$},\ }\href {https://doi.org/10.1038/s41467-022-30206-w}
  {\bibfield  {journal} {\bibinfo  {journal} {Nature Commun.}\ }\textbf
  {\bibinfo {volume} {13}},\ \bibinfo {pages} {3351} (\bibinfo {year}
  {2022}{\natexlab{a}})},\ \Eprint {https://arxiv.org/abs/2109.01056}
  {arXiv:2109.01056 [hep-ex]} \BibitemShut {NoStop}%
\bibitem [{\citenamefont {Aaij}\ \emph
  {et~al.}(2022{\natexlab{b}})\citenamefont {Aaij} \emph
  {et~al.}}]{LHCb:2021vvq}%
  \BibitemOpen
  \bibfield  {author} {\bibinfo {author} {\bibfnamefont {R.}~\bibnamefont
  {Aaij}} \emph {et~al.} (\bibinfo {collaboration} {LHCb}),\ }\bibfield
  {title} {\bibinfo {title} {Observation of an exotic narrow doubly charmed
  tetraquark},\ }\href {https://doi.org/10.1038/s41567-022-01614-y} {\bibfield
  {journal} {\bibinfo  {journal} {Nature Phys.}\ }\textbf {\bibinfo {volume}
  {18}},\ \bibinfo {pages} {751} (\bibinfo {year} {2022}{\natexlab{b}})},\
  \Eprint {https://arxiv.org/abs/2109.01038} {arXiv:2109.01038 [hep-ex]}
  \BibitemShut {NoStop}%
\bibitem [{\citenamefont {Aaij}\ \emph
  {et~al.}(2015{\natexlab{a}})\citenamefont {Aaij} \emph
  {et~al.}}]{LHCb:2015yax}%
  \BibitemOpen
  \bibfield  {author} {\bibinfo {author} {\bibfnamefont {R.}~\bibnamefont
  {Aaij}} \emph {et~al.} (\bibinfo {collaboration} {LHCb}),\ }\bibfield
  {title} {\bibinfo {title} {{Observation of ${J/\psi p}$ Resonances Consistent
  with Pentaquark States in ${\Lambda_b^0 \to {J}/\psi K^- p}$ Decays}},\
  }\href {https://doi.org/10.1103/PhysRevLett.115.072001} {\bibfield  {journal}
  {\bibinfo  {journal} {Phys. Rev. Lett.}\ }\textbf {\bibinfo {volume} {115}},\
  \bibinfo {pages} {072001} (\bibinfo {year} {2015}{\natexlab{a}})},\ \Eprint
  {https://arxiv.org/abs/1507.03414} {arXiv:1507.03414 [hep-ex]} \BibitemShut
  {NoStop}%
\bibitem [{\citenamefont {Aaij}\ \emph {et~al.}(2019)\citenamefont {Aaij} \emph
  {et~al.}}]{LHCb:2019kea}%
  \BibitemOpen
  \bibfield  {author} {\bibinfo {author} {\bibfnamefont {R.}~\bibnamefont
  {Aaij}} \emph {et~al.} (\bibinfo {collaboration} {LHCb}),\ }\bibfield
  {title} {\bibinfo {title} {Observation of a narrow pentaquark state,
  ${{P}}_c(4312)^+$, and of two-peak structure of the ${{P}}_c(4450)^+$},\
  }\href {https://doi.org/10.1103/PhysRevLett.122.222001} {\bibfield  {journal}
  {\bibinfo  {journal} {Phys. Rev. Lett.}\ }\textbf {\bibinfo {volume} {122}},\
  \bibinfo {pages} {222001} (\bibinfo {year} {2019})},\ \Eprint
  {https://arxiv.org/abs/1904.03947} {arXiv:1904.03947 [hep-ex]} \BibitemShut
  {NoStop}%
\bibitem [{\citenamefont {Aaij}\ \emph {et~al.}(2021)\citenamefont {Aaij} \emph
  {et~al.}}]{LHCb:2020jpq}%
  \BibitemOpen
  \bibfield  {author} {\bibinfo {author} {\bibfnamefont {R.}~\bibnamefont
  {Aaij}} \emph {et~al.} (\bibinfo {collaboration} {LHCb}),\ }\bibfield
  {title} {\bibinfo {title} {{Evidence of a $J/\psi\Lambda$ structure and
  observation of excited $\Xi^-$ states in the $\Xi^-_b \to J/\psi\Lambda K^-$
  decay}},\ }\href {https://doi.org/10.1016/j.scib.2021.02.030} {\bibfield
  {journal} {\bibinfo  {journal} {Sci. Bull.}\ }\textbf {\bibinfo {volume}
  {66}},\ \bibinfo {pages} {1278} (\bibinfo {year} {2021})},\ \Eprint
  {https://arxiv.org/abs/2012.10380} {arXiv:2012.10380 [hep-ex]} \BibitemShut
  {NoStop}%
\bibitem [{\citenamefont {Aaij}\ \emph {et~al.}(2020)\citenamefont {Aaij} \emph
  {et~al.}}]{LHCb:2020bwg}%
  \BibitemOpen
  \bibfield  {author} {\bibinfo {author} {\bibfnamefont {R.}~\bibnamefont
  {Aaij}} \emph {et~al.} (\bibinfo {collaboration} {LHCb}),\ }\bibfield
  {title} {\bibinfo {title} {{Observation of structure in the $J /\psi$ -pair
  mass spectrum}},\ }\href {https://doi.org/10.1016/j.scib.2020.08.032}
  {\bibfield  {journal} {\bibinfo  {journal} {Sci. Bull.}\ }\textbf {\bibinfo
  {volume} {65}},\ \bibinfo {pages} {1983} (\bibinfo {year} {2020})},\ \Eprint
  {https://arxiv.org/abs/2006.16957} {arXiv:2006.16957 [hep-ex]} \BibitemShut
  {NoStop}%
\bibitem [{\citenamefont {Hayrapetyan}\ \emph {et~al.}(2024)\citenamefont
  {Hayrapetyan} \emph {et~al.}}]{CMS:2023owd}%
  \BibitemOpen
  \bibfield  {author} {\bibinfo {author} {\bibfnamefont {A.}~\bibnamefont
  {Hayrapetyan}} \emph {et~al.} (\bibinfo {collaboration} {CMS}),\ }\bibfield
  {title} {\bibinfo {title} {{New Structures in the $J/\psi J/\psi$ Mass
  Spectrum in Proton-Proton Collisions at $\sqrt{s}=13$ TeV}},\ }\href
  {https://doi.org/10.1103/PhysRevLett.132.111901} {\bibfield  {journal}
  {\bibinfo  {journal} {Phys. Rev. Lett.}\ }\textbf {\bibinfo {volume} {132}},\
  \bibinfo {pages} {111901} (\bibinfo {year} {2024})},\ \Eprint
  {https://arxiv.org/abs/2306.07164} {arXiv:2306.07164 [hep-ex]} \BibitemShut
  {NoStop}%
\bibitem [{\citenamefont {Aad}\ \emph {et~al.}(2023)\citenamefont {Aad} \emph
  {et~al.}}]{ATLAS:2023bft}%
  \BibitemOpen
  \bibfield  {author} {\bibinfo {author} {\bibfnamefont {G.}~\bibnamefont
  {Aad}} \emph {et~al.} (\bibinfo {collaboration} {ATLAS}),\ }\bibfield
  {title} {\bibinfo {title} {{Observation of an Excess of Dicharmonium Events
  in the Four-Muon Final State with the ATLAS Detector}},\ }\href
  {https://doi.org/10.1103/PhysRevLett.131.151902} {\bibfield  {journal}
  {\bibinfo  {journal} {Phys. Rev. Lett.}\ }\textbf {\bibinfo {volume} {131}},\
  \bibinfo {pages} {151902} (\bibinfo {year} {2023})},\ \Eprint
  {https://arxiv.org/abs/2304.08962} {arXiv:2304.08962 [hep-ex]} \BibitemShut
  {NoStop}%
\bibitem [{\citenamefont {Wang}(2025)}]{Wang:2025sic}%
  \BibitemOpen
  \bibfield  {author} {\bibinfo {author} {\bibfnamefont {Z.-G.}\ \bibnamefont
  {Wang}},\ }\bibfield  {title} {\bibinfo {title} {Review of the {{QCD}} sum
  rules for exotic states},\ }\bibfield  {journal} {\bibinfo  {journal}
  {arXiv:2502.11351 [hep-ph]}\ }\href
  {https://doi.org/10.48550/arXiv.2502.11351} {10.48550/arXiv.2502.11351}
  (\bibinfo {year} {2025}),\ \Eprint {https://arxiv.org/abs/2502.11351}
  {arXiv:2502.11351 [hep-ph]} \BibitemShut {NoStop}%
\bibitem [{\citenamefont {Chen}\ \emph
  {et~al.}(2025{\natexlab{a}})\citenamefont {Chen}, \citenamefont {Chen},
  \citenamefont {Guo}, \citenamefont {Ma}, \citenamefont {Shen}, \citenamefont
  {Shou}, \citenamefont {Shou}, \citenamefont {Wang}, \citenamefont {Wu},\ and\
  \citenamefont {Zou}}]{Chen:2024eaq}%
  \BibitemOpen
  \bibfield  {author} {\bibinfo {author} {\bibfnamefont {J.-H.}\ \bibnamefont
  {Chen}}, \bibinfo {author} {\bibfnamefont {J.}~\bibnamefont {Chen}}, \bibinfo
  {author} {\bibfnamefont {F.-K.}\ \bibnamefont {Guo}}, \bibinfo {author}
  {\bibfnamefont {Y.-G.}\ \bibnamefont {Ma}}, \bibinfo {author} {\bibfnamefont
  {C.-P.}\ \bibnamefont {Shen}}, \bibinfo {author} {\bibfnamefont {Q.-Y.}\
  \bibnamefont {Shou}}, \bibinfo {author} {\bibfnamefont {Q.}~\bibnamefont
  {Shou}}, \bibinfo {author} {\bibfnamefont {Q.}~\bibnamefont {Wang}}, \bibinfo
  {author} {\bibfnamefont {J.-J.}\ \bibnamefont {Wu}},\ and\ \bibinfo {author}
  {\bibfnamefont {B.-S.}\ \bibnamefont {Zou}},\ }\bibfield  {title} {\bibinfo
  {title} {Production of exotic hadrons in pp and nuclear collisions},\ }\href
  {https://doi.org/10.1007/s41365-025-01664-w} {\bibfield  {journal} {\bibinfo
  {journal} {Nucl. Sci. Tech.}\ }\textbf {\bibinfo {volume} {36}},\ \bibinfo
  {pages} {55} (\bibinfo {year} {2025}{\natexlab{a}})},\ \Eprint
  {https://arxiv.org/abs/2411.18257} {arXiv:2411.18257 [hep-ph]} \BibitemShut
  {NoStop}%
\bibitem [{\citenamefont {Liu}\ \emph {et~al.}(2025{\natexlab{a}})\citenamefont
  {Liu}, \citenamefont {Pan}, \citenamefont {Liu}, \citenamefont {Wu},
  \citenamefont {Lu},\ and\ \citenamefont {Geng}}]{Liu:2024uxn}%
  \BibitemOpen
  \bibfield  {author} {\bibinfo {author} {\bibfnamefont {M.-Z.}\ \bibnamefont
  {Liu}}, \bibinfo {author} {\bibfnamefont {Y.-W.}\ \bibnamefont {Pan}},
  \bibinfo {author} {\bibfnamefont {Z.-W.}\ \bibnamefont {Liu}}, \bibinfo
  {author} {\bibfnamefont {T.-W.}\ \bibnamefont {Wu}}, \bibinfo {author}
  {\bibfnamefont {J.-X.}\ \bibnamefont {Lu}},\ and\ \bibinfo {author}
  {\bibfnamefont {L.-S.}\ \bibnamefont {Geng}},\ }\bibfield  {title} {\bibinfo
  {title} {Three ways to decipher the nature of exotic hadrons: {{Multiplets}},
  three-body hadronic molecules, and correlation functions},\ }\href
  {https://doi.org/10.1016/j.physrep.2024.12.001} {\bibfield  {journal}
  {\bibinfo  {journal} {Phys. Rept.}\ }\textbf {\bibinfo {volume} {1108}},\
  \bibinfo {pages} {2368} (\bibinfo {year} {2025}{\natexlab{a}})},\ \Eprint
  {https://arxiv.org/abs/2404.06399} {arXiv:2404.06399 [hep-ph]} \BibitemShut
  {NoStop}%
\bibitem [{\citenamefont {Chen}\ \emph
  {et~al.}(2016{\natexlab{a}})\citenamefont {Chen}, \citenamefont {Chen},
  \citenamefont {Liu},\ and\ \citenamefont {Zhu}}]{Chen:2016qju}%
  \BibitemOpen
  \bibfield  {author} {\bibinfo {author} {\bibfnamefont {H.-X.}\ \bibnamefont
  {Chen}}, \bibinfo {author} {\bibfnamefont {W.}~\bibnamefont {Chen}}, \bibinfo
  {author} {\bibfnamefont {X.}~\bibnamefont {Liu}},\ and\ \bibinfo {author}
  {\bibfnamefont {S.-L.}\ \bibnamefont {Zhu}},\ }\bibfield  {title} {\bibinfo
  {title} {The hidden-charm pentaquark and tetraquark states},\ }\href
  {https://doi.org/10.1016/j.physrep.2016.05.004} {\bibfield  {journal}
  {\bibinfo  {journal} {Phys. Rept.}\ }\textbf {\bibinfo {volume} {639}},\
  \bibinfo {pages} {1} (\bibinfo {year} {2016}{\natexlab{a}})},\ \Eprint
  {https://arxiv.org/abs/1601.02092} {arXiv:1601.02092 [hep-ph]} \BibitemShut
  {NoStop}%
\bibitem [{\citenamefont {Chen}\ \emph {et~al.}(2022)\citenamefont {Chen},
  \citenamefont {Chen}, \citenamefont {Liu}, \citenamefont {Liu},\ and\
  \citenamefont {Zhu}}]{Chen:2022asf}%
  \BibitemOpen
  \bibfield  {author} {\bibinfo {author} {\bibfnamefont {H.-X.}\ \bibnamefont
  {Chen}}, \bibinfo {author} {\bibfnamefont {W.}~\bibnamefont {Chen}}, \bibinfo
  {author} {\bibfnamefont {X.}~\bibnamefont {Liu}}, \bibinfo {author}
  {\bibfnamefont {Y.-R.}\ \bibnamefont {Liu}},\ and\ \bibinfo {author}
  {\bibfnamefont {S.-L.}\ \bibnamefont {Zhu}},\ }\bibfield  {title} {\bibinfo
  {title} {An updated review of the new hadron states},\ }\href
  {https://doi.org/10.1088/1361-6633/aca3b6} {\bibfield  {journal} {\bibinfo
  {journal} {Rept. Prog. Phys.}\ }\textbf {\bibinfo {volume} {86}},\ \bibinfo
  {pages} {026201} (\bibinfo {year} {2022})},\ \Eprint
  {https://arxiv.org/abs/2204.02649} {arXiv:2204.02649 [hep-ph]} \BibitemShut
  {NoStop}%
\bibitem [{\citenamefont {Meng}\ \emph {et~al.}(2023)\citenamefont {Meng},
  \citenamefont {Wang}, \citenamefont {Wang},\ and\ \citenamefont
  {Zhu}}]{Meng:2022ozq}%
  \BibitemOpen
  \bibfield  {author} {\bibinfo {author} {\bibfnamefont {L.}~\bibnamefont
  {Meng}}, \bibinfo {author} {\bibfnamefont {B.}~\bibnamefont {Wang}}, \bibinfo
  {author} {\bibfnamefont {G.-J.}\ \bibnamefont {Wang}},\ and\ \bibinfo
  {author} {\bibfnamefont {S.-L.}\ \bibnamefont {Zhu}},\ }\bibfield  {title}
  {\bibinfo {title} {Chiral perturbation theory for heavy hadrons and chiral
  effective field theory for heavy hadronic molecules},\ }\href
  {https://doi.org/10.1016/j.physrep.2023.04.003} {\bibfield  {journal}
  {\bibinfo  {journal} {Phys. Rept.}\ }\textbf {\bibinfo {volume} {1019}},\
  \bibinfo {pages} {1} (\bibinfo {year} {2023})},\ \Eprint
  {https://arxiv.org/abs/2204.08716} {arXiv:2204.08716 [hep-ph]} \BibitemShut
  {NoStop}%
\bibitem [{\citenamefont {Ablikim}\ \emph {et~al.}(2020)\citenamefont {Ablikim}
  \emph {et~al.}}]{BESIII:2020nme}%
  \BibitemOpen
  \bibfield  {author} {\bibinfo {author} {\bibfnamefont {M.}~\bibnamefont
  {Ablikim}} \emph {et~al.} (\bibinfo {collaboration} {BESIII}),\ }\bibfield
  {title} {\bibinfo {title} {Future {{Physics Programme}} of {{BESIII}}},\
  }\href {https://doi.org/10.1088/1674-1137/44/4/040001} {\bibfield  {journal}
  {\bibinfo  {journal} {Chin. Phys. C}\ }\textbf {\bibinfo {volume} {44}},\
  \bibinfo {pages} {040001} (\bibinfo {year} {2020})},\ \Eprint
  {https://arxiv.org/abs/1912.05983} {arXiv:1912.05983 [hep-ex]} \BibitemShut
  {NoStop}%
\bibitem [{\citenamefont {Guo}\ \emph {et~al.}(2020)\citenamefont {Guo},
  \citenamefont {Liu},\ and\ \citenamefont {Sakai}}]{Guo:2019twa}%
  \BibitemOpen
  \bibfield  {author} {\bibinfo {author} {\bibfnamefont {F.-K.}\ \bibnamefont
  {Guo}}, \bibinfo {author} {\bibfnamefont {X.-H.}\ \bibnamefont {Liu}},\ and\
  \bibinfo {author} {\bibfnamefont {S.}~\bibnamefont {Sakai}},\ }\bibfield
  {title} {\bibinfo {title} {Threshold cusps and triangle singularities in
  hadronic reactions},\ }\href {https://doi.org/10.1016/j.ppnp.2020.103757}
  {\bibfield  {journal} {\bibinfo  {journal} {Prog. Part. Nucl. Phys.}\
  }\textbf {\bibinfo {volume} {112}},\ \bibinfo {pages} {103757} (\bibinfo
  {year} {2020})},\ \Eprint {https://arxiv.org/abs/1912.07030}
  {arXiv:1912.07030 [hep-ph]} \BibitemShut {NoStop}%
\bibitem [{\citenamefont {Brambilla}\ \emph {et~al.}(2020)\citenamefont
  {Brambilla}, \citenamefont {Eidelman}, \citenamefont {Hanhart}, \citenamefont
  {Nefediev}, \citenamefont {Shen}, \citenamefont {Thomas}, \citenamefont
  {Vairo},\ and\ \citenamefont {Yuan}}]{Brambilla:2019esw}%
  \BibitemOpen
  \bibfield  {author} {\bibinfo {author} {\bibfnamefont {N.}~\bibnamefont
  {Brambilla}}, \bibinfo {author} {\bibfnamefont {S.}~\bibnamefont {Eidelman}},
  \bibinfo {author} {\bibfnamefont {C.}~\bibnamefont {Hanhart}}, \bibinfo
  {author} {\bibfnamefont {A.}~\bibnamefont {Nefediev}}, \bibinfo {author}
  {\bibfnamefont {C.-P.}\ \bibnamefont {Shen}}, \bibinfo {author}
  {\bibfnamefont {C.~E.}\ \bibnamefont {Thomas}}, \bibinfo {author}
  {\bibfnamefont {A.}~\bibnamefont {Vairo}},\ and\ \bibinfo {author}
  {\bibfnamefont {C.-Z.}\ \bibnamefont {Yuan}},\ }\bibfield  {title} {\bibinfo
  {title} {The xyz states: Experimental and theoretical status and
  perspectives},\ }\href {https://doi.org/10.1016/j.physrep.2020.05.001}
  {\bibfield  {journal} {\bibinfo  {journal} {Phys. Rept.}\ }\textbf {\bibinfo
  {volume} {873}},\ \bibinfo {pages} {1} (\bibinfo {year} {2020})},\ \Eprint
  {https://arxiv.org/abs/1907.07583} {arXiv:1907.07583 [hep-ex]} \BibitemShut
  {NoStop}%
\bibitem [{\citenamefont {Liu}\ \emph {et~al.}(2019)\citenamefont {Liu},
  \citenamefont {Chen}, \citenamefont {Chen}, \citenamefont {Liu},\ and\
  \citenamefont {Zhu}}]{Liu:2019zoy}%
  \BibitemOpen
  \bibfield  {author} {\bibinfo {author} {\bibfnamefont {Y.-R.}\ \bibnamefont
  {Liu}}, \bibinfo {author} {\bibfnamefont {H.-X.}\ \bibnamefont {Chen}},
  \bibinfo {author} {\bibfnamefont {W.}~\bibnamefont {Chen}}, \bibinfo {author}
  {\bibfnamefont {X.}~\bibnamefont {Liu}},\ and\ \bibinfo {author}
  {\bibfnamefont {S.-L.}\ \bibnamefont {Zhu}},\ }\bibfield  {title} {\bibinfo
  {title} {Pentaquark and tetraquark states},\ }\href
  {https://doi.org/10.1016/j.ppnp.2019.04.003} {\bibfield  {journal} {\bibinfo
  {journal} {Prog. Part. Nucl. Phys.}\ }\textbf {\bibinfo {volume} {107}},\
  \bibinfo {pages} {237} (\bibinfo {year} {2019})},\ \Eprint
  {https://arxiv.org/abs/1903.11976} {arXiv:1903.11976 [hep-ph]} \BibitemShut
  {NoStop}%
\bibitem [{\citenamefont {Guo}\ \emph {et~al.}(2018)\citenamefont {Guo},
  \citenamefont {Hanhart}, \citenamefont {Meißner}, \citenamefont {Wang},
  \citenamefont {Zhao},\ and\ \citenamefont {Zou}}]{Guo:2017jvc}%
  \BibitemOpen
  \bibfield  {author} {\bibinfo {author} {\bibfnamefont {F.-K.}\ \bibnamefont
  {Guo}}, \bibinfo {author} {\bibfnamefont {C.}~\bibnamefont {Hanhart}},
  \bibinfo {author} {\bibfnamefont {U.-G.}\ \bibnamefont {Meißner}}, \bibinfo
  {author} {\bibfnamefont {Q.}~\bibnamefont {Wang}}, \bibinfo {author}
  {\bibfnamefont {Q.}~\bibnamefont {Zhao}},\ and\ \bibinfo {author}
  {\bibfnamefont {B.-S.}\ \bibnamefont {Zou}},\ }\bibfield  {title} {\bibinfo
  {title} {Hadronic molecules},\ }\href
  {https://doi.org/10.1103/RevModPhys.90.015004} {\bibfield  {journal}
  {\bibinfo  {journal} {Rev. Mod. Phys.}\ }\textbf {\bibinfo {volume} {90}},\
  \bibinfo {pages} {015004} (\bibinfo {year} {2018})},\ \Eprint
  {https://arxiv.org/abs/1705.00141} {arXiv:1705.00141 [hep-ph]} \BibitemShut
  {NoStop}%
\bibitem [{\citenamefont {Liu}(2014)}]{Liu:2013waa}%
  \BibitemOpen
  \bibfield  {author} {\bibinfo {author} {\bibfnamefont {X.}~\bibnamefont
  {Liu}},\ }\bibfield  {title} {\bibinfo {title} {An overview of {{XYZ}} new
  particles},\ }\href {https://doi.org/10.1007/s11434-014-0407-2} {\bibfield
  {journal} {\bibinfo  {journal} {Chin. Sci. Bull.}\ }\textbf {\bibinfo
  {volume} {59}},\ \bibinfo {pages} {3815} (\bibinfo {year} {2014})},\ \Eprint
  {https://arxiv.org/abs/1312.7408} {arXiv:1312.7408 [hep-ph]} \BibitemShut
  {NoStop}%
\bibitem [{\citenamefont {Brambilla}\ \emph {et~al.}(2011)\citenamefont
  {Brambilla}, \citenamefont {Eidelman},\ and\ \citenamefont {Heltsley~et
  al}}]{Brambilla:2010cs}%
  \BibitemOpen
  \bibfield  {author} {\bibinfo {author} {\bibfnamefont {N.}~\bibnamefont
  {Brambilla}}, \bibinfo {author} {\bibfnamefont {S.}~\bibnamefont
  {Eidelman}},\ and\ \bibinfo {author} {\bibfnamefont {B.~K.}\ \bibnamefont
  {Heltsley~et al}},\ }\bibfield  {title} {\bibinfo {title} {Heavy quarkonium:
  progress, puzzles, and opportunities},\ }\href
  {https://doi.org/10.1140/epjc/s10052-010-1534-9} {\bibfield  {journal}
  {\bibinfo  {journal} {Eur. Phys. J. C}\ }\textbf {\bibinfo {volume} {71}},\
  \bibinfo {pages} {1534} (\bibinfo {year} {2011})},\ \Eprint
  {https://arxiv.org/abs/1010.5827} {arXiv:1010.5827 [hep-ph]} \BibitemShut
  {NoStop}%
\bibitem [{\citenamefont {Cleven}(2014)}]{Cleven:2013rkf}%
  \BibitemOpen
  \bibfield  {author} {\bibinfo {author} {\bibfnamefont {M.}~\bibnamefont
  {Cleven}},\ }\bibfield  {title} {\bibinfo {title} {Systematic study of
  hadronic molecules in the heavy-quark sector},\ }\bibfield  {journal}
  {\bibinfo  {journal} {arXiv:1405.4195 [hep-ph]}\ }\href
  {https://doi.org/10.48550/arXiv.1405.4195} {10.48550/arXiv.1405.4195}
  (\bibinfo {year} {2014}),\ \Eprint {https://arxiv.org/abs/1405.4195}
  {arXiv:1405.4195 [hep-ph]} \BibitemShut {NoStop}%
\bibitem [{\citenamefont {Hanhart}(2025)}]{Hanhart:2025bun}%
  \BibitemOpen
  \bibfield  {author} {\bibinfo {author} {\bibfnamefont {C.}~\bibnamefont
  {Hanhart}},\ }\href@noop {} {\bibinfo {title} {{Hadronic molecules and
  multiquark states}}} (\bibinfo {year} {2025}),\ \Eprint
  {https://arxiv.org/abs/2504.06043} {arXiv:2504.06043 [hep-ph]} \BibitemShut
  {NoStop}%
\bibitem [{\citenamefont {Du}\ \emph {et~al.}(2016)\citenamefont {Du},
  \citenamefont {Meißner},\ and\ \citenamefont {Wang}}]{Du:2016qcr}%
  \BibitemOpen
  \bibfield  {author} {\bibinfo {author} {\bibfnamefont {M.-L.}\ \bibnamefont
  {Du}}, \bibinfo {author} {\bibfnamefont {U.-G.}\ \bibnamefont {Meißner}},\
  and\ \bibinfo {author} {\bibfnamefont {Q.}~\bibnamefont {Wang}},\ }\bibfield
  {title} {\bibinfo {title} {{$P$-wave coupled channel effects in
  electron-positron annihilation}},\ }\href
  {https://doi.org/10.1103/PhysRevD.94.096006} {\bibfield  {journal} {\bibinfo
  {journal} {Phys. Rev. D}\ }\textbf {\bibinfo {volume} {94}},\ \bibinfo
  {pages} {096006} (\bibinfo {year} {2016})},\ \Eprint
  {https://arxiv.org/abs/1608.02537} {arXiv:1608.02537 [hep-ph]} \BibitemShut
  {NoStop}%
\bibitem [{\citenamefont {Guo}\ \emph {et~al.}(2015)\citenamefont {Guo},
  \citenamefont {Hanhart}, \citenamefont {Kalashnikova}, \citenamefont
  {Meißner},\ and\ \citenamefont {Nefediev}}]{Guo:2014taa}%
  \BibitemOpen
  \bibfield  {author} {\bibinfo {author} {\bibfnamefont {F.-K.}\ \bibnamefont
  {Guo}}, \bibinfo {author} {\bibfnamefont {C.}~\bibnamefont {Hanhart}},
  \bibinfo {author} {\bibfnamefont {Y.~S.}\ \bibnamefont {Kalashnikova}},
  \bibinfo {author} {\bibfnamefont {U.-G.}\ \bibnamefont {Meißner}},\ and\
  \bibinfo {author} {\bibfnamefont {A.~V.}\ \bibnamefont {Nefediev}},\
  }\bibfield  {title} {\bibinfo {title} {{What can radiative decays of the
  $X(3872)$ teach us about its nature?}},\ }\href
  {https://doi.org/10.1016/j.physletb.2015.02.013} {\bibfield  {journal}
  {\bibinfo  {journal} {Phys. Lett. B}\ }\textbf {\bibinfo {volume} {742}},\
  \bibinfo {pages} {394} (\bibinfo {year} {2015})},\ \Eprint
  {https://arxiv.org/abs/1410.6712} {arXiv:1410.6712 [hep-ph]} \BibitemShut
  {NoStop}%
\bibitem [{\citenamefont {Dong}\ \emph {et~al.}(2017)\citenamefont {Dong},
  \citenamefont {Faessler},\ and\ \citenamefont {Lyubovitskij}}]{Dong:2017gaw}%
  \BibitemOpen
  \bibfield  {author} {\bibinfo {author} {\bibfnamefont {Y.}~\bibnamefont
  {Dong}}, \bibinfo {author} {\bibfnamefont {A.}~\bibnamefont {Faessler}},\
  and\ \bibinfo {author} {\bibfnamefont {V.~E.}\ \bibnamefont {Lyubovitskij}},\
  }\bibfield  {title} {\bibinfo {title} {Description of heavy exotic resonances
  as molecular states using phenomenological lagrangians},\ }\href
  {https://doi.org/10.1016/j.ppnp.2017.01.002} {\bibfield  {journal} {\bibinfo
  {journal} {Prog. Part. Nucl. Phys.}\ }\textbf {\bibinfo {volume} {94}},\
  \bibinfo {pages} {282} (\bibinfo {year} {2017})}\BibitemShut {NoStop}%
\bibitem [{\citenamefont {Cleven}\ \emph {et~al.}(2014)\citenamefont {Cleven},
  \citenamefont {Wang}, \citenamefont {Guo}, \citenamefont {Hanhart},
  \citenamefont {Meißner},\ and\ \citenamefont {Zhao}}]{Cleven:2013mka}%
  \BibitemOpen
  \bibfield  {author} {\bibinfo {author} {\bibfnamefont {M.}~\bibnamefont
  {Cleven}}, \bibinfo {author} {\bibfnamefont {Q.}~\bibnamefont {Wang}},
  \bibinfo {author} {\bibfnamefont {F.-K.}\ \bibnamefont {Guo}}, \bibinfo
  {author} {\bibfnamefont {C.}~\bibnamefont {Hanhart}}, \bibinfo {author}
  {\bibfnamefont {U.-G.}\ \bibnamefont {Meißner}},\ and\ \bibinfo {author}
  {\bibfnamefont {Q.}~\bibnamefont {Zhao}},\ }\bibfield  {title} {\bibinfo
  {title} {{$Y(4260)$ as the first $S$-wave open charm vector molecular
  state?}},\ }\href {https://doi.org/10.1103/PhysRevD.90.074039} {\bibfield
  {journal} {\bibinfo  {journal} {Phys. Rev. D}\ }\textbf {\bibinfo {volume}
  {90}},\ \bibinfo {pages} {074039} (\bibinfo {year} {2014})},\ \Eprint
  {https://arxiv.org/abs/1310.2190} {arXiv:1310.2190 [hep-ph]} \BibitemShut
  {NoStop}%
\bibitem [{\citenamefont {Guo}\ \emph {et~al.}(2013)\citenamefont {Guo},
  \citenamefont {Hanhart}, \citenamefont {Meißner}, \citenamefont {Wang},\
  and\ \citenamefont {Zhao}}]{Guo:2013zbw}%
  \BibitemOpen
  \bibfield  {author} {\bibinfo {author} {\bibfnamefont {F.-K.}\ \bibnamefont
  {Guo}}, \bibinfo {author} {\bibfnamefont {C.}~\bibnamefont {Hanhart}},
  \bibinfo {author} {\bibfnamefont {U.-G.}\ \bibnamefont {Meißner}}, \bibinfo
  {author} {\bibfnamefont {Q.}~\bibnamefont {Wang}},\ and\ \bibinfo {author}
  {\bibfnamefont {Q.}~\bibnamefont {Zhao}},\ }\bibfield  {title} {\bibinfo
  {title} {{Production of the $X(3872)$ in charmonia radiative decays}},\
  }\href {https://doi.org/10.1016/j.physletb.2013.06.053} {\bibfield  {journal}
  {\bibinfo  {journal} {Phys. Lett. B}\ }\textbf {\bibinfo {volume} {725}},\
  \bibinfo {pages} {127} (\bibinfo {year} {2013})},\ \Eprint
  {https://arxiv.org/abs/1306.3096} {arXiv:1306.3096 [hep-ph]} \BibitemShut
  {NoStop}%
\bibitem [{\citenamefont {Li}\ and\ \citenamefont {Liu}(2013)}]{Li:2013yla}%
  \BibitemOpen
  \bibfield  {author} {\bibinfo {author} {\bibfnamefont {G.}~\bibnamefont
  {Li}}\ and\ \bibinfo {author} {\bibfnamefont {X.-H.}\ \bibnamefont {Liu}},\
  }\bibfield  {title} {\bibinfo {title} {{Investigating possible decay modes of
  $Y(4260)$ under ${D}_{1}(2420)\overline{D}+c.c.$ molecular state ansatz}},\
  }\href {https://doi.org/10.1103/PhysRevD.88.094008} {\bibfield  {journal}
  {\bibinfo  {journal} {Phys. Rev. D}\ }\textbf {\bibinfo {volume} {88}},\
  \bibinfo {pages} {094008} (\bibinfo {year} {2013})},\ \Eprint
  {https://arxiv.org/abs/1307.2622} {arXiv:1307.2622 [hep-ph]} \BibitemShut
  {NoStop}%
\bibitem [{\citenamefont {Wang}\ \emph {et~al.}(2013)\citenamefont {Wang},
  \citenamefont {Hanhart},\ and\ \citenamefont {Zhao}}]{Wang:2013cya}%
  \BibitemOpen
  \bibfield  {author} {\bibinfo {author} {\bibfnamefont {Q.}~\bibnamefont
  {Wang}}, \bibinfo {author} {\bibfnamefont {C.}~\bibnamefont {Hanhart}},\ and\
  \bibinfo {author} {\bibfnamefont {Q.}~\bibnamefont {Zhao}},\ }\bibfield
  {title} {\bibinfo {title} {{Decoding the Riddle of $Y(4260)$ and
  ${Z}_{c}(3900)$}},\ }\href {https://doi.org/10.1103/PhysRevLett.111.132003}
  {\bibfield  {journal} {\bibinfo  {journal} {Phys. Rev. Lett.}\ }\textbf
  {\bibinfo {volume} {111}},\ \bibinfo {pages} {132003} (\bibinfo {year}
  {2013})},\ \Eprint {https://arxiv.org/abs/1303.6355} {arXiv:1303.6355
  [hep-ph]} \BibitemShut {NoStop}%
\bibitem [{\citenamefont {Du}\ \emph {et~al.}(2022)\citenamefont {Du},
  \citenamefont {Baru}, \citenamefont {Dong}, \citenamefont {Filin},
  \citenamefont {Guo}, \citenamefont {Hanhart}, \citenamefont {Nefediev},
  \citenamefont {Nieves},\ and\ \citenamefont {Wang}}]{Du:2021zzh}%
  \BibitemOpen
  \bibfield  {author} {\bibinfo {author} {\bibfnamefont {M.-L.}\ \bibnamefont
  {Du}}, \bibinfo {author} {\bibfnamefont {V.}~\bibnamefont {Baru}}, \bibinfo
  {author} {\bibfnamefont {X.-K.}\ \bibnamefont {Dong}}, \bibinfo {author}
  {\bibfnamefont {A.}~\bibnamefont {Filin}}, \bibinfo {author} {\bibfnamefont
  {F.-K.}\ \bibnamefont {Guo}}, \bibinfo {author} {\bibfnamefont
  {C.}~\bibnamefont {Hanhart}}, \bibinfo {author} {\bibfnamefont
  {A.}~\bibnamefont {Nefediev}}, \bibinfo {author} {\bibfnamefont
  {J.}~\bibnamefont {Nieves}},\ and\ \bibinfo {author} {\bibfnamefont
  {Q.}~\bibnamefont {Wang}},\ }\bibfield  {title} {\bibinfo {title}
  {{Coupled-channel approach to $T_{cc}^+$ including three-body effects}},\
  }\href {https://doi.org/10.1103/PhysRevD.105.014024} {\bibfield  {journal}
  {\bibinfo  {journal} {Phys. Rev. D}\ }\textbf {\bibinfo {volume} {105}},\
  \bibinfo {pages} {014024} (\bibinfo {year} {2022})},\ \Eprint
  {https://arxiv.org/abs/2110.13765} {arXiv:2110.13765 [hep-ph]} \BibitemShut
  {NoStop}%
\bibitem [{\citenamefont {Feijoo}\ \emph {et~al.}(2021)\citenamefont {Feijoo},
  \citenamefont {Liang},\ and\ \citenamefont {Oset}}]{Feijoo:2021ppq}%
  \BibitemOpen
  \bibfield  {author} {\bibinfo {author} {\bibfnamefont {A.}~\bibnamefont
  {Feijoo}}, \bibinfo {author} {\bibfnamefont {W.}~\bibnamefont {Liang}},\ and\
  \bibinfo {author} {\bibfnamefont {E.}~\bibnamefont {Oset}},\ }\bibfield
  {title} {\bibinfo {title} {${D}^{0}{D}^{0}{\ensuremath{\pi}}^{+}$ mass
  distribution in the production of the ${T}_{cc}$ exotic state},\ }\href
  {https://doi.org/10.1103/PhysRevD.104.114015} {\bibfield  {journal} {\bibinfo
   {journal} {Phys. Rev. D}\ }\textbf {\bibinfo {volume} {104}},\ \bibinfo
  {pages} {114015} (\bibinfo {year} {2021})},\ \Eprint
  {https://arxiv.org/abs/2108.02730} {arXiv:2108.02730 [hep-ph]} \BibitemShut
  {NoStop}%
\bibitem [{\citenamefont {Meng}\ \emph {et~al.}(2021)\citenamefont {Meng},
  \citenamefont {Wang}, \citenamefont {Wang},\ and\ \citenamefont
  {Zhu}}]{Meng:2021jnw}%
  \BibitemOpen
  \bibfield  {author} {\bibinfo {author} {\bibfnamefont {L.}~\bibnamefont
  {Meng}}, \bibinfo {author} {\bibfnamefont {G.-J.}\ \bibnamefont {Wang}},
  \bibinfo {author} {\bibfnamefont {B.}~\bibnamefont {Wang}},\ and\ \bibinfo
  {author} {\bibfnamefont {S.-L.}\ \bibnamefont {Zhu}},\ }\bibfield  {title}
  {\bibinfo {title} {{Probing the long-range structure of the $T_{cc}^+$ with
  the strong and electromagnetic decays}},\ }\href
  {https://doi.org/10.1103/PhysRevD.104.L051502} {\bibfield  {journal}
  {\bibinfo  {journal} {Phys. Rev. D}\ }\textbf {\bibinfo {volume} {104}},\
  \bibinfo {pages} {051502} (\bibinfo {year} {2021})},\ \Eprint
  {https://arxiv.org/abs/2107.14784} {arXiv:2107.14784 [hep-ph]} \BibitemShut
  {NoStop}%
\bibitem [{\citenamefont {Ling}\ \emph {et~al.}(2022)\citenamefont {Ling},
  \citenamefont {Liu}, \citenamefont {Geng}, \citenamefont {Wang},\ and\
  \citenamefont {Xie}}]{Ling:2021bir}%
  \BibitemOpen
  \bibfield  {author} {\bibinfo {author} {\bibfnamefont {X.-Z.}\ \bibnamefont
  {Ling}}, \bibinfo {author} {\bibfnamefont {M.-Z.}\ \bibnamefont {Liu}},
  \bibinfo {author} {\bibfnamefont {L.-S.}\ \bibnamefont {Geng}}, \bibinfo
  {author} {\bibfnamefont {E.}~\bibnamefont {Wang}},\ and\ \bibinfo {author}
  {\bibfnamefont {J.-J.}\ \bibnamefont {Xie}},\ }\bibfield  {title} {\bibinfo
  {title} {{Can we understand the decay width of the $T_{cc}^+$ state?}},\
  }\href {https://doi.org/10.1016/j.physletb.2022.136897} {\bibfield  {journal}
  {\bibinfo  {journal} {Phys. Lett. B}\ }\textbf {\bibinfo {volume} {826}},\
  \bibinfo {pages} {136897} (\bibinfo {year} {2022})},\ \Eprint
  {https://arxiv.org/abs/2108.00947} {arXiv:2108.00947 [hep-ph]} \BibitemShut
  {NoStop}%
\bibitem [{\citenamefont {Song}\ \emph {et~al.}(2024)\citenamefont {Song},
  \citenamefont {Zhang}, \citenamefont {Baru}, \citenamefont {Guo},
  \citenamefont {Hanhart},\ and\ \citenamefont {Nefediev}}]{Song:2024ykq}%
  \BibitemOpen
  \bibfield  {author} {\bibinfo {author} {\bibfnamefont {Y.-L.}\ \bibnamefont
  {Song}}, \bibinfo {author} {\bibfnamefont {Y.}~\bibnamefont {Zhang}},
  \bibinfo {author} {\bibfnamefont {V.}~\bibnamefont {Baru}}, \bibinfo {author}
  {\bibfnamefont {F.-K.}\ \bibnamefont {Guo}}, \bibinfo {author} {\bibfnamefont
  {C.}~\bibnamefont {Hanhart}},\ and\ \bibinfo {author} {\bibfnamefont
  {A.}~\bibnamefont {Nefediev}},\ }\bibfield  {title} {\bibinfo {title}
  {{Towards a Precision Determination of the $X(6200)$ Parameters from Data}},\
  }\href@noop {} {\bibfield  {journal} {\bibinfo  {journal} {arXiv:2411.12062
  [hep-ph]}\ } (\bibinfo {year} {2024})},\ \Eprint
  {https://arxiv.org/abs/2411.12062} {arXiv:2411.12062 [hep-ph]} \BibitemShut
  {NoStop}%
\bibitem [{\citenamefont {Lin}\ \emph {et~al.}(2024)\citenamefont {Lin},
  \citenamefont {Wang}, \citenamefont {Cheng}, \citenamefont {Meng},\ and\
  \citenamefont {Zhu}}]{Lin:2024qcq}%
  \BibitemOpen
  \bibfield  {author} {\bibinfo {author} {\bibfnamefont {Z.-Y.}\ \bibnamefont
  {Lin}}, \bibinfo {author} {\bibfnamefont {J.-Z.}\ \bibnamefont {Wang}},
  \bibinfo {author} {\bibfnamefont {J.-B.}\ \bibnamefont {Cheng}}, \bibinfo
  {author} {\bibfnamefont {L.}~\bibnamefont {Meng}},\ and\ \bibinfo {author}
  {\bibfnamefont {S.-L.}\ \bibnamefont {Zhu}},\ }\bibfield  {title} {\bibinfo
  {title} {{Identification of the $G(3900)$ as the P-wave
  $D\bar{D}^*/\bar{D}D^*$ resonance}},\ }\href
  {https://doi.org/10.1103/PhysRevLett.133.241903} {\bibfield  {journal}
  {\bibinfo  {journal} {Phys. Rev. Lett.}\ }\textbf {\bibinfo {volume} {133}},\
  \bibinfo {pages} {241903} (\bibinfo {year} {2024})},\ \Eprint
  {https://arxiv.org/abs/2403.01727} {arXiv:2403.01727 [hep-ph]} \BibitemShut
  {NoStop}%
\bibitem [{\citenamefont {Ablikim}\ \emph {et~al.}(2024)\citenamefont {Ablikim}
  \emph {et~al.}}]{BESIII:2024ths}%
  \BibitemOpen
  \bibfield  {author} {\bibinfo {author} {\bibfnamefont {M.}~\bibnamefont
  {Ablikim}} \emph {et~al.} (\bibinfo {collaboration} {BESIII}),\ }\bibfield
  {title} {\bibinfo {title} {{Precise Measurement of Born Cross Sections for
  $e^+e^-\to D\bar{D}$ at $\sqrt{s}=3.80\text{-}4.95$ GeV}},\ }\href
  {https://doi.org/10.1103/PhysRevLett.133.081901} {\bibfield  {journal}
  {\bibinfo  {journal} {Phys. Rev. Lett.}\ }\textbf {\bibinfo {volume} {133}},\
  \bibinfo {pages} {081901} (\bibinfo {year} {2024})},\ \Eprint
  {https://arxiv.org/abs/2402.03829} {arXiv:2402.03829 [hep-ex]} \BibitemShut
  {NoStop}%
\bibitem [{\citenamefont {Aubert}\ \emph {et~al.}(2007)\citenamefont {Aubert}
  \emph {et~al.}}]{BaBar:2006qlj}%
  \BibitemOpen
  \bibfield  {author} {\bibinfo {author} {\bibfnamefont {B.}~\bibnamefont
  {Aubert}} \emph {et~al.} (\bibinfo {collaboration} {BaBar}),\ }\bibfield
  {title} {\bibinfo {title} {{Study of the Exclusive Initial-State Radiation
  Production of the $D$ anti-$D$ System}},\ }\href
  {https://doi.org/10.1103/PhysRevD.76.111105} {\bibfield  {journal} {\bibinfo
  {journal} {Phys. Rev. D}\ }\textbf {\bibinfo {volume} {76}},\ \bibinfo
  {pages} {111105} (\bibinfo {year} {2007})},\ \Eprint
  {https://arxiv.org/abs/hep-ex/0607083} {arXiv:hep-ex/0607083} \BibitemShut
  {NoStop}%
\bibitem [{\citenamefont {Aubert}\ \emph {et~al.}(2009)\citenamefont {Aubert}
  \emph {et~al.}}]{BaBar:2009elc}%
  \BibitemOpen
  \bibfield  {author} {\bibinfo {author} {\bibfnamefont {B.}~\bibnamefont
  {Aubert}} \emph {et~al.} (\bibinfo {collaboration} {BaBar}),\ }\bibfield
  {title} {\bibinfo {title} {{Exclusive Initial-State-Radiation Production of
  the $D\bar{D}$ , $D^*\bar{D}$ , and $D^\ast\bar{D}$ Systems}},\ }\href
  {https://doi.org/10.1103/PhysRevD.79.092001} {\bibfield  {journal} {\bibinfo
  {journal} {Phys. Rev. D}\ }\textbf {\bibinfo {volume} {79}},\ \bibinfo
  {pages} {092001} (\bibinfo {year} {2009})},\ \Eprint
  {https://arxiv.org/abs/0903.1597} {arXiv:0903.1597 [hep-ex]} \BibitemShut
  {NoStop}%
\bibitem [{\citenamefont {Pakhlova}\ \emph {et~al.}(2008)\citenamefont
  {Pakhlova} \emph {et~al.}}]{Belle:2007qxm}%
  \BibitemOpen
  \bibfield  {author} {\bibinfo {author} {\bibfnamefont {G.}~\bibnamefont
  {Pakhlova}} \emph {et~al.} (\bibinfo {collaboration} {Belle}),\ }\bibfield
  {title} {\bibinfo {title} {{Measurement of the near-threshold $e^+e^-\to
  D\bar{D}$ cross section using initial-state radiation}},\ }\href
  {https://doi.org/10.1103/PhysRevD.77.011103} {\bibfield  {journal} {\bibinfo
  {journal} {Phys. Rev. D}\ }\textbf {\bibinfo {volume} {77}},\ \bibinfo
  {pages} {011103} (\bibinfo {year} {2008})},\ \Eprint
  {https://arxiv.org/abs/0708.0082} {arXiv:0708.0082 [hep-ex]} \BibitemShut
  {NoStop}%
\bibitem [{\citenamefont {Hüsken}\ \emph {et~al.}(2024)\citenamefont
  {Hüsken}, \citenamefont {Lebed}, \citenamefont {Mitchell}, \citenamefont
  {Swanson}, \citenamefont {Wang},\ and\ \citenamefont
  {Yuan}}]{Husken:2024hmi}%
  \BibitemOpen
  \bibfield  {author} {\bibinfo {author} {\bibfnamefont {N.}~\bibnamefont
  {Hüsken}}, \bibinfo {author} {\bibfnamefont {R.~F.}\ \bibnamefont {Lebed}},
  \bibinfo {author} {\bibfnamefont {R.~E.}\ \bibnamefont {Mitchell}}, \bibinfo
  {author} {\bibfnamefont {E.~S.}\ \bibnamefont {Swanson}}, \bibinfo {author}
  {\bibfnamefont {Y.-Q.}\ \bibnamefont {Wang}},\ and\ \bibinfo {author}
  {\bibfnamefont {C.-Z.}\ \bibnamefont {Yuan}},\ }\bibfield  {title} {\bibinfo
  {title} {{Poles and poltergeists in
  ${e}^{+}{e}^{\ensuremath{-}}\ensuremath{\rightarrow}D\bar{D}$ data}},\ }\href
  {https://doi.org/10.1103/PhysRevD.109.114010} {\bibfield  {journal} {\bibinfo
   {journal} {Phys. Rev. D}\ }\textbf {\bibinfo {volume} {109}},\ \bibinfo
  {pages} {114010} (\bibinfo {year} {2024})},\ \Eprint
  {https://arxiv.org/abs/2404.03896} {arXiv:2404.03896 [hep-ph]} \BibitemShut
  {NoStop}%
\bibitem [{\citenamefont {Zhang}\ and\ \citenamefont
  {Zhao}(2010{\natexlab{a}})}]{Zhang:2009gy}%
  \BibitemOpen
  \bibfield  {author} {\bibinfo {author} {\bibfnamefont {Y.-J.}\ \bibnamefont
  {Zhang}}\ and\ \bibinfo {author} {\bibfnamefont {Q.}~\bibnamefont {Zhao}},\
  }\bibfield  {title} {\bibinfo {title} {{The Lineshape of $\psi(3770)$ and
  low-lying vector charmonium resonance parameters in $e^+e^-\to D\bar{D}$}},\
  }\href {https://doi.org/10.1103/PhysRevD.81.034011} {\bibfield  {journal}
  {\bibinfo  {journal} {Phys. Rev. D}\ }\textbf {\bibinfo {volume} {81}},\
  \bibinfo {pages} {034011} (\bibinfo {year} {2010}{\natexlab{a}})},\ \Eprint
  {https://arxiv.org/abs/0911.5651} {arXiv:0911.5651 [hep-ph]} \BibitemShut
  {NoStop}%
\bibitem [{\citenamefont {Zhang}\ and\ \citenamefont
  {Zhao}(2010{\natexlab{b}})}]{Zhang:2010zv}%
  \BibitemOpen
  \bibfield  {author} {\bibinfo {author} {\bibfnamefont {Y.-J.}\ \bibnamefont
  {Zhang}}\ and\ \bibinfo {author} {\bibfnamefont {Q.}~\bibnamefont {Zhao}},\
  }\bibfield  {title} {\bibinfo {title} {{Lineshape of $e^+ e^-\to D^* \bar
  D+c.c.$ and electromagnetic form factor of $D^*\to D$ transition in the
  time-like region}},\ }\href {https://doi.org/10.1103/PhysRevD.81.074016}
  {\bibfield  {journal} {\bibinfo  {journal} {Phys. Rev. D}\ }\textbf {\bibinfo
  {volume} {81}},\ \bibinfo {pages} {074016} (\bibinfo {year}
  {2010}{\natexlab{b}})},\ \Eprint {https://arxiv.org/abs/1002.1612}
  {arXiv:1002.1612 [hep-ph]} \BibitemShut {NoStop}%
\bibitem [{\citenamefont {Cao}\ and\ \citenamefont
  {Lenske}(2014)}]{Cao:2014qna}%
  \BibitemOpen
  \bibfield  {author} {\bibinfo {author} {\bibfnamefont {X.}~\bibnamefont
  {Cao}}\ and\ \bibinfo {author} {\bibfnamefont {H.}~\bibnamefont {Lenske}},\
  }\href@noop {} {\bibinfo {title} {{The nature and line shapes of charmonium
  in the $e^+e^- \to D\bar{D}$ reactions}}} (\bibinfo {year} {2014}),\ \Eprint
  {https://arxiv.org/abs/1410.1375} {arXiv:1410.1375 [nucl-th]} \BibitemShut
  {NoStop}%
\bibitem [{\citenamefont {Ye}\ \emph {et~al.}(2025)\citenamefont {Ye},
  \citenamefont {Zhang}, \citenamefont {Du}, \citenamefont {Mei\ss{}ner},
  \citenamefont {Niu},\ and\ \citenamefont {Wang}}]{Ye:2025ywy}%
  \BibitemOpen
  \bibfield  {author} {\bibinfo {author} {\bibfnamefont {Q.}~\bibnamefont
  {Ye}}, \bibinfo {author} {\bibfnamefont {Z.}~\bibnamefont {Zhang}}, \bibinfo
  {author} {\bibfnamefont {M.-L.}\ \bibnamefont {Du}}, \bibinfo {author}
  {\bibfnamefont {U.-G.}\ \bibnamefont {Mei\ss{}ner}}, \bibinfo {author}
  {\bibfnamefont {P.-Y.}\ \bibnamefont {Niu}},\ and\ \bibinfo {author}
  {\bibfnamefont {Q.}~\bibnamefont {Wang}},\ }\href@noop {} {\bibinfo {title}
  {{The resonance parameters of the vector charmonium-like state $G(3900)$}}}
  (\bibinfo {year} {2025}),\ \Eprint {https://arxiv.org/abs/2504.17431}
  {arXiv:2504.17431 [hep-ph]} \BibitemShut {NoStop}%
\bibitem [{\citenamefont {Chen}\ \emph
  {et~al.}(2025{\natexlab{b}})\citenamefont {Chen}, \citenamefont {Ding},\ and\
  \citenamefont {He}}]{Chen:2025gxe}%
  \BibitemOpen
  \bibfield  {author} {\bibinfo {author} {\bibfnamefont {X.-X.}\ \bibnamefont
  {Chen}}, \bibinfo {author} {\bibfnamefont {Z.-M.}\ \bibnamefont {Ding}},\
  and\ \bibinfo {author} {\bibfnamefont {J.}~\bibnamefont {He}},\ }\bibfield
  {title} {\bibinfo {title} {{Pole trajectories from S- and P-wave $D {\bar
  D}^*$ interactions}},\ }\href {https://doi.org/10.1103/36vy-r2z8} {\bibfield
  {journal} {\bibinfo  {journal} {Phys. Rev. D}\ }\textbf {\bibinfo {volume}
  {111}},\ \bibinfo {pages} {114008} (\bibinfo {year} {2025}{\natexlab{b}})},\
  \Eprint {https://arxiv.org/abs/2504.15534} {arXiv:2504.15534 [hep-ph]}
  \BibitemShut {NoStop}%
\bibitem [{\citenamefont {Nakamura}\ \emph {et~al.}(2024)\citenamefont
  {Nakamura}, \citenamefont {Li}, \citenamefont {Peng}, \citenamefont {Sun},\
  and\ \citenamefont {Zhou}}]{Nakamura:2023obk}%
  \BibitemOpen
  \bibfield  {author} {\bibinfo {author} {\bibfnamefont {S.~X.}\ \bibnamefont
  {Nakamura}}, \bibinfo {author} {\bibfnamefont {X.-H.}\ \bibnamefont {Li}},
  \bibinfo {author} {\bibfnamefont {H.-P.}\ \bibnamefont {Peng}}, \bibinfo
  {author} {\bibfnamefont {Z.-T.}\ \bibnamefont {Sun}},\ and\ \bibinfo {author}
  {\bibfnamefont {X.-R.}\ \bibnamefont {Zhou}},\ }\bibfield  {title} {\bibinfo
  {title} {{Global coupled-channel analysis of $e^+e^-\to c\bar{c}$ processes
  in $\sqrt{s}=3.75-4.7$ GeV}},\ }\bibfield  {journal} {\bibinfo  {journal}
  {arXiv:2312.17658 [hep-ph]}\ }\href
  {https://doi.org/10.48550/arXiv.2312.17658} {10.48550/arXiv.2312.17658}
  (\bibinfo {year} {2024}),\ \Eprint {https://arxiv.org/abs/2312.17658}
  {arXiv:2312.17658 [hep-ph]} \BibitemShut {NoStop}%
\bibitem [{\citenamefont {Cao}\ \emph {et~al.}(2025)\citenamefont {Cao},
  \citenamefont {Zhang}, \citenamefont {She}, \citenamefont {Lei},
  \citenamefont {Zhang}, \citenamefont {Zheng}, \citenamefont {Zhou},
  \citenamefont {Yan}, \citenamefont {Wang},\ and\ \citenamefont
  {Sa}}]{BESIII:2025wlf}%
  \BibitemOpen
  \bibfield  {author} {\bibinfo {author} {\bibfnamefont {J.}~\bibnamefont
  {Cao}}, \bibinfo {author} {\bibfnamefont {W.-C.}\ \bibnamefont {Zhang}},
  \bibinfo {author} {\bibfnamefont {Z.-L.}\ \bibnamefont {She}}, \bibinfo
  {author} {\bibfnamefont {A.-K.}\ \bibnamefont {Lei}}, \bibinfo {author}
  {\bibfnamefont {J.-P.}\ \bibnamefont {Zhang}}, \bibinfo {author}
  {\bibfnamefont {H.}~\bibnamefont {Zheng}}, \bibinfo {author} {\bibfnamefont
  {D.-M.}\ \bibnamefont {Zhou}}, \bibinfo {author} {\bibfnamefont {Y.-L.}\
  \bibnamefont {Yan}}, \bibinfo {author} {\bibfnamefont {Z.-Q.}\ \bibnamefont
  {Wang}},\ and\ \bibinfo {author} {\bibfnamefont {B.-H.}\ \bibnamefont {Sa}}
  (\bibinfo {collaboration} {BESIII}),\ }\bibfield  {title} {\bibinfo {title}
  {{Charmonium-like exotic hadron productions in $e^+e^-$ collisions at the
  BESIII energy with the PACIAE model}},\ }\href
  {https://doi.org/10.1103/1gtz-xkgh} {\bibfield  {journal} {\bibinfo
  {journal} {Phys. Rev. D}\ }\textbf {\bibinfo {volume} {112}},\ \bibinfo
  {pages} {014033} (\bibinfo {year} {2025})},\ \Eprint
  {https://arxiv.org/abs/2502.16822} {arXiv:2502.16822 [hep-ph]} \BibitemShut
  {NoStop}%
\bibitem [{\citenamefont {Aaij}\ \emph {et~al.}(2023)\citenamefont {Aaij} \emph
  {et~al.}}]{LHCb:2022jez}%
  \BibitemOpen
  \bibfield  {author} {\bibinfo {author} {\bibfnamefont {R.}~\bibnamefont
  {Aaij}} \emph {et~al.} (\bibinfo {collaboration} {LHCb}),\ }\bibfield
  {title} {\bibinfo {title} {{Observation of sizeable $\omega$ contribution to
  $\chi_{c1}(3872)\to\pi^+\pi^- J/\psi$ decays}},\ }\href
  {https://doi.org/10.1103/PhysRevD.108.L011103} {\bibfield  {journal}
  {\bibinfo  {journal} {Phys. Rev. D}\ }\textbf {\bibinfo {volume} {108}},\
  \bibinfo {pages} {L011103} (\bibinfo {year} {2023})},\ \Eprint
  {https://arxiv.org/abs/2204.12597} {arXiv:2204.12597 [hep-ex]} \BibitemShut
  {NoStop}%
\bibitem [{\citenamefont {Zhang}\ \emph {et~al.}(2024)\citenamefont {Zhang},
  \citenamefont {Ji}, \citenamefont {Dong}, \citenamefont {Guo}, \citenamefont
  {Hanhart}, \citenamefont {Mei\ss{}ner},\ and\ \citenamefont
  {Rusetsky}}]{Zhang:2024fxy}%
  \BibitemOpen
  \bibfield  {author} {\bibinfo {author} {\bibfnamefont {Z.-H.}\ \bibnamefont
  {Zhang}}, \bibinfo {author} {\bibfnamefont {T.}~\bibnamefont {Ji}}, \bibinfo
  {author} {\bibfnamefont {X.-K.}\ \bibnamefont {Dong}}, \bibinfo {author}
  {\bibfnamefont {F.-K.}\ \bibnamefont {Guo}}, \bibinfo {author} {\bibfnamefont
  {C.}~\bibnamefont {Hanhart}}, \bibinfo {author} {\bibfnamefont {U.-G.}\
  \bibnamefont {Mei\ss{}ner}},\ and\ \bibinfo {author} {\bibfnamefont
  {A.}~\bibnamefont {Rusetsky}},\ }\bibfield  {title} {\bibinfo {title}
  {{Predicting isovector charmonium-like states from $X(3872)$ properties}},\
  }\href {https://doi.org/10.1007/JHEP08(2024)130} {\bibfield  {journal}
  {\bibinfo  {journal} {JHEP}\ }\textbf {\bibinfo {volume} {08}},\ \bibinfo
  {pages} {130}},\ \Eprint {https://arxiv.org/abs/2404.11215} {arXiv:2404.11215
  [hep-ph]} \BibitemShut {NoStop}%
\bibitem [{\citenamefont {Zhu}\ and\ \citenamefont
  {Huang}(2022)}]{Zhu:2021exs}%
  \BibitemOpen
  \bibfield  {author} {\bibinfo {author} {\bibfnamefont {H.~Q.}\ \bibnamefont
  {Zhu}}\ and\ \bibinfo {author} {\bibfnamefont {Y.}~\bibnamefont {Huang}},\
  }\bibfield  {title} {\bibinfo {title} {{Possible $P$-wave $D_s D_{s0}(2317)$
  molecular state $Y'(4274)$}},\ }\href
  {https://doi.org/10.1103/PhysRevD.105.056011} {\bibfield  {journal} {\bibinfo
   {journal} {Phys. Rev. D}\ }\textbf {\bibinfo {volume} {105}},\ \bibinfo
  {pages} {056011} (\bibinfo {year} {2022})},\ \Eprint
  {https://arxiv.org/abs/2110.14253} {arXiv:2110.14253 [hep-ph]} \BibitemShut
  {NoStop}%
\bibitem [{\citenamefont {Ma}(2010)}]{Ma:2010xx}%
  \BibitemOpen
  \bibfield  {author} {\bibinfo {author} {\bibfnamefont {Y.-L.}\ \bibnamefont
  {Ma}},\ }\bibfield  {title} {\bibinfo {title} {Estimates for x(4350) decays
  from the effective lagrangian approach},\ }\href
  {https://doi.org/10.1103/PhysRevD.82.015013} {\bibfield  {journal} {\bibinfo
  {journal} {Phys. Rev. D}\ }\textbf {\bibinfo {volume} {82}},\ \bibinfo
  {pages} {015013} (\bibinfo {year} {2010})},\ \Eprint
  {https://arxiv.org/abs/1006.1276} {arXiv:1006.1276 [hep-ph]} \BibitemShut
  {NoStop}%
\bibitem [{\citenamefont {Yue}\ \emph {et~al.}(2024)\citenamefont {Yue},
  \citenamefont {Pan},\ and\ \citenamefont {Chen}}]{Yue:2024bvy}%
  \BibitemOpen
  \bibfield  {author} {\bibinfo {author} {\bibfnamefont {Z.-L.}\ \bibnamefont
  {Yue}}, \bibinfo {author} {\bibfnamefont {Y.}~\bibnamefont {Pan}},\ and\
  \bibinfo {author} {\bibfnamefont {D.-Y.}\ \bibnamefont {Chen}},\ }\bibfield
  {title} {\bibinfo {title} {{Hidden charm decays of Y(4626) in a
  $D_s^{*+}D_{s1}^-(2536)$ molecular frame}},\ }\href
  {https://doi.org/10.1103/PhysRevD.110.074013} {\bibfield  {journal} {\bibinfo
   {journal} {Phys. Rev. D}\ }\textbf {\bibinfo {volume} {110}},\ \bibinfo
  {pages} {074013} (\bibinfo {year} {2024})},\ \Eprint
  {https://arxiv.org/abs/2408.08546} {arXiv:2408.08546 [hep-ph]} \BibitemShut
  {NoStop}%
\bibitem [{\citenamefont {Xiao}\ \emph {et~al.}(2016)\citenamefont {Xiao},
  \citenamefont {Chen},\ and\ \citenamefont {Ma}}]{Xiao:2016hoa}%
  \BibitemOpen
  \bibfield  {author} {\bibinfo {author} {\bibfnamefont {C.-J.}\ \bibnamefont
  {Xiao}}, \bibinfo {author} {\bibfnamefont {D.-Y.}\ \bibnamefont {Chen}},\
  and\ \bibinfo {author} {\bibfnamefont {Y.-L.}\ \bibnamefont {Ma}},\
  }\bibfield  {title} {\bibinfo {title} {{Radiative and pionic transitions from
  the $D_{s1}(2460)$ to the $D_{s0}^\ast(2317)$}},\ }\href
  {https://doi.org/10.1103/PhysRevD.93.094011} {\bibfield  {journal} {\bibinfo
  {journal} {Phys. Rev. D}\ }\textbf {\bibinfo {volume} {93}},\ \bibinfo
  {pages} {094011} (\bibinfo {year} {2016})},\ \Eprint
  {https://arxiv.org/abs/1601.06399} {arXiv:1601.06399 [hep-ph]} \BibitemShut
  {NoStop}%
\bibitem [{\citenamefont {Weinberg}(1963)}]{Weinberg:1962hj}%
  \BibitemOpen
  \bibfield  {author} {\bibinfo {author} {\bibfnamefont {S.}~\bibnamefont
  {Weinberg}},\ }\bibfield  {title} {\bibinfo {title} {Elementary particle
  theory of composite particles},\ }\href
  {https://doi.org/10.1103/PhysRev.130.776} {\bibfield  {journal} {\bibinfo
  {journal} {Phys. Rev.}\ }\textbf {\bibinfo {volume} {130}},\ \bibinfo {pages}
  {776} (\bibinfo {year} {1963})}\BibitemShut {NoStop}%
\bibitem [{\citenamefont {Salam}(1962)}]{Salam:1962ap}%
  \BibitemOpen
  \bibfield  {author} {\bibinfo {author} {\bibfnamefont {A.}~\bibnamefont
  {Salam}},\ }\bibfield  {title} {\bibinfo {title} {Lagrangian theory of
  composite particles},\ }\href {https://doi.org/10.1007/BF02733330} {\bibfield
   {journal} {\bibinfo  {journal} {Nuovo Cim.}\ }\textbf {\bibinfo {volume}
  {25}},\ \bibinfo {pages} {224} (\bibinfo {year} {1962})}\BibitemShut
  {NoStop}%
\bibitem [{\citenamefont {Colangelo}\ \emph {et~al.}(2004)\citenamefont
  {Colangelo}, \citenamefont {De~Fazio},\ and\ \citenamefont
  {Pham}}]{Colangelo:2003sa}%
  \BibitemOpen
  \bibfield  {author} {\bibinfo {author} {\bibfnamefont {P.}~\bibnamefont
  {Colangelo}}, \bibinfo {author} {\bibfnamefont {F.}~\bibnamefont
  {De~Fazio}},\ and\ \bibinfo {author} {\bibfnamefont {T.}~\bibnamefont
  {Pham}},\ }\bibfield  {title} {\bibinfo {title} {Nonfactorizable
  contributions in b decays to charmonium: The case of
  ${B}^{\ensuremath{-}}\ensuremath{\rightarrow}{K}^{\ensuremath{-}}{h}_{c}$},\
  }\href {https://doi.org/10.1103/PhysRevD.69.054023} {\bibfield  {journal}
  {\bibinfo  {journal} {Phys. Rev. D}\ }\textbf {\bibinfo {volume} {69}},\
  \bibinfo {pages} {054023} (\bibinfo {year} {2004})},\ \Eprint
  {https://arxiv.org/abs/hep-ph/0310084} {arXiv:hep-ph/0310084} \BibitemShut
  {NoStop}%
\bibitem [{\citenamefont {Casalbuoni}\ \emph {et~al.}(1997)\citenamefont
  {Casalbuoni}, \citenamefont {Deandrea}, \citenamefont {Di~Bartolomeo},
  \citenamefont {Gatto}, \citenamefont {Feruglio},\ and\ \citenamefont
  {Nardulli}}]{Casalbuoni:1996pg}%
  \BibitemOpen
  \bibfield  {author} {\bibinfo {author} {\bibfnamefont {R.}~\bibnamefont
  {Casalbuoni}}, \bibinfo {author} {\bibfnamefont {A.}~\bibnamefont
  {Deandrea}}, \bibinfo {author} {\bibfnamefont {N.}~\bibnamefont
  {Di~Bartolomeo}}, \bibinfo {author} {\bibfnamefont {R.}~\bibnamefont
  {Gatto}}, \bibinfo {author} {\bibfnamefont {F.}~\bibnamefont {Feruglio}},\
  and\ \bibinfo {author} {\bibfnamefont {G.}~\bibnamefont {Nardulli}},\
  }\bibfield  {title} {\bibinfo {title} {Phenomenology of heavy meson chiral
  lagrangians},\ }\href {https://doi.org/10.1016/S0370-1573(96)00027-0}
  {\bibfield  {journal} {\bibinfo  {journal} {Phys. Rept.}\ }\textbf {\bibinfo
  {volume} {281}},\ \bibinfo {pages} {145} (\bibinfo {year} {1997})},\ \Eprint
  {https://arxiv.org/abs/hep-ph/9605342} {arXiv:hep-ph/9605342} \BibitemShut
  {NoStop}%
\bibitem [{\citenamefont {Li}\ \emph {et~al.}(2021)\citenamefont {Li},
  \citenamefont {Bai}, \citenamefont {Huang},\ and\ \citenamefont
  {Liu}}]{Li:2021jjt}%
  \BibitemOpen
  \bibfield  {author} {\bibinfo {author} {\bibfnamefont {Y.-S.}\ \bibnamefont
  {Li}}, \bibinfo {author} {\bibfnamefont {Z.-Y.}\ \bibnamefont {Bai}},
  \bibinfo {author} {\bibfnamefont {Q.}~\bibnamefont {Huang}},\ and\ \bibinfo
  {author} {\bibfnamefont {X.}~\bibnamefont {Liu}},\ }\bibfield  {title}
  {\bibinfo {title} {Hidden-bottom hadronic decays of
  $\mathrm{\ensuremath{\Upsilon}}(10753)$ with a
  ${\ensuremath{\eta}}^{(\prime)}$ or $\ensuremath{\omega}$ emission},\ }\href
  {https://doi.org/10.1103/PhysRevD.104.034036} {\bibfield  {journal} {\bibinfo
   {journal} {Phys. Rev. D}\ }\textbf {\bibinfo {volume} {104}},\ \bibinfo
  {pages} {034036} (\bibinfo {year} {2021})},\ \Eprint
  {https://arxiv.org/abs/2106.14123} {arXiv:2106.14123 [hep-ph]} \BibitemShut
  {NoStop}%
\bibitem [{\citenamefont {Xu}\ \emph {et~al.}(2016)\citenamefont {Xu},
  \citenamefont {Liu},\ and\ \citenamefont {Matsuki}}]{Xu:2016kbn}%
  \BibitemOpen
  \bibfield  {author} {\bibinfo {author} {\bibfnamefont {H.}~\bibnamefont
  {Xu}}, \bibinfo {author} {\bibfnamefont {X.}~\bibnamefont {Liu}},\ and\
  \bibinfo {author} {\bibfnamefont {T.}~\bibnamefont {Matsuki}},\ }\bibfield
  {title} {\bibinfo {title} {Understanding
  ${B}^{\ensuremath{-}}\ensuremath{\rightarrow}x(3823){K}^{\ensuremath{-}}$ via
  rescattering mechanism and predicting
  ${B}^{\ensuremath{-}}\ensuremath{\rightarrow}{\ensuremath{\eta}}_{c2}({^{1}D}_{2})/{\ensuremath{\psi}}_{3}({^{3}D}_{3}){K}^{\ensuremath{-}}$},\
  }\href {https://doi.org/10.1103/PhysRevD.94.034005} {\bibfield  {journal}
  {\bibinfo  {journal} {Phys. Rev. D}\ }\textbf {\bibinfo {volume} {94}},\
  \bibinfo {pages} {034005} (\bibinfo {year} {2016})},\ \Eprint
  {https://arxiv.org/abs/1605.04776} {arXiv:1605.04776 [hep-ph]} \BibitemShut
  {NoStop}%
\bibitem [{\citenamefont {Manohar}\ and\ \citenamefont
  {Wise}(2000)}]{Manohar:2000dt}%
  \BibitemOpen
  \bibfield  {author} {\bibinfo {author} {\bibfnamefont {A.~V.}\ \bibnamefont
  {Manohar}}\ and\ \bibinfo {author} {\bibfnamefont {M.~B.}\ \bibnamefont
  {Wise}},\ }\href {https://doi.org/10.1017/9781009402125} {\emph {\bibinfo
  {title} {{Heavy quark physics}}}},\ Vol.~\bibinfo {volume} {10}\ (\bibinfo
  {year} {2000})\BibitemShut {NoStop}%
\bibitem [{\citenamefont {Navas}\ \emph {et~al.}(2024)\citenamefont {Navas}
  \emph {et~al.}}]{ParticleDataGroup:2024cfk}%
  \BibitemOpen
  \bibfield  {author} {\bibinfo {author} {\bibfnamefont {S.}~\bibnamefont
  {Navas}} \emph {et~al.} (\bibinfo {collaboration} {Particle Data Group}),\
  }\bibfield  {title} {\bibinfo {title} {{Review of particle physics}},\ }\href
  {https://doi.org/10.1103/PhysRevD.110.030001} {\bibfield  {journal} {\bibinfo
   {journal} {Phys. Rev. D}\ }\textbf {\bibinfo {volume} {110}},\ \bibinfo
  {pages} {030001} (\bibinfo {year} {2024})}\BibitemShut {NoStop}%
\bibitem [{\citenamefont {Deandrea}\ \emph {et~al.}(2003)\citenamefont
  {Deandrea}, \citenamefont {Nardulli},\ and\ \citenamefont
  {Polosa}}]{Deandrea:2003pv}%
  \BibitemOpen
  \bibfield  {author} {\bibinfo {author} {\bibfnamefont {A.}~\bibnamefont
  {Deandrea}}, \bibinfo {author} {\bibfnamefont {G.}~\bibnamefont {Nardulli}},\
  and\ \bibinfo {author} {\bibfnamefont {A.~D.}\ \bibnamefont {Polosa}},\
  }\bibfield  {title} {\bibinfo {title} {{$J/\ensuremath{\psi}$} couplings to
  charmed resonances and to $\ensuremath{\pi}$},\ }\href
  {https://doi.org/10.1103/PhysRevD.68.034002} {\bibfield  {journal} {\bibinfo
  {journal} {Phys. Rev. D}\ }\textbf {\bibinfo {volume} {68}},\ \bibinfo
  {pages} {034002} (\bibinfo {year} {2003})},\ \Eprint
  {https://arxiv.org/abs/hep-ph/0302273} {arXiv:hep-ph/0302273} \BibitemShut
  {NoStop}%
\bibitem [{\citenamefont {Badalian}\ \emph {et~al.}(2010)\citenamefont
  {Badalian}, \citenamefont {Bakker},\ and\ \citenamefont
  {Danilkin}}]{Badalian:2009bu}%
  \BibitemOpen
  \bibfield  {author} {\bibinfo {author} {\bibfnamefont {A.~M.}\ \bibnamefont
  {Badalian}}, \bibinfo {author} {\bibfnamefont {B.~L.~G.}\ \bibnamefont
  {Bakker}},\ and\ \bibinfo {author} {\bibfnamefont {I.~V.}\ \bibnamefont
  {Danilkin}},\ }\bibfield  {title} {\bibinfo {title} {Dielectron widths of the
  {$S$-}, {$D$-}vector bottomonium states},\ }\href
  {https://doi.org/10.1134/S1063778810010163} {\bibfield  {journal} {\bibinfo
  {journal} {Phys. Atom. Nucl.}\ }\textbf {\bibinfo {volume} {73}},\ \bibinfo
  {pages} {138} (\bibinfo {year} {2010})},\ \Eprint
  {https://arxiv.org/abs/0903.3643} {arXiv:0903.3643 [hep-ph]} \BibitemShut
  {NoStop}%
\bibitem [{\citenamefont {Li}\ \emph {et~al.}(2013)\citenamefont {Li},
  \citenamefont {Shao}, \citenamefont {Zhao},\ and\ \citenamefont
  {Zhao}}]{Li:2012as}%
  \BibitemOpen
  \bibfield  {author} {\bibinfo {author} {\bibfnamefont {G.}~\bibnamefont
  {Li}}, \bibinfo {author} {\bibfnamefont {F.-l.}\ \bibnamefont {Shao}},
  \bibinfo {author} {\bibfnamefont {C.-W.}\ \bibnamefont {Zhao}},\ and\
  \bibinfo {author} {\bibfnamefont {Q.}~\bibnamefont {Zhao}},\ }\bibfield
  {title} {\bibinfo {title}
  {${Z}_{b}/{Z}_{b}^{\ensuremath{'}}\ensuremath{\rightarrow}\ensuremath{\Upsilon}\ensuremath{\pi}$
  and ${h}_{b}\ensuremath{\pi}$ decays in intermediate meson loops model},\
  }\href {https://doi.org/10.1103/PhysRevD.87.034020} {\bibfield  {journal}
  {\bibinfo  {journal} {Phys. Rev. D}\ }\textbf {\bibinfo {volume} {87}},\
  \bibinfo {pages} {034020} (\bibinfo {year} {2013})},\ \Eprint
  {https://arxiv.org/abs/1212.3784} {arXiv:1212.3784 [hep-ph]} \BibitemShut
  {NoStop}%
\bibitem [{\citenamefont {Liu}\ \emph {et~al.}(2024)\citenamefont {Liu},
  \citenamefont {Cai}, \citenamefont {Jia}, \citenamefont {Li},\ and\
  \citenamefont {Xie}}]{Liu:2023gtx}%
  \BibitemOpen
  \bibfield  {author} {\bibinfo {author} {\bibfnamefont {S.}~\bibnamefont
  {Liu}}, \bibinfo {author} {\bibfnamefont {Z.}~\bibnamefont {Cai}}, \bibinfo
  {author} {\bibfnamefont {Z.}~\bibnamefont {Jia}}, \bibinfo {author}
  {\bibfnamefont {G.}~\bibnamefont {Li}},\ and\ \bibinfo {author}
  {\bibfnamefont {J.}~\bibnamefont {Xie}},\ }\bibfield  {title} {\bibinfo
  {title} {Hidden-bottom hadronic transitions of {$\Upsilon(10753)$}},\ }\href
  {https://doi.org/10.1103/PhysRevD.109.014039} {\bibfield  {journal} {\bibinfo
   {journal} {Phys. Rev. D}\ }\textbf {\bibinfo {volume} {109}},\ \bibinfo
  {pages} {014039} (\bibinfo {year} {2024})},\ \Eprint
  {https://arxiv.org/abs/2312.02761} {arXiv:2312.02761 [hep-ph]} \BibitemShut
  {NoStop}%
\bibitem [{\citenamefont {Be\v{c}irevi\'c}\ \emph {et~al.}(2014)\citenamefont
  {Be\v{c}irevi\'c}, \citenamefont {Duplan\v{c}i\'c}, \citenamefont {Klajn},
  \citenamefont {Meli\'c},\ and\ \citenamefont
  {Sanfilippo}}]{Becirevic:2013bsa}%
  \BibitemOpen
  \bibfield  {author} {\bibinfo {author} {\bibfnamefont {D.}~\bibnamefont
  {Be\v{c}irevi\'c}}, \bibinfo {author} {\bibfnamefont {G.}~\bibnamefont
  {Duplan\v{c}i\'c}}, \bibinfo {author} {\bibfnamefont {B.}~\bibnamefont
  {Klajn}}, \bibinfo {author} {\bibfnamefont {B.}~\bibnamefont {Meli\'c}},\
  and\ \bibinfo {author} {\bibfnamefont {F.}~\bibnamefont {Sanfilippo}},\
  }\bibfield  {title} {\bibinfo {title} {{Lattice QCD and QCD sum rule
  determination of the decay constants of $\eta_c$, J/$\psi$ and $h_c$
  states}},\ }\href {https://doi.org/10.1016/j.nuclphysb.2014.03.024}
  {\bibfield  {journal} {\bibinfo  {journal} {Nucl. Phys. B}\ }\textbf
  {\bibinfo {volume} {883}},\ \bibinfo {pages} {306} (\bibinfo {year}
  {2014})},\ \Eprint {https://arxiv.org/abs/1312.2858} {arXiv:1312.2858
  [hep-ph]} \BibitemShut {NoStop}%
\bibitem [{\citenamefont {Liu}\ \emph {et~al.}(2025{\natexlab{b}})\citenamefont
  {Liu}, \citenamefont {Wu},\ and\ \citenamefont {Li}}]{Liu:2025bjm}%
  \BibitemOpen
  \bibfield  {author} {\bibinfo {author} {\bibfnamefont {S.}~\bibnamefont
  {Liu}}, \bibinfo {author} {\bibfnamefont {Q.}~\bibnamefont {Wu}},\ and\
  \bibinfo {author} {\bibfnamefont {G.}~\bibnamefont {Li}},\ }\href@noop {}
  {\bibinfo {title} {{Dipionic transitions of $Y(4500)$ to $J/\psi$}}}
  (\bibinfo {year} {2025}{\natexlab{b}}),\ \Eprint
  {https://arxiv.org/abs/2504.14792} {arXiv:2504.14792 [hep-ph]} \BibitemShut
  {NoStop}%
\bibitem [{\citenamefont {Veliev}\ \emph {et~al.}(2010)\citenamefont {Veliev},
  \citenamefont {Sundu}, \citenamefont {Azizi},\ and\ \citenamefont
  {Bayar}}]{Veliev:2010gb}%
  \BibitemOpen
  \bibfield  {author} {\bibinfo {author} {\bibfnamefont {E.~V.}\ \bibnamefont
  {Veliev}}, \bibinfo {author} {\bibfnamefont {H.}~\bibnamefont {Sundu}},
  \bibinfo {author} {\bibfnamefont {K.}~\bibnamefont {Azizi}},\ and\ \bibinfo
  {author} {\bibfnamefont {M.}~\bibnamefont {Bayar}},\ }\bibfield  {title}
  {\bibinfo {title} {{Scalar Quarkonia at Finite Temperature}},\ }\href
  {https://doi.org/10.1103/PhysRevD.82.056012} {\bibfield  {journal} {\bibinfo
  {journal} {Phys. Rev. D}\ }\textbf {\bibinfo {volume} {82}},\ \bibinfo
  {pages} {056012} (\bibinfo {year} {2010})},\ \Eprint
  {https://arxiv.org/abs/1003.0119} {arXiv:1003.0119 [hep-ph]} \BibitemShut
  {NoStop}%
\bibitem [{\citenamefont {Isola}\ \emph {et~al.}(2003)\citenamefont {Isola},
  \citenamefont {Ladisa}, \citenamefont {Nardulli},\ and\ \citenamefont
  {Santorelli}}]{Isola:2003fh}%
  \BibitemOpen
  \bibfield  {author} {\bibinfo {author} {\bibfnamefont {C.}~\bibnamefont
  {Isola}}, \bibinfo {author} {\bibfnamefont {M.}~\bibnamefont {Ladisa}},
  \bibinfo {author} {\bibfnamefont {G.}~\bibnamefont {Nardulli}},\ and\
  \bibinfo {author} {\bibfnamefont {P.}~\bibnamefont {Santorelli}},\ }\bibfield
   {title} {\bibinfo {title} {Charming penguin contributions in {$B\to K^\ast
  \pi, K(\rho,\,\omega,\,\phi)$} decays},\ }\href
  {https://doi.org/10.1103/PhysRevD.68.114001} {\bibfield  {journal} {\bibinfo
  {journal} {Phys. Rev. D}\ }\textbf {\bibinfo {volume} {68}},\ \bibinfo
  {pages} {114001} (\bibinfo {year} {2003})},\ \Eprint
  {https://arxiv.org/abs/hep-ph/0307367} {arXiv:hep-ph/0307367} \BibitemShut
  {NoStop}%
\bibitem [{\citenamefont {Casalbuoni}\ \emph {et~al.}(1993)\citenamefont
  {Casalbuoni}, \citenamefont {Deandrea}, \citenamefont {Di~Bartolomeo},
  \citenamefont {Gatto}, \citenamefont {Feruglio},\ and\ \citenamefont
  {Nardulli}}]{Casalbuoni:1992dx}%
  \BibitemOpen
  \bibfield  {author} {\bibinfo {author} {\bibfnamefont {R.}~\bibnamefont
  {Casalbuoni}}, \bibinfo {author} {\bibfnamefont {A.}~\bibnamefont
  {Deandrea}}, \bibinfo {author} {\bibfnamefont {N.}~\bibnamefont
  {Di~Bartolomeo}}, \bibinfo {author} {\bibfnamefont {R.}~\bibnamefont
  {Gatto}}, \bibinfo {author} {\bibfnamefont {F.}~\bibnamefont {Feruglio}},\
  and\ \bibinfo {author} {\bibfnamefont {G.}~\bibnamefont {Nardulli}},\
  }\bibfield  {title} {\bibinfo {title} {Effective lagrangian for heavy and
  light mesons. semileptonic decays},\ }\href
  {https://doi.org/10.1016/0370-2693(93)90895-O} {\bibfield  {journal}
  {\bibinfo  {journal} {Phys. Lett. B}\ }\textbf {\bibinfo {volume} {299}},\
  \bibinfo {pages} {139} (\bibinfo {year} {1993})},\ \Eprint
  {https://arxiv.org/abs/hep-ph/9211248} {arXiv:hep-ph/9211248} \BibitemShut
  {NoStop}%
\bibitem [{\citenamefont {Gamermann}\ and\ \citenamefont
  {Oset}(2009)}]{Gamermann:2009fv}%
  \BibitemOpen
  \bibfield  {author} {\bibinfo {author} {\bibfnamefont {D.}~\bibnamefont
  {Gamermann}}\ and\ \bibinfo {author} {\bibfnamefont {E.}~\bibnamefont
  {Oset}},\ }\bibfield  {title} {\bibinfo {title} {{Isospin breaking effects in
  the $X(3872)$ resonance}},\ }\href
  {https://doi.org/10.1103/PhysRevD.80.014003} {\bibfield  {journal} {\bibinfo
  {journal} {Phys. Rev. D}\ }\textbf {\bibinfo {volume} {80}},\ \bibinfo
  {pages} {014003} (\bibinfo {year} {2009})},\ \Eprint
  {https://arxiv.org/abs/0905.0402} {arXiv:0905.0402 [hep-ph]} \BibitemShut
  {NoStop}%
\bibitem [{\citenamefont {Ablikim}\ \emph {et~al.}(2019)\citenamefont
  {Ablikim}, \citenamefont {Achasov}, \citenamefont {Adlarson} \emph
  {et~al.}}]{BESIII:2019qvy}%
  \BibitemOpen
  \bibfield  {author} {\bibinfo {author} {\bibfnamefont {M.}~\bibnamefont
  {Ablikim}}, \bibinfo {author} {\bibfnamefont {M.}~\bibnamefont {Achasov}},
  \bibinfo {author} {\bibfnamefont {P.}~\bibnamefont {Adlarson}}, \emph
  {et~al.} (\bibinfo {collaboration} {BESIII}),\ }\bibfield  {title} {\bibinfo
  {title} {{Study of
  ${e}^{+}{e}^{\ensuremath{-}}\ensuremath{\rightarrow}\ensuremath{\gamma}\ensuremath{\omega}J/\ensuremath{\psi}$
  and Observation of
  $X(3872)\ensuremath{\rightarrow}\ensuremath{\omega}J/\ensuremath{\psi}$}},\
  }\href {https://doi.org/10.1103/PhysRevLett.122.232002} {\bibfield  {journal}
  {\bibinfo  {journal} {Phys. Rev. Lett.}\ }\textbf {\bibinfo {volume} {122}},\
  \bibinfo {pages} {232002} (\bibinfo {year} {2019})},\ \Eprint
  {https://arxiv.org/abs/1903.04695} {arXiv:1903.04695 [hep-ex]} \BibitemShut
  {NoStop}%
\bibitem [{\citenamefont {Aaij}\ \emph
  {et~al.}(2015{\natexlab{b}})\citenamefont {Aaij}, \citenamefont {Adeva},
  \citenamefont {Adinolfi} \emph {et~al.}}]{LHCb:2015jfc}%
  \BibitemOpen
  \bibfield  {author} {\bibinfo {author} {\bibfnamefont {R.}~\bibnamefont
  {Aaij}}, \bibinfo {author} {\bibfnamefont {B.}~\bibnamefont {Adeva}},
  \bibinfo {author} {\bibfnamefont {M.}~\bibnamefont {Adinolfi}}, \emph
  {et~al.} (\bibinfo {collaboration} {LHCb}),\ }\bibfield  {title} {\bibinfo
  {title} {{Quantum numbers of the $X(3872)$ state and orbital angular momentum
  in its ${\ensuremath{\rho}}^{0}J/\ensuremath{\psi}$ decay}},\ }\href
  {https://doi.org/10.1103/PhysRevD.92.011102} {\bibfield  {journal} {\bibinfo
  {journal} {Phys. Rev. D}\ }\textbf {\bibinfo {volume} {92}},\ \bibinfo
  {pages} {011102} (\bibinfo {year} {2015}{\natexlab{b}})},\ \Eprint
  {https://arxiv.org/abs/1504.06339} {arXiv:1504.06339 [hep-ex]} \BibitemShut
  {NoStop}%
\bibitem [{\citenamefont {Wu}\ \emph {et~al.}(2021)\citenamefont {Wu},
  \citenamefont {Chen},\ and\ \citenamefont {Matsuki}}]{Wu:2021udi}%
  \BibitemOpen
  \bibfield  {author} {\bibinfo {author} {\bibfnamefont {Q.}~\bibnamefont
  {Wu}}, \bibinfo {author} {\bibfnamefont {D.-Y.}\ \bibnamefont {Chen}},\ and\
  \bibinfo {author} {\bibfnamefont {T.}~\bibnamefont {Matsuki}},\ }\bibfield
  {title} {\bibinfo {title} {{A phenomenological analysis on isospin-violating
  decay of $X(3872)$}},\ }\href
  {https://doi.org/10.1140/epjc/s10052-021-08984-2} {\bibfield  {journal}
  {\bibinfo  {journal} {Eur. Phys. J. C}\ }\textbf {\bibinfo {volume} {81}},\
  \bibinfo {pages} {193} (\bibinfo {year} {2021})},\ \Eprint
  {https://arxiv.org/abs/2102.08637} {arXiv:2102.08637 [hep-ph]} \BibitemShut
  {NoStop}%
\bibitem [{\citenamefont {Guo}\ \emph {et~al.}(2014)\citenamefont {Guo},
  \citenamefont {Hidalgo-Duque}, \citenamefont {Nieves}, \citenamefont
  {Ozpineci},\ and\ \citenamefont {Valderrama}}]{Guo:2014hqa}%
  \BibitemOpen
  \bibfield  {author} {\bibinfo {author} {\bibfnamefont {F.-K.}\ \bibnamefont
  {Guo}}, \bibinfo {author} {\bibfnamefont {C.}~\bibnamefont {Hidalgo-Duque}},
  \bibinfo {author} {\bibfnamefont {J.}~\bibnamefont {Nieves}}, \bibinfo
  {author} {\bibfnamefont {A.}~\bibnamefont {Ozpineci}},\ and\ \bibinfo
  {author} {\bibfnamefont {M.~P.}\ \bibnamefont {Valderrama}},\ }\bibfield
  {title} {\bibinfo {title} {Detecting the long-distance structure of the
  $x(3872)$},\ }\href {https://doi.org/10.1140/epjc/s10052-014-2885-4}
  {\bibfield  {journal} {\bibinfo  {journal} {Eur. Phys. J. C}\ }\textbf
  {\bibinfo {volume} {74}},\ \bibinfo {pages} {2885} (\bibinfo {year}
  {2014})},\ \Eprint {https://arxiv.org/abs/1404.1776} {arXiv:1404.1776
  [hep-ph]} \BibitemShut {NoStop}%
\bibitem [{\citenamefont {Meng}\ and\ \citenamefont
  {Chao}(2008)}]{Meng:2008dd}%
  \BibitemOpen
  \bibfield  {author} {\bibinfo {author} {\bibfnamefont {C.}~\bibnamefont
  {Meng}}\ and\ \bibinfo {author} {\bibfnamefont {K.-T.}\ \bibnamefont
  {Chao}},\ }\bibfield  {title} {\bibinfo {title} {{Peak shifts due to
  ${B}^{(*)}\ensuremath{-}{\overline{B}}^{(*)}$ rescattering in
  $\ensuremath{\Upsilon}(5S)$ dipion transitions}},\ }\href
  {https://doi.org/10.1103/PhysRevD.78.034022} {\bibfield  {journal} {\bibinfo
  {journal} {Phys. Rev. D}\ }\textbf {\bibinfo {volume} {78}},\ \bibinfo
  {pages} {034022} (\bibinfo {year} {2008})},\ \Eprint
  {https://arxiv.org/abs/0805.0143} {arXiv:0805.0143 [hep-ph]} \BibitemShut
  {NoStop}%
\bibitem [{\citenamefont {Bai}\ \emph {et~al.}(2022)\citenamefont {Bai},
  \citenamefont {Li}, \citenamefont {Huang}, \citenamefont {Liu},\ and\
  \citenamefont {Matsuki}}]{Bai:2022cfz}%
  \BibitemOpen
  \bibfield  {author} {\bibinfo {author} {\bibfnamefont {Z.-Y.}\ \bibnamefont
  {Bai}}, \bibinfo {author} {\bibfnamefont {Y.-S.}\ \bibnamefont {Li}},
  \bibinfo {author} {\bibfnamefont {Q.}~\bibnamefont {Huang}}, \bibinfo
  {author} {\bibfnamefont {X.}~\bibnamefont {Liu}},\ and\ \bibinfo {author}
  {\bibfnamefont {T.}~\bibnamefont {Matsuki}},\ }\bibfield  {title} {\bibinfo
  {title}
  {{$\mathrm{\ensuremath{\Upsilon}}(10753)\ensuremath{\rightarrow}\mathrm{\ensuremath{\Upsilon}}(\mathrm{n}\mathrm{S}){\ensuremath{\pi}}^{+}{\ensuremath{\pi}}^{\ensuremath{-}}$
  decays induced by hadronic loop mechanism}},\ }\href
  {https://doi.org/10.1103/PhysRevD.105.074007} {\bibfield  {journal} {\bibinfo
   {journal} {Phys. Rev. D}\ }\textbf {\bibinfo {volume} {105}},\ \bibinfo
  {pages} {074007} (\bibinfo {year} {2022})},\ \Eprint
  {https://arxiv.org/abs/2201.12715} {arXiv:2201.12715 [hep-ph]} \BibitemShut
  {NoStop}%
\bibitem [{\citenamefont {Chen}\ \emph {et~al.}(2011)\citenamefont {Chen},
  \citenamefont {He}, \citenamefont {Li},\ and\ \citenamefont
  {Liu}}]{Chen:2011qx}%
  \BibitemOpen
  \bibfield  {author} {\bibinfo {author} {\bibfnamefont {D.-Y.}\ \bibnamefont
  {Chen}}, \bibinfo {author} {\bibfnamefont {J.}~\bibnamefont {He}}, \bibinfo
  {author} {\bibfnamefont {X.-Q.}\ \bibnamefont {Li}},\ and\ \bibinfo {author}
  {\bibfnamefont {X.}~\bibnamefont {Liu}},\ }\bibfield  {title} {\bibinfo
  {title} {{Dipion invariant mass distribution of the anomalous
  $\ensuremath{\Upsilon}(1S){\ensuremath{\pi}}^{+}{\ensuremath{\pi}}^{\ensuremath{-}}$
  and
  $\ensuremath{\Upsilon}(2S){\ensuremath{\pi}}^{+}{\ensuremath{\pi}}^{\ensuremath{-}}$
  production near the peak of $\ensuremath{\Upsilon}(10860)$}},\ }\href
  {https://doi.org/10.1103/PhysRevD.84.074006} {\bibfield  {journal} {\bibinfo
  {journal} {Phys. Rev. D}\ }\textbf {\bibinfo {volume} {84}},\ \bibinfo
  {pages} {074006} (\bibinfo {year} {2011})},\ \Eprint
  {https://arxiv.org/abs/1105.1672} {arXiv:1105.1672 [hep-ph]} \BibitemShut
  {NoStop}%
\bibitem [{\citenamefont {Chen}\ \emph
  {et~al.}(2016{\natexlab{b}})\citenamefont {Chen}, \citenamefont {Liu},\ and\
  \citenamefont {Matsuki}}]{Chen:2015bma}%
  \BibitemOpen
  \bibfield  {author} {\bibinfo {author} {\bibfnamefont {D.-Y.}\ \bibnamefont
  {Chen}}, \bibinfo {author} {\bibfnamefont {X.}~\bibnamefont {Liu}},\ and\
  \bibinfo {author} {\bibfnamefont {T.}~\bibnamefont {Matsuki}},\ }\bibfield
  {title} {\bibinfo {title} {{Search for missing $\ensuremath{\psi}(4S)$ in the
  ${e}^{+}{e}^{\ensuremath{-}}\ensuremath{\rightarrow}{\ensuremath{\pi}}^{+}{\ensuremath{\pi}}^{\ensuremath{-}}\ensuremath{\psi}(2S)$
  process}},\ }\href {https://doi.org/10.1103/PhysRevD.93.034028} {\bibfield
  {journal} {\bibinfo  {journal} {Phys. Rev. D}\ }\textbf {\bibinfo {volume}
  {93}},\ \bibinfo {pages} {034028} (\bibinfo {year} {2016}{\natexlab{b}})},\
  \Eprint {https://arxiv.org/abs/1509.00736} {arXiv:1509.00736 [hep-ph]}
  \BibitemShut {NoStop}%
\bibitem [{\citenamefont {Dias}\ \emph {et~al.}(2025)\citenamefont {Dias},
  \citenamefont {Ji}, \citenamefont {Dong}, \citenamefont {Guo}, \citenamefont
  {Hanhart}, \citenamefont {Mei{\ss}ner}, \citenamefont {Zhang},\ and\
  \citenamefont {Zhang}}]{Dias:2024zfh}%
  \BibitemOpen
  \bibfield  {author} {\bibinfo {author} {\bibfnamefont {J.~M.}\ \bibnamefont
  {Dias}}, \bibinfo {author} {\bibfnamefont {T.}~\bibnamefont {Ji}}, \bibinfo
  {author} {\bibfnamefont {X.-K.}\ \bibnamefont {Dong}}, \bibinfo {author}
  {\bibfnamefont {F.-K.}\ \bibnamefont {Guo}}, \bibinfo {author} {\bibfnamefont
  {C.}~\bibnamefont {Hanhart}}, \bibinfo {author} {\bibfnamefont {U.-G.}\
  \bibnamefont {Mei{\ss}ner}}, \bibinfo {author} {\bibfnamefont
  {Y.}~\bibnamefont {Zhang}},\ and\ \bibinfo {author} {\bibfnamefont {Z.-H.}\
  \bibnamefont {Zhang}},\ }\bibfield  {title} {\bibinfo {title} {{Dispersive
  analysis of the isospin breaking in the $X(3872)\to J/\psi \pi^+\pi^-$ and
  $X(3872)\to J/\psi \pi^+\pi^0\pi^-$ decays}},\ }\href
  {https://doi.org/10.1103/PhysRevD.111.014031} {\bibfield  {journal} {\bibinfo
   {journal} {Phys. Rev. D}\ }\textbf {\bibinfo {volume} {111}},\ \bibinfo
  {pages} {014031} (\bibinfo {year} {2025})},\ \Eprint
  {https://arxiv.org/abs/2409.13245} {arXiv:2409.13245 [hep-ph]} \BibitemShut
  {NoStop}%
\end{thebibliography}%
\end{document}